\documentclass[11pt]{article}
\usepackage{amsmath,young,youngtab,amsfonts,amssymb,epsfig,color,psfrag}
\usepackage{rotating}
\begin{document}
\title{On P vs. NP, Geometric Complexity Theory,
and \\ The Flip I: a high-level view}
\author{
Dedicated to Sri Ramakrishna \\ \\
Ketan D. Mulmuley \footnote{Part of this work was done while
the author was visiting I.I.T. Mumbai}
 \\
The University of Chicago
\\  \\
http://ramakrishnadas.cs.uchicago.edu \\ \\
Technical Report TR-2007-13, Computer Science Department,\\
The University of Chicago \\
September, 2007
}

\maketitle

\newtheorem{prop}{Proposition}[section]
\newtheorem{claim}[prop]{Claim}
\newtheorem{goal}[prop]{Goal}
\newtheorem{theorem}[prop]{Theorem}
\newtheorem{metathesis}[prop]{Metathesis}
\newtheorem{mainpoint}{Main Point}
\newtheorem{hypo}[prop]{Hypothesis}
\newtheorem{guess}[prop]{Guess}
\newtheorem{problem}[prop]{Problem}
\newtheorem{axiom}[prop]{Axiom}
\newtheorem{question}[prop]{Question}
\newtheorem{remark}[prop]{Remark}
\newtheorem{lemma}[prop]{Lemma}
\newtheorem{claimedlemma}[prop]{Claimed Lemma}
\newtheorem{claimedtheorem}[prop]{Claimed Theorem}
\newtheorem{cor}[prop]{Corollary}
\newtheorem{defn}[prop]{Definition}
\newtheorem{ex}[prop]{Example}
\newtheorem{conj}[prop]{Conjecture}
\newtheorem{obs}[prop]{Observation}
\newtheorem{phyp}[prop]{Positivity Hypothesis}
\newcommand{\bitlength}[1]{\langle #1 \rangle}
\newcommand{\ca}[1]{{\cal #1}}
\newcommand{\floor}[1]{{\lfloor #1 \rfloor}}
\newcommand{\ceil}[1]{{\lceil #1 \rceil}}
\newcommand{\gt}[1]{{\langle  #1 |}}
\newcommand{\C}{\mathbb{C}}
\newcommand{\N}{\mathbb{N}}
\newcommand{\R}{\mathbb{R}}
\newcommand{\Z}{\mathbb{Z}}
\newcommand{\frcgc}[5]{\left(\begin{array}{ll} #1 &  \\ #2 & | #4 \\ #3 & | #5
\end{array}\right)}

\newcommand{\cgc}[6]{\left(\begin{array}{ll} #1 ;& \quad #3\\ #2 ; & \quad #4
\end{array}\right| \left. \begin{array}{l} #5 \\ #6 \end{array} \right)}

\newcommand{\wigner}[6]
{\left(\begin{array}{ll} #1 ;& \quad #3\\ #2 ; & \quad #4
\end{array}\right| \left. \begin{array}{l} #5 \\ #6 \end{array} \right)}

\newcommand{\rcgc}[9]{\left(\begin{array}{ll} #1 & \quad #4\\ #2  & \quad #5
\\ #3 &\quad #6
\end{array}\right| \left. \begin{array}{l} #7 \\ #8 \\#9 \end{array} \right)}

\newcommand{\srcgc}[4]{\left(\begin{array}{ll} #1 & \\ #2 & | #4  \\ #3 & |
\end{array}\right)}

\newcommand{\arr}[2]{\left(\begin{array}{l} #1 \\ #2   \end{array} \right)}
\newcommand{\unshuffle}[1]{\langle #1 \rangle}
\newcommand{\ignore}[1]{}
\newcommand{\f}[2]{{\frac {#1} {#2}}}
\newcommand{\tableau}[5]{
\begin{array}{ccc} 
#1 & #2  &#3 \\
#4 & #5 
\end{array}}
\newcommand{\embed}[1]{{#1}^\phi}
\newcommand{\stab}{{\mbox {stab}}}
\newcommand{\perm}{{\mbox {perm}}}
\newcommand{\trace}{{\mbox {trace}}}
\newcommand{\polylog}{{\mbox {polylog}}}
\newcommand{\sign}{{\mbox {sign}}}
\newcommand{\proj}{{\mbox {Proj}}}
\newcommand{\poly}{{\mbox {poly}}}
\newcommand{\std}{{\mbox {std}}}
\newcommand{\m}{\mbox}
\newcommand{\formula}{{\mbox {Formula}}}
\newcommand{\circuit}{{\mbox {Circuit}}}
\newcommand{\sgn}{{\mbox {sgn}}}
\newcommand{\core}{{\mbox {core}}}
\newcommand{\orbit}{{\mbox {orbit}}}
\newcommand{\cycle}{{\mbox {cycle}}}
\newcommand{\ideal}{{\mbox {ideal}}}
\newcommand{\qed}{{\mbox {Q.E.D.}}}
\newcommand{\proof}{\noindent {\em Proof: }}
\newcommand{\weight}{{\mbox {wt}}}
\newcommand{\tab}{{\mbox {Tab}}}
\newcommand{\level}{{\mbox {level}}}
\newcommand{\vol}{{\mbox {vol}}}
\newcommand{\vect}{{\mbox {Vect}}}
\newcommand{\val}{{\mbox {wt}}}
\newcommand{\sym}{{\mbox {Sym}}}
\newcommand{\convex}{{\mbox {convex}}}
\newcommand{\spec}{{\mbox {spec}}}
\newcommand{\strong}{{\mbox {strong}}}
\newcommand{\adm}{{\mbox {Adm}}}
\newcommand{\eval}{{\mbox {eval}}}
\newcommand{\for}{{\quad \mbox {for}\ }}
\newcommand{\Q}{Q}
\newcommand{\mand}{{\quad \mbox {and}\ }}
\newcommand{\invlim}{{\mbox {lim}_\leftarrow}}
\newcommand{\directlim}{{\mbox {lim}_\rightarrow}}
\newcommand{\sformal}{{\cal S}^{\mbox f}}
\newcommand{\vformal}{{\cal V}^{\mbox f}}
\newcommand{\crystal}{\mbox{crystal}}
\newcommand{\conje}{\mbox{\bf Conj}}
\newcommand{\graph}{\mbox{graph}}
\newcommand{\ind}{\mbox{index}}

\newcommand{\rank}{\mbox{rank}}
\newcommand{\id}{\mbox{id}}
\newcommand{\str}{\mbox{string}}
\newcommand{\RSK}{\mbox{RSK}}
\newcommand{\wt}{\mbox{wt}}
\setlength{\unitlength}{.75in}

\begin{abstract} 
Geometric complexity theory (GCT) is an approach to the $P$ vs. $NP$ and
related problems through algebraic geometry and representation theory.
This article gives a high-level exposition 
of the basic plan of GCT based on
the  principle, called the 
{\em flip},  without   assuming
any background in algebraic geometry
or representation theory.
\end{abstract}
\tableofcontents

\section{Introduction} \label{sintro}
Geometric complexity theory (GCT)  is
a plausible
approach to the  $P$ vs $NP$ \cite{cook,karp,levin} and related problems in
complexity theory via algebraic geometry and representation theory. 
The goal of this paper is to give a  high-level  overview of its
basic plan and the underlying principle  called the {\em flip}, 
without assuming any background in algebraic geometry 
or representation theory. A detailed exposition for mathematicians
will appear in  \cite{GCTflip2}. A brief proposal and announcement 
appeared earlier in  cf.\cite{GCTabs}.
The flip has been partially 
implemented in a series of papers \cite{GCT1}-\cite{GCT11}.
This article, followed by \cite{GCTintro},
should provide an introduction to the
overall structure of GCT for computer scientists who wish to get a high-level
picture before going any further. We assume a  few elementary notions of 
algebraic geometry and representation theory in this introduction. 
They are described in full detail in 
Section~\ref{sbasicdefn}, which can be referred to if necessary.
For the readers looking for  a quick overview, the article \cite{GCTAbs},
which gives a nontechnical synopsis of this paper, followed by just this
introduction, which has been written to read as a short paper, should suffice.

In this article, the underlying field of computation is taken be $\C$. 
In \cite{GCT11},
the problems that arise in the context  of the flip over
an algebraically closed field 
of positive characteristic, or a finite field are discussed. The usual
$P\not = NP$ conjecture is over a finite field of which the one over
$\C$ is in a sense  the crux and, being 
also a formal implication \cite{GCT1}, has to be proved first anyway.

\begin{figure}
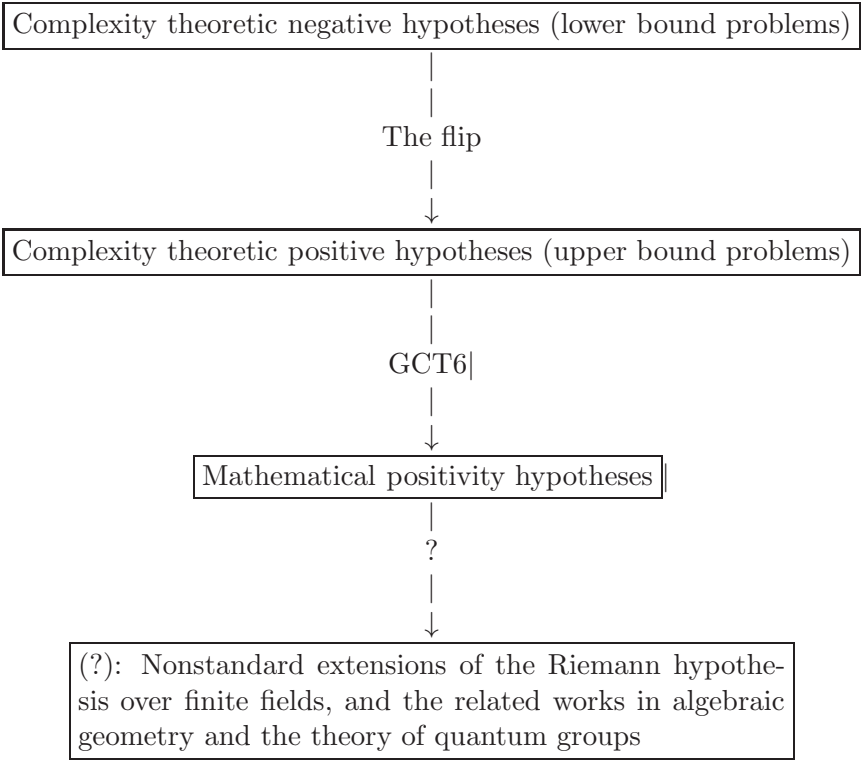
 
\[ \begin{array}{c}
\fbox{Complexity theoretic negative hypotheses (lower bound problems)} \\
 | \\
 | \\
\mbox{The flip} \\
 | \\
\downarrow \\
\fbox{Complexity theoretic positive hypotheses (upper  bound problems)} \\
 | \\
 | \\
\mbox{GCT6}
 | \\
 | \\
\downarrow \\
\fbox{Mathematical positivity hypotheses}
 | \\
 | \\
\mbox{?}\\
 | \\
\downarrow \\
\fbox{\parbox{3.7in}{(?): Nonstandard extensions of the Riemann hypothesis 
over finite fields, and 
the related works in algebraic geometry and the theory of quantum groups}}
\end{array} \]
\caption{The basic plan of GCT}
\label{fbasicintro}
\end{figure}

The flip, in essence,  ``reduces''
the negative hypotheses (lower bound problems) in 
complexity theory,
such as the $P\not =?NP$  conjecture  over $\C$,
to  positive hypotheses in complexity theory
(upper bound problems): specifically, to 
showing  that  a series of decision problems 
in representation theory and
algebraic geometry belong to the complexity class $P$.
The ``reduction'' here  is only ``in essence''. 
It is not a formal Turing machine reduction. If it were, 
it would be relativizable. It is described briefly
in Section~\ref{sreductionintro} below, and in detail in
Section~\ref{sreduction} later.
This reduction basically constitutes 
a flip from {\em hard, nonexistence} to {\em easy existence}.
In \cite{GCT6}, these complexity-theoretic positive hypotheses 
are further reduced to mathematical positivity hypotheses, supported by
the theoretical and experimental evidence therein.
The mathematical positivity hypotheses
roughly say that certain nonnegative 
structural functions in algebraic geometry 
and representation theory have  {\em positive}
formulae--i.e., formulae without alternating signs--akin to 
the usual formula for the permanent (in contrast, the usual formula for
the determinant has alternating signs). 
It turns out that the 
validity of these mathematical positivity hypotheses 
is intimately linked to the Riemann hypothesis over
finite fields--proved in  \cite{weil2} as a culmination of extensive 
effort in mathematics--and the 
related works in algebraic geometry and the theory of 
quantum groups 
\cite{beilinson,kazhdan1,kashiwaraglobal,lusztigcanonical,lusztigbook}.
In \cite{GCT6}, a plan is suggested for proving them via the theory of
quantum groups.  Generalizations of the standard quantum group
 \cite{drinfeld,jimbo,rtf} 
needed for this purpose, which we   call {\em nonstandard 
quantum groups}, are constructed in \cite{GCT4,GCT7}, with further 
conjectural extensions pointed out in \cite{GCT10}. 
All papers of GCT together suggest that if the 
Riemann hypothesis over finite fields and the related works in
the theory of standard quantum groups mentioned 
above can be systematically extended to 
the setting of the nonstandard quantum groups that arise in GCT, then
this may lead to the proof of the $P \not = NP$ conjecture over
$\C$. This basic plan of GCT is summarized in 
Figure~\ref{fbasicintro}. Question marks indicate the main open problems.

\begin{figure}[!h]
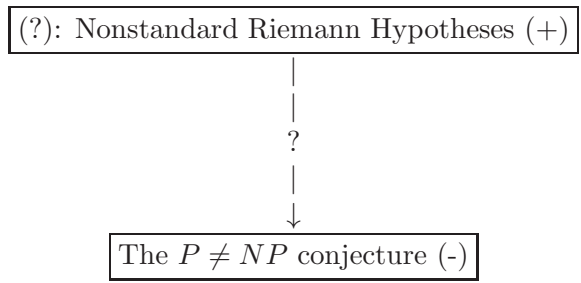

\centering
\[\begin{array} {c} 
\fbox{(?): Nonstandard Riemann Hypotheses (+)} \\
|\\
| \\
?\\
|\\
\downarrow\\
\fbox{The  $P  \not = NP$ conjecture (-)}
\end{array}\]
\caption{The ultimate goal of the  flip} 
\label{fgctultimate}
\end{figure}

The proof in characteristic zero may eventually extend to finite fields,
as in the usual form of the conjecture, along the lines suggested in 
\cite{GCT11}.
Thus the ultimate goal  of the GCT flip is to deduce 
the ultimate negative hypothesis of mathematics,
the $P\not = NP$ conjecture, in essence, from the
ultimate positive hypotheses in mathematics,  (nonstandard) Riemann 
Hypotheses, thereby giving the ultimate flip shown in
Figure~\ref{fgctultimate}.

In the rest of this introduction,
we elaborate Figure~\ref{fbasicintro} further.

\subsection*{Acknowledgement} 
The author is deeply grateful to Madhav Nori, who taught him algebraic 
geometry,  Milind Sohoni, who collaborated in GCT 1-4,
and  Manju the source of energy  behind this work.  
The author is also grateful to A. Razborov for pointing out the need for 
a high-level  account.
This article is essentially an elaboration of the
answers to his questions. 
A part of this work was done while the author was visiting I.I.T. Mumbai
to which the author is grateful for its  hospitality.
It is also a pleasure to thank the graduate students who took the
accompanying introductory course \cite{GCTintro}  on GCT for their feedback.

\subsection{The flip}
We begin with the top arrow in Figure~\ref{fbasicintro}: the flip.
It is motivated by  the 
{\em classical flip}--from the undecidable (negative) to
the decidable (positive)--that occurs in G\"odel's incompleteness theorem.
All known lower bound results--e.g. the 
hierarchy theorems in complexity theory or  the lower bound 
results in the constant depth \cite{sipser}  or the  
PRAM model without bit operations \cite{PRAM}--depend on  flips from 
lower bounds to upper bounds of some sort. But such  variations of the 
classical flip 
cannot work in the context of the $P$ vs. $NP$ problem because they
are either relativizable \cite{solovay} or naturalizable \cite{rudich}.
In contrast, the flip here should be  nonrelativizable
and nonnaturalizable (Section~\ref{snatural}).

There are actually two flips within this flip: (1) from nonexistence to
existence, and (2) from hard to easy. Here hard  means:
the problem of deciding if a  computational circuit of size 
$m$ exists for a given  function $f(x)=f(x_1,\ldots,x_n)$ is hard.
Accordingly, the flip
from hard nonexistence to easy existence goes in two stages.

\subsubsection{From nonexistence to existence}
The flip from nonexistence to existence is addressed in \cite{GCT1,GCT2}.
Here the nonexistence (lower bound) problem is reduced to
an existence problem: specifically,  to the problem of proving existence 
of {\em obstructions}, which serve as ``proofs'' or ``witnesses'' for 
nonexistence of an efficient computational circuit  for the
explicit  hard function in the lower bound problem under consideration.
Just as existence of a forbidden Kurotowoski minor in a graph serves 
as an obstruction, i.e., a ``proof'' for nonexistence of a planar embedding.

An obstruction in \cite{GCT1,GCT2} is intuitively defined as follows.
First a specific (co)-NP-complete function
$E(X)=E(x_1,\ldots,x_n)$, and a specific
$P$-complete function $H(Y)=H(y_1,\ldots,y_l)$ are constructed 
in \cite{GCT1} so as to have  special properties that we shall
describe in a moment. 
Using $H(Y)$,  a projective algebraic variety 
$X_{P}(l)=X_P(H;l)$, for every positive integer $l$,
is associated with the complexity class $P$,
called the {\em class variety} 
associated with  $P$, or the simply the {\em $P$-variety}.
Here, a projective  algebraic variety means the zero set of a system of
homogeneous polynomial equations (cf. Section~\ref{sbasicdefnalg}).
These are generalizations of  the familar 
curves and  surfaces.
It will turn out that 
$X_P(l)$ is a $G$-variety for $G=GL_l(\C)$,
the group of invertible $l\times l$ 
complex matrices. This means elements of $G$ act on this variety as its
transformations--i.e., move its
points around--just as $G$ acts on $\C^l$ in the usual way.
Similarly, using $E(X)$, 
a projective variety $X_{NP}(n,l)=X_{NP}(E;n,l)$, for every positive integer
$n$ and   $l \ge n$,
is associated with the complexity class $NP$.
It is called the {\em class variety} associated with $NP$,
or simply the {\em $NP$-variety}.
It will again be a $G$-variety.
The  functions $E(X)$ and $H(Y)$ have been specially chosen 
so that these class
varieties are  exceptional and  their algebraic geometry can be 
analyzed in depth. 
If $E(X)$ can be computed by a circuit of size $m$,
then it would turn out that 
$X_{NP}(n,l)$ can be embedded in $X_P(l)$ as a $G$-subvariety for $l=O(m^2)$.
 Pictorially:
\begin{equation} \label{eqembasic} 
 X_{NP}(n,l) \hookrightarrow X_{P}(l).
\end{equation}
We want to show that this embedding is impossible if $m=\poly(n)$,
as $n\rightarrow \infty$. 
This would show that $E(X)$ cannot be computed by a circuit of  
$m=\poly(n)$ size, and hence, $P\not = NP$ over $\C$. 

Let $R(n,l)=R(E;n,l)$ and  $S(l)=S(H;l)$ denote the
 homogeneous coordinate rings
of $X_{NP}(E;n,l)$ and $X_P(H;l)$, respectively. Here by
the  coordinate
ring of a variety, we mean the ring of polynomial ``functions'' (of some kind)
on the variety as defined in  Section~\ref{sbasicdefnalg}. 
These are akin to the ring of polynomial functions on $\C^l$.
Since the class varieties are $G$-varieties, these 
homogeneous coordinate rings will be $G$-representations
(Section~\ref{sbasicdefnrepr}). By a $G$-representation we mean a 
vector space on which the  elements
of $G$ act as linear transformations, just as they do on $\C^l$.
If the embedding (\ref{eqembasic})
exists, then it would turn out that $R(n,l)$ is a $G$-subrepresentation of
$S(l)$. We say that an irreducible, i.e., a minimal, nonzero
representation $W$ of
$G$ is an {\em obstruction}, for given $n$ and $l$,
if it occurs as a $G$-subrepresentation of 
$R(n,l)$, but not as a $G$-subrepresentation of $S(l)$.
Existence of such a $W$,  for given $n$ and $l$, implies that 
$R(n,l)$ cannot be embedded as a $G$-subrepresentation of $S(l)$, and
hence, the embedding (\ref{eqembasic}) cannot exist. Thus an obstruction
serves as a ``witness'' or a ``proof'' that
the embedding (\ref{eqembasic}) cannot exist. 

We now reformulate this notion of obstruction using a few 
basic notions in representation theory described in 
Section~\ref{sbasicdefnrepr}.
It is known that (polynomial)
irreducible representations of $G$ are in one-to-one correspondence
with the set of sequences, also called {\em partitions},
$\lambda: \lambda_1 \ge \lambda_2 \cdots \lambda_k>0$ of positive integers
of length $k\le l$.
The irreducible representation of $G$ labelled by $\lambda$ 
is  called a {\em Weyl-module}, and 
is denoted by $V_\lambda(G)$. It is also known 
that each finite
dimensional representation $V$ of $G$ can be written as a direct sum 
of irreducible representations:
\[ V= \bigoplus_\lambda m_\lambda V_\lambda(G),\] 
where $m_\lambda V_\lambda(G)$ denotes the direct sum of $m_\lambda$
copies of $V_\lambda(G)$, and each $m_\lambda$, called the {\em multiplicity}
of $V_\lambda(G)$ in $V$, is uniquely defined. 
Thus $V_\lambda(G)$ occurs in $V$ as a subrepresentation iff
the multiplicity $m_\lambda$ is nonzero.

Let $R(E;n,l)_d$  and $S(H;l)_d$ denote 
the subspaces in $R(E;n,l)$ and $S(H;l)$, respectively,
of forms of degree $d$. 
Let $s_d^\lambda(H;l)$ denote the multiplicity of $V_\lambda(G)$ in 
$S(H;l)_d$.
Let $s_d^\lambda(E;n,l)$ denote the multiplicity of $V_\lambda(G)$
in $R(E;n,l)_d$.
Then $V_\lambda(G)$ is an obstruction for given $n$ and $l$ iff for some $d$ 
$s_d^\lambda(E;n,l)$ is nonzero but $s_d^\lambda(H;l)$ is zero.
Here $d$ is uniquely  determined by the size $\sum_i \lambda_i$ of $\lambda$.
We also say that $V_\lambda(G)$ is an {\em  obstruction of degree $d$}, and
by an abuse of language, also that the label $\lambda$ is an obstruction 
of degree $d$.

The main algebro-geometric result of \cite{GCT2} (Theorem~\ref{tweaksft})
indicates that such obstructions should  exist
in the context of the $P$ vs. $NP$ problem, when $m=\poly(n)$, assuming that
$P\not = NP$, as we expect.
The goal  then is to show that obstructions indeed 
exist, as expected,  for all 
$n\rightarrow \infty$, assuming $m=\poly(n)$.
The story is similar  for  other related lower bound problems.
This addresses the easier half of the flip from
nonexistence to existence.

\subsubsection{From hard to easy}
But how should one prove that obstructions actually exist?
The main hypothesis governing the flip, which addresses this question,
is the following one that constitutes the harder half of the flip:
from hard to easy.

\begin{hypo} {\bf (PHflip1)} \label{hphflipintro}
Consider the $P$ vs. $NP$ problem over $\C$. Let $E(X)$ be 
the explicit function in \cite{GCT1} mentioned above.
Then  the following problems are
``easy''; i.e.,  belong to $P$.
Specifically,

\noindent (a) {\bf Verification of an obstruction}:
given $n$, $l$
and the partition $\lambda$,
whether $V_\lambda(G)$ 
is an obstruction for given $n$ and $l$  can be decided in
$\poly(n,l,\bitlength{\lambda})$ time, where $\bitlength{\lambda}$ 
denotes the bitlength of the specification of $\lambda$.  

\noindent (b) {\bf Explicit construction of  obstructions}:
Suppose $l=n^{\log n}$ (say).
Then, for every $n\rightarrow \infty$, a label  $\lambda(n)$ of an 
obstruction $V_\lambda(G)$ for $n$ and  $l$ 
can be constructed explicitly  in 
$\poly(n,l)$ time, thereby proving existence of an obstruction 
for every such $n$ and $l$.
\end{hypo}

In view of the definition of an obstruction,
the statement (a) for verification clearly follows from:

\begin{hypo}\label{hdecisionintro} {\bf (PHflip2)}
The following 
the decision problems are easy; i.e., belong to $P$. Specifically, 

\noindent (a)
 Given $d,n,l$ and  a partition
 $\lambda$,  whether $s_d^\lambda(E;n,l)$ is nonzero, 
i.e., whether $V_\lambda(G)$  occurs as a $G$-subrepresentation 
of $R(n,l)_d$ can be decided in $\poly(\bitlength{d},\bitlength{\lambda},n,l)$
time.  Here $\bitlength{d}$ denotes the bitlength of $d$.

\noindent (b) 
Given $d,l$ and  a partition $\lambda$, whether $s_d^\lambda(H;l)$ is nonzero,
i.e., whether $V_\lambda(G)$
occurs as a $G$-subrepresentation 
of $S(l)_d$ can be decided in $\poly(\bitlength{d},\bitlength{\lambda},l)$
time.
\end{hypo}

The decision problems in Hypothesis~\ref{hdecisionintro}
are the crux of the matter. Once easy algorithms for these decision 
problems are found, the goal is 
to prove existence of an obstruction for every $n\rightarrow 
\infty$, when $l=n^{\log n}$ (say), by constructing such an obstruction
{\em explicitly}, as per Hypothesis~\ref{hphflipintro} (b). 
We shall discuss how this is to done  in Section~\ref{sreductionintro} below.
Assuming for the moment that this transformation of easy  algorithms
for the decision problems in Hypothesis~\ref{hdecisionintro} 
into an easy procedure for {\em explicit construction of obstructions}
(Hypothesis~\ref{hphflipintro}(b)) for all $n\rightarrow \infty$,
when $l=n^{\log n}$, works,
we get  the ``reduction'' shown in the top arrow of Figure~\ref{fbasicintro}:
from the original hard nonexistence (lower bound)
problem to the basic upper bound problems in Hypothesis~\ref{hdecisionintro}.

\subsection{The $P$-barrier  and its crossing} \label{spbarrierintro}
But, by divine justice, 
the task of showing that the problems in Hypothesis~\ref{hdecisionintro}
are  easy 
turned out to be extremely hard. Thus, paradoxically, the hardest aspect 
of the flip is
just to prove  that the basic decision problems that arise in
the  construction of obstructions are actually {\em easy}; i.e., belong to $P$.
The best algorithms for these decision problems  obtained 
using the general purpose 
algorithms in algebraic geometry and representation theory take
space that is double exponential in $m$ and time that is triple
exponential in $m$. This means even  verification of  an obstruction,  
let alone its discovery, takes time that is triple exponential in $m$
if one were to use the general purpose techniques.

The gap between this triple exponential time bound and the polynomial
time bound sought in Hypothesis~\ref{hdecisionintro} 
is so huge that, at the surface, this hypothesis may seem impossible.
This was the main
barrier, called the $P$-barrier (Section~\ref{seasybarrier}),
on this path towards the $P$ vs $NP$ problem when the 
flip was briefly announced in \cite{GCTabs}.

The article  \cite{GCT6} says that it can be crossed under 
reasonable mathematical assumptions. We now turn to a brief
description of these results.

For that we need a few definitions.

We say that a function
$f(k)$, $k$ a nonnegative integer, 
is a {\em quasi-polynomial}
if for some integer $l\ge 1$ there exist polynomials $f_i(k)$, $1 \le i \le l$,
such that $f(k)=f_i(k)$ if $k=i$ modulo $l$. Here $l$ is called 
the period of the
quasi-polynomial.
An important example of a quasi-polynomial is the {\em Ehrhart
quasi-polynomial}
$f_P(k)$ of a polytope $P$.
By definition, it is the number of integer points in the dilated polytope 
$k P$. This is known to be a quasi-polynomial \cite{stanleyenu}.

We say that a quasi-polynomial $f(k)$ is {\em positive}, if the coefficients 
of all $f_i(k)$ are nonnegative. We say that it is {\em saturated} if either
$f_1(k)$ is  identically zero as a polynomial, or if not, 
$f(1)=f_1(1)\not =0$.
If $f(k)$ is positive, it is clearly saturated.

Next, let us associate with the multiplicities 
$s_d^\lambda(H;l)$   and $s_d^\lambda(E;n,l)$
the following stretching functions:

\begin{equation} 
\tilde s_d^\lambda(H;l)(k) = s_{k d}^{k \lambda}(H;l),
\end{equation}
and 
\begin{equation} 
\tilde s_d^\lambda(E;n,l)(k) = s_{k d}^{k \lambda}(E;n,l).
\end{equation}

The following is the main algebro-geometric result in \cite{GCT6}.

\begin{theorem} (cf. Theorem 3.4.11 in \cite{GCT6}) 
\label{tmainquasipolyintro}

\noindent (Rationality Hypothesis):
Assume that the singularities of the class varieties 
$X_P(H;m)$ and $X_NP(E;n,l)$ are  ``nice'' (rational).

Then  the stretching functions $\tilde s_d^\lambda(H;l)(k)$ and 
$\tilde s_d^\lambda(E;n,l)(k)$ 
are quasi-polynomials.
\end{theorem} 

We do not need to know the 
exact definition of a rational singularity here, which can be found in
\cite{kempf}. It just means that the singularities are nice. This 
depends on the exceptional nature of the class varieties 
(cf. Section~\ref{sclass})
and   is supported by the algebro-geometric results and arguments
in \cite{GCT2,GCT10}.

Using Theorem~\ref{tmainquasipolyintro}, we can now formulate the
conjectural  mathematical 
positivity hypotheses mentioned in the third box from above in 
Figure~\ref{fbasicintro}. Assume the rationality hypothesis above.

\begin{hypo} \label{hph1genintro}
 {\bf (PH1:)} The structural constant $s_d^\lambda(H;l)$
can be expressed as the
number of integer points in a polytope $P_d^\lambda(H;l)$ 
of $\poly(l,\bitlength{d},\bitlength{\lambda})$ dimension,
whose Ehrhart quasi-polynomial coincides with the 
stretching quasi-polynomial $\tilde s_d^\lambda(H;l)(k)$ in 
Theorem~\ref{tmainquasipolyintro}. Furthermore, $P_d^\lambda(H;l)$ 
can be given in the
form of a $\poly(l,\bitlength{d},\bitlength{\lambda})$-time separation oracle 
as in \cite{lovasz}.

There exists a polytope $P_d^\lambda(E;n,l)$ for the structural
constant  $s_d^\lambda(E;n,l)$ with similar properties.
\end{hypo}

This, in particular, implies that $s_d^\lambda(H;l)$ and 
$s_d^\lambda(E;n,l)$  belong to $\#P$.

\begin{hypo} \label{hph2genintro}
{\bf PH2:} The quasi-polynomials $\tilde s_d^\lambda(H;l)$ and
$\tilde s_d^\lambda(E;n,l)$ in Theorem~\ref{tmainquasipolyintro}  are positive.
\end{hypo}

Its weaker form is:

\begin{hypo}  \label{hshgenintro}
 {\bf (SH:)} These quasi-polynomials are  saturated.
\end{hypo}

PH1 and SH (PH2) together say that each decision problem in 
Hypothesis~\ref{hdecisionintro} can be transformed in polynomial time 
into a special kind of an integer programming problem called 
{\em saturated (resp. positive) integer programming problem}
(Section~\ref{sgct6satpgm}).

\begin{theorem} \label{tgct6intro}  (cf. \cite{GCT6}) 
The  decision problems in 
Hypothesis~\ref{hdecisionintro} 
are indeed in $P$, assuming PH1 and SH (or more strongly PH2) above.
\end{theorem}
This follows from a polynomial time algorithm in \cite{GCT6} 
for saturated (positive)
integer programming.

This result reduces the positive 
complexity-theoretic hypotheses in Hypothesis~\ref{hdecisionintro} 
to  the mathematical positivity hypotheses PH1 and SH,
as shown in  the  middle arrow  in Figure~\ref{fbasicintro}. 
The algorithms in Theorem~\ref{tgct6intro} are 
conceptually  extremely simple. They just need 
linear programming \cite{lovasz} and 
computation of Smith normal forms \cite{kannan}.

But their correctness depends on the positivity hypotheses PH1 and SH (PH2),
whose validity, in turn, is intimately linked to deep phenomena in
algebraic geometry and the theory of quantum groups as we shall soon see.
An indication of such a link is already here.
Since the proof of Theorem~\ref{tmainquasipolyintro},
which is necessary to even
formulate these hypotheses,   needs a few fundamental  results
in algebraic geometry; namely, \cite{boutot} (which in turn is based on
\cite{hironaka} and other results), and \cite{kempf,flenner}.
It should not then be surprising if the proofs 
the hypotheses  need far more.
Indeed, the quantum-group-theoretic
and algebro-geometric machinery is needed in GCT essentially
to  prove  these hypotheses, and hence, that these
extremely simple algorithms are actually correct.

\subsection{Why should PH1 and PH2 hold?}
But first, we need to  justify why these hypotheses should hold in the
first place. 
For that, let us consider the simplest analogue of the decision problems
in Hypothesis~\ref{hdecisionintro} in representation theory:

\begin{problem} (Littlewood-Richardson problem) \label{plittleintro}
Given partitions $\alpha,\beta$ and $\lambda$, 
decide 
if the Littlewood-Richardson coefficient $c_{\alpha,\beta}^\lambda$
(cf. Section~\ref{stensor}) is positive (nonzero). 
This is defined to 
be the multiplicity of the irreducible representation $V_\lambda(G)$ 
in the tensor product $V_\alpha(G) \otimes V_\beta(G)$ (which becomes a
$G$-representation by letting the elements of $G$ act on its two factors
simultaneously).
\end{problem}

The analogous mathematical positivity 
hypotheses in this setting are as follows.

Define the  stretching function 
\[ \tilde c_{\alpha,\beta}^\lambda(k) = c_{k\alpha,k\beta}^{k\lambda},
\quad k\ge 0, \] 
which is obtained by stretching 
the  Littlewood-Richardson coefficient  by a factor of $k$. It is
known to be a polynomial \cite{derkesen,kirillov,rassart}. Then

\begin{hypo} \label{htlittleph1}
{\bf (PH1)} The Littlewood-Richardson coefficient 
$c_{\alpha,\beta}^\lambda$ can be 
expressed as the number of integer points in a polytope
 $P=P_{\alpha,\beta}^\lambda$ of dimension 
polynomial in the total length  
of $\alpha,\beta$ and $\lambda$. Furthermore, 
the Ehrhart quasi-polynomial of $P$ coincides with the stretching
polynomial $\tilde c_{\alpha,\beta}^\lambda(k)$ and the membership 
function of $P$ is computable in time that is polynomial
in the bit lengths of $\alpha,\beta$ and $\lambda$.
\end{hypo} 

This is shown, for example, in \cite{berenstein}. 
There are many choices for $P_{\alpha,\beta}^\lambda$. One choice is
called a hive polytope \cite{knutson}. 

\begin{hypo} 
{(\bf PH2)} The coefficients of $\tilde c_{\alpha,\beta}^\lambda(k)$ are
nonnegative.
\end{hypo} 

This implies:
\begin{hypo}  \label{htlittlesh}
{\bf (SH)} The stretching polynomial
$\tilde c_{\alpha,\beta}^\lambda(k)$
is saturated. 
\end{hypo}
Since  $\tilde c_{\alpha,\beta}^\lambda(k)$ is a polynomial, this simply
means if $c_{k\alpha,k \beta}^{k\lambda}$ is nonzero for some $k\ge 1$
then $c_{\alpha,\beta}^{\lambda}$ is also nonzero.
PH2 is still open, but has a considerable experimental evidence in its
support \cite{king}.
That SH holds  is the
saturation theorem in \cite{knutson}. PH1 and SH  in
conjunction with linear programming leads  \cite{loera,GCT3,knutson2}  to a 
polynomial time algorithm for the 
Littlewood-Richardson problem (Problem~\ref{plittleintro}), and 
a polynomial time algorithm 
\cite{GCT5} for a certain generalized 
Littlewood-Richardson problem assuming SH. These results
were indeed a starting motivation for Theorem~\ref{tgct6intro}.

The Littlewood-Richardson coefficient is a special case of a far-reaching 
class of fundamental constants in  representation theory, called
{\em plethysm constants}, described in Section~\ref{sreprdec}.
The structural constants 
$s_d^\lambda(H;l)$ and $s_d^\lambda(E;n,l)$ can be considered to be
``hyped up'' versions of the plethysm constant. 
Considerable theoretical and experimental evidence in support of the analogous
positivity hypotheses PH1 and PH2 for the plethysm constants 
is given in \cite{GCT6}; cf. Section~\ref{sreprdec}. This constitutes the main
evidence in support of PH1 and PH2 for 
$s_d^\lambda(H;l)$, $s_d^\lambda(E;n,l)$ and other similar 
algebro-geometric structural constants that arise in GCT.

\subsection{The reduction} \label{sreductionintro}
Before we turn to the plan suggested in \cite{GCT6} for proving PH1 and
SH, we explain the nature of the 
reduction in the top arrow of Figure~\ref{fbasicintro}.

For this, the easy algorithms in Theorem~\ref{tgct6intro} have 
to be transformed into an easy procedure for explicit 
construction of obstructions
as per Hypothesis~\ref{hphflipintro} (b).
This transformation  cannot be carried out at present since we do not
have explicit descriptions of the polytopes 
$P_d^\lambda(H;l)$ and $P_d^\lambda(E;n,l)$ in PH1. 
But it is explained in Section~\ref{sreduction} and in detail in \cite{GCT6} 
why it should be possible to
carry out this transformation 
if PH1 and SH can be  proved and  explicit descriptions of the polytopes 
therein become
available. The scheme for transformation suggested there goes 
in two steps: 

First, the easy algorithms in Theorem~~\ref{tgct6intro}
have to be used to 
get an easy $\poly(n,l)$ procedure for {\em discovering} an obstruction (label)
for given $n$ and $l$, if one exists.

Second, this easy algorithm for discovering an obstruction,
or rather its structure and the underlying techniques have to be used
to prove that an obstruction always exists for every $n\rightarrow \infty$,
assuming $l=n^{\log n}$, say.  That is, to prove that 
this easy algorithm  always says ``yes'' for such $n$ and $l$.
Just as the structure of 
the easy Hungarian method for discovering
a perfect matching in a bipartite graph can be used to prove
Hall's theorem that every $d$-regular bipartite graph always has a perfect
matching.

This transformation of an easy algorithm for discovery into an
easy (i.e. feasible) 
constructive proof--which we shall call a {\em $P$-constructive
proof}--also gives, as a side product, an easy, i.e., polynomial time
algorithm for explicit construction of obstructions (labels), as in
Hypothesis~\ref{hphflipintro} (b). 
One may wonder why we are going for
explicit construction of obstructions, when just their existence would have
sufficed.  Because the nature of obstructions here is such that
the complexity  deciding their existence and of
constructing them explicitly, if they do, should  be
more or less the same; cf.   Section~\ref{sfromto}.
Just as the complexity
of deciding if a bipartite graph has a perfect matching is more or less the
same as that  of  constructing one, if it exists,

In the context of these transformations it is crucial that 
the  algorithms in Theorem~\ref{tgct6intro}
are not only easy, i.e., polynomial-time
algorithms, but  also have a genuinely
simple structure of the right kind, being just variations
of linear programming.
Of course, we can not hope to use the ellipsoid algorithm for linear
programming--which though simple  is intricate--for
a constructive proof of existence of obstructions.
Rather we have to use the structure
of the underlying polytopes. The analogues of the polytopes 
$P_d^\lambda(H;l)$ and $P_d^\lambda(E;n,l)$ in PH1 in the simplified
setting of the Littlewood-Richardson problem (Problem~\ref{plittleintro})
are called hive polytopes \cite{knutson}. These  have extremely regular 
structure. The same is expected to be the case for the polytopes 
$P_d^\lambda(H;l)$ and $P_d^\lambda(E;n,l)$ that actually arise here.
For this and other reasons given in \cite{GCT6}, it is expected that,
once explicit descriptions of the 
polytopes $P_d^\lambda(H;l)$ and $P_d^\lambda(E;n,l)$ become available,
the algorithms in Theorem~\ref{tgct6intro} can be transformed 
into simple greedy Hungarian-type algorithms which do not even need 
linear programming. This is the main reason why the transformation of
these easy, polynomial time  algorithms 
into an easy (feasible) proof of existence of
obstructions is expected to work in our setting, just as it does 
in the case of Hall's theorem that we mentioned above.

Assuming that this works,
we    would  get an
explicit family $\{\lambda(n)\}$ of obstructions (rather their labels),
as $n\rightarrow \infty$, and $l=n^{\log n}$.
The existence of such an  obstruction family would imply 
that $P\not = NP$ over $\C$. 

\subsection{Towards PH1 and SH via PH0} \label{stowardsph1intro}
Now we turn to the basic plan  suggested  in \cite{GCT6} for proving 
PH1 and SH. This will explain   the bottom arrow in Figure~\ref{fbasicintro}.

This plan is motivated by the proof of PH1 
(Hypothesis~\ref{htlittleph1}) in the 
simplified setting of the Littlewood-Richardson problem 
via the theory of quantum groups \cite{kashiwara1,littelmann,lusztigbook}.
Specifically, it is known that this PH1  is 
a consequence, in a nontrivial way, 
of a deep positivity statement in the theory of {\em standard quantum groups}
\cite{drinfeld,jimbo,rtf}--whose
intuitive description is given later in Section~\ref{sstdquantum}--namely:
their representations and coordinate rings have {\em canonical bases} 
\cite{kashiwara2,lusztigcanonical,lusztigbook},
 whose structural constants determining their representation-theoretic
and multiplicative structure are all nonnegative. 
We shall refer to the existence of a canonical basis with this  positivity
property as PH0, the zeroth positivity hypothesis (property).

Motivated by this work,
certain positivity hypotheses, again called {\em PH0}, are
formulated in \cite{GCT6}, and it is pointed out how and why these may
similarly lead to  the proof 
of the required PH1 and also 
SH (Hypotheses~\ref{hph1genintro} and \ref{hshgenintro}).
The PH0 hypotheses  in \cite{GCT6} may be thought of as
generalizations of  PH0 in  the 
theory of standard quantum groups.
PH1 and SH for Littlewood-Richardson coefficients 
(Hypotheses~\ref{htlittleph1} and
\ref{htlittlesh})
have purely combinatorial proofs \cite{fultonyoung,knutson}, and hence,
PH0 is strictly  speaking not required in this context.
But in the context of the 
PH1 that we are finally interested in  (Hypothesis~\ref{hph1genintro})
the full power of  PH0 seems
needed for the plan in  \cite{GCT8,GCT10} to work.

A natural approach to prove PH0 in \cite{GCT6} in the context of
this PH1 is to somehow 
generalize the proof of PH0 in the theory of the standard quantum group.
But the theory of standard quantum groups does not work, as expected, in
this context. The reason is briefly as follows.

One can associate a complexity class
with  each structural constant that arises in GCT, which we  call
its {\em index class}. 
Roughly, if a structural constant is associated with a  class variety
for a complexity class $C$, then its index class is defined to $C$.
For example, the index  classes of the multiplicities  $s_d^\lambda(H;l)$ and
$s_d^\lambda(E;n,l)$ are $P$ and $NP$ (over $\C$),
 since they are associated with $P$- and $NP$-varieties, respectively.
Similarly,
the index class of the Littlewood-Richardson
coefficient is the class of  circuits (of restricted kinds) of depth two;
cf. Section~\ref{slittlegct6}. The index class of
the {\em Kronecker coefficient} (Section~\ref{stensor}), which is the analogue 
of the Littlewood-Richardson coefficient 
in the representation theory of the symmetric group, is $NC^2$, the 
class of problems that can be solved by circuits of $\log^2 n$ depth and
polynomial size.
The Littlewood-Richardson coefficient as well as the Kronecker 
coefficient are special cases of the 
plethysm constants (Section~\ref{splethysmproblem})
which we mentioned earlier. 
The generalized plethysm constant  is not associated with any class variety,
but it is qualitatively
similar to, though much simpler than $s_d^\lambda(E;n,l)$. 
Hence, we  define its index class to be $NP$, with the understanding that
this is to  be taken  only in a rough sense.
The index classes of the structural constants here are  not be confused 
with their usual computational complexity classes: they are all (conjecturally)
in $\#P$ by PH1.

The standard quantum group is the quantum group that occurs in the context 
of PH1 for Littlewood-Richardson coefficients
(Hypothesis~\ref{htlittleph1}). Hence, we define its
{\em index class}  to be the same as that of Littlewood-Richardson 
coefficients, i.e., the class of circuits of depth two. 
Thus the standard quantum group is the quantum group 
attached to  constant-depth  (depth-two) circuits.

Given a big  difference between the lower bound problems for
constant and nonconstant depth circuits, 
it should not be a surprise if  the standard quantum group cannot  be
used in the context of PH1 for the structural constants that actually arise
in GCT; cf. Section~\ref{sriemann} for an intuitive mathematical explanation
for why this is so.

\subsection{Nonstandard quantum groups} \label{snonstdintro}
What is needed then are quantum groups that can play the role of the
standard quantum group in the context of the decision problems
and  positivity hypotheses for these structural constants.
The main result in this context is the following:

\begin{theorem} \label{tquantumintro}

\noindent \cite{GCT4} There exists a quantum group, which is 
qualitatively similar to the standard quantum group, that can play such
a role in the context of  the Kronecker coefficients.

\noindent \cite{GCT7} More generally,
there exists a (possibly singular) quantum group that can 
play such a role in the context of the  generalized plethysm constants.
\end{theorem}
A less informal statement will be given later (Theorem~\ref{tnonstdquantum}).
A conjectural scheme for generalizing these quantum groups
to the ones that can  play such a role in the context of
$s_d^{\lambda}(E;n,l), s_d^\lambda(H;l)$ and other structural constants  in
GCT is suggested in \cite{GCT10}.
We shall call the new  quantum groups in Theorem~\ref{tquantumintro}
{\em nonstandard}, because,
though they are qualitatively similar to the standard quantum group,
they are also fundamentally different, as expected.

Thus, standard corresponds to constant depth and nonstandard 
to nonconstant depth circuits.

The article \cite{GCT8} gives 
a conjecturally correct algorithm to  construct canonical bases
of the irreducible representations and
 coordinate rings of the nonstandard quantum groups in 
\cite{GCT4,GCT7} with the required positivity properties (PH0).
These are natural
generalizations of the canonical basis due to Kashiwara and Lusztig  
\cite{kashiwara2,lusztigcanonical,lusztigbook} mentioned
above  for the irreducible representations and the coordinate ring of the
standard quantum group. \cite{GCT8}
also gives a conjecturally correct algorithm
to construct canonical bases with similar positivity properties (PH0)  for
the nonstandard deformations of the 
symmtric group algebra that are dually paired with the nonstandard
quantum groups--these generalize the Kazhdan-Lusztig basis \cite{kazhdan} 
 of the Hecke algebra. It is also shown in \cite{GCT7,GCT8} that
PH1 for the plethysm constants follows from PH0 and other conjectural
properties of these nonstandard canonical bases and quantum objects.
The story for the
general constants $s_d^\lambda(E;n,l)$ and $s_d^\lambda(H;l)$ can 
be expected to be similar \cite{GCT10}.

At present we can neither   prove correctness of the algorithms
in \cite{GCT8} for constructing nonstandard canonical bases nor the
required conjectural properties for the reasons that we shall describe in
a moment. But  a considerable evidence is given in \cite{GCT8} in
support of PH0 for the nonstandard quantum group in \cite{GCT4}.
%So if PH0 indeed holds for the nonstandard quantum groups in \cite{GCT4,GCT7},
%then one can expect PH1 for the plethysm constants
%to follow from it just as in the standard case, and 

In the standard case, PH1 follows
from PH0 in a more or less rigid  way 
\cite{dehy,kashiwara1,littelmann,lusztigbook}. This means 
the polytope that occurs in  PH1 for the Littlewood-Richardson coefficient
(Hypothesis~\ref{htlittleph1}) 
is more or less determined by the canonical basis for the standard 
quantum group--not completely, since there are a few choices 
for this polytope; e.g. a hive
polytope in \cite{knutson}, or a polytope in \cite{berenstein}.
But all these 
choices are intimately related. A common feature is that they all have
extremely regular structures. The same can be expected for the polytopes
that should arise in the nonstandard setting. 
This regularity is crucial for the
final transformation of easy algorithms for the basic decision problems
in Hypothesis~\ref{hdecisionintro} into
easy algorithms for  explicit construction of obstructions;
cf. Sections~\ref{sreductionintro} and \ref{sreduction}.

Existence of  nonstandard quantum groups of polylogarithmic \cite{GCT4} 
and superpolynomial \cite{GCT7} depth complexity, the
conjecturally correct algorithm in   \cite{GCT8} for constructing
canonical bases (PH0)  of their coordinate rings and irreducible 
representations,
and the principle that is suggested by
the theory of standard quantum groups--namely, 
once a canonical basis is there (PH0), everything else in the story 
more or less follows a rigid path--is 
the main reason why GCT may be expected to deliver lower bounds 
for circuits of superpolynomial depth and size eventually.

\subsection{Nonstandard Riemann hypotheses?}
But for this plan to work,
PH0 for the nonstandard quantum groups has to be proved.
This brings us  to the  main open question in this story:
how can we prove correctness of the algorithm
 in \cite{GCT8} for constructing the canonical bases (PH0) 
of the coordinate rings of the nonstandard quantum
groups?

There are two constructions of the canonical basis in the standard setting.
An  algebraic construction in \cite{kashiwaraglobal},
where it is called global crystal
basis, and a topological construction in \cite{lusztigcanonical,lusztigbook}.
Both constructions give rise to  the same  basis \cite{gro}.
In fact, both constructions follow the same basic scheme. Only
the proofs of correctness of  this basic scheme are  different.
The topological proof   is based on the theory of
perverse sheaves \cite{beilinson},
which in turn, is based on Riemann hypothesis over
finite fields \cite{weil2}. In essence,
PH0 is thus ultimately deduced in the topological proof
from the Riemann Hypothesis over finite fields, which 
is again a deep positive statement. 
Because its usual statement 
is, after all,  a positive statement, and
it can also be reformulated as stipulating 
positivity (nonnegativity) of some mathematical quantities (cf. page 458 in
\cite{hartshorne}). The topological proof
also  gives, as a side product, the only known proof of 
nonnegativity of the structural constants associated with the 
canonical basis in the standard setting.
Though this nonnegativity is not needed for proving
PH1 for the Littlewood-Richardson coefficients, it is
crucial in the nonstandard setting for the reasons given in \cite{GCT8,GCT10}.

For this reason, 
the  topological  approach seems to be the only viable   option in the
nonstandard setting, as far as we can see.
Besides, the 
algebraic complexity of the nonstandard quantum groups is so huge--as
to be expected in view of the huge gap between constant and nonconstant depth
circuits--that a purely algebraic proof of correctness of the algorithm 
in \cite{GCT8}  for constructing canonical bases  in the nonstandard setting
seems difficult.

But the standard Riemann hypothesis over finite fields and 
the related techniques cannot be expected to work in the
the nonstandard setting for the reasons given in \cite{GCT7,GCT8}.
Again this should not be surprising given the big difference between
constant and nonconstant depth circuits.
Hence what seem to be needed \cite{GCT8} to make the topological approach
work in the nonstandard setting  are 
{\em nonstandard extensions} of the Riemann hypothesis over finite 
fields and the related work on perverse sheaves. By nonstandard, we mean
the extensions that will work in the context of the nonstandard quantum
groups.

\begin{figure}[!p]
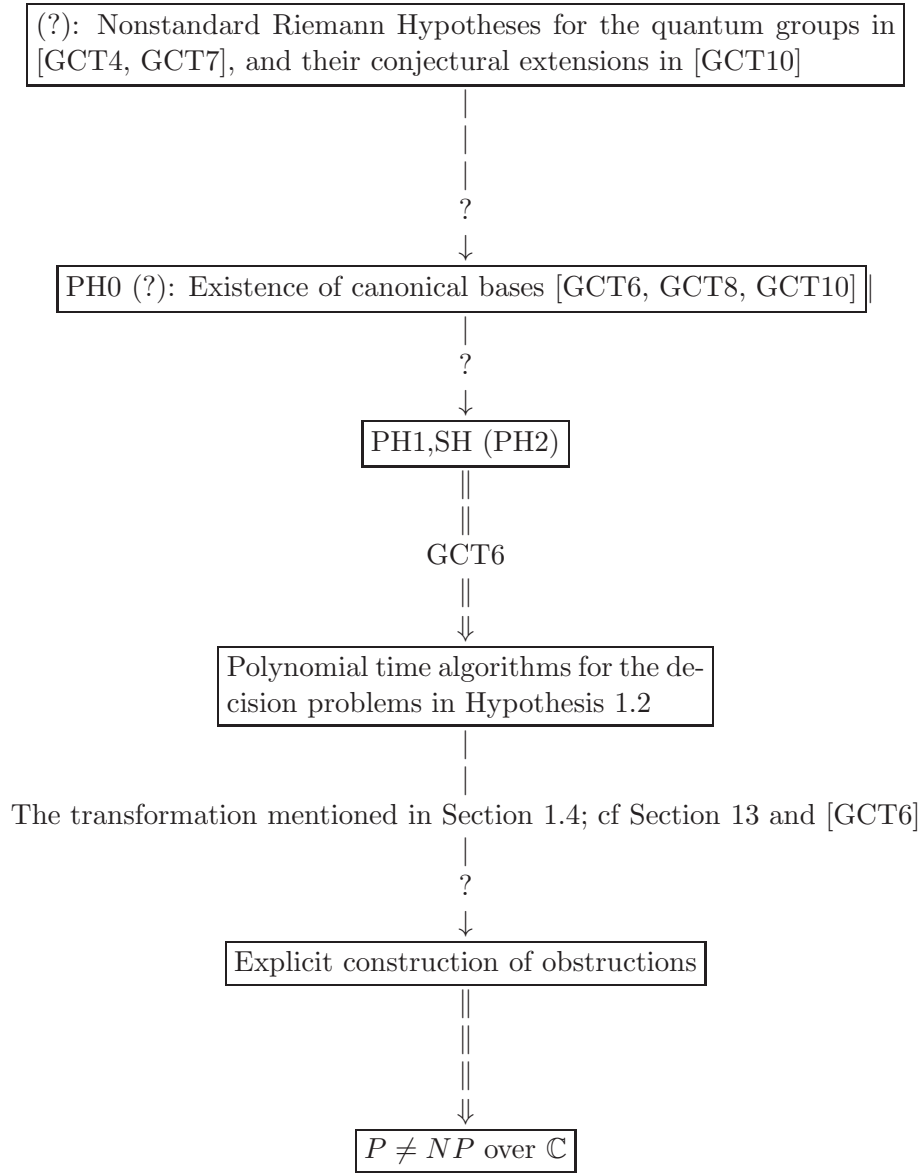

\centering
\[\begin{array} {c} 
\fbox{\parbox{4.5in}{(?): Nonstandard Riemann Hypotheses for the quantum groups in
\cite{GCT4,GCT7}, and their conjectural extensions in \cite{GCT10}}} \\
|\\
| \\
|\\
?\\
\downarrow \\
\fbox{PH0 (?): Existence of canonical bases \cite{GCT6,GCT8,GCT10}}
|\\
|\\
?\\
\downarrow\\
\fbox{PH1,SH (PH2)} \\
\|\\
\|\\
\mbox{GCT6}\\
\|\\
\Downarrow \\
\fbox{\parbox{2.5in}{Polynomial time algorithms for the decision problems in
 Hypothesis~\ref{hdecisionintro}}} \\
|\\
|\\
\mbox{The transformation mentioned in Section~\ref{sreductionintro}; cf
Section~\ref{sreduction} and \cite{GCT6}}\\
|\\
? \\
\downarrow \\
\fbox{Explicit construction of obstructions} \\
\|\\
\|\\
\|\\
\Downarrow \\
\fbox{$P\not = NP$ over $\C$}
\end{array}\]
\caption{The basic plan for implementing the flip in GCT6} 
\label{fgct6}
\end{figure}

The author does  not have the mathematical expertize
to even formulate such  hypotheses, let alone prove them. But 
the theoretical and experimental evidence in \cite{GCT4,GCT7,GCT8}
(cf. Section~\ref{sriemann}) suggests
that such extensions exist, and that they   ought to 
be provable by a systematic extension of the theory of
standard quantum groups  to the  nonstandard setting.
Hence it is reasonable to hope  that the experts would be able to do so
eventually,
leading to the proof of PH0 hypotheses along the topological lines,
and finally, to the 
explicit construction of obstructions as outlined above,
which would then imply that $P \not = NP$ over $\C$.
The whole picture is 
summarized in Figure~\ref{fgct6}, which is an elaboration of the
earlier Figure~\ref{fbasicintro}. The  arrows with question marks are 
conjectural, the double arrows are unconditional.
The ? signs indicate the main open problems at the heart of this approach.
The story over $\C$ may eventually lift to the story over finite fields
along lines suggested in \cite{GCT11}.

\subsection{Obstructions vs. expanders}  \label{sobsvsexpintro}
\begin{figure}[!h]
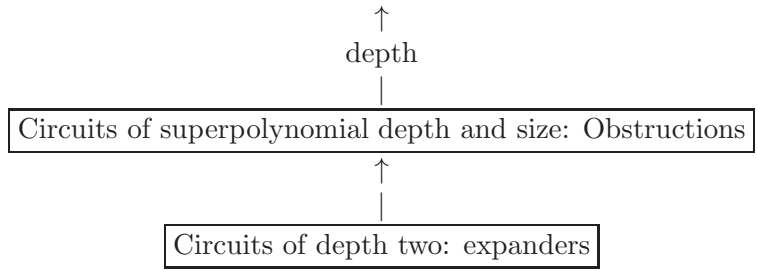

\centering
\[\begin{array} {c} 
\uparrow \\
\mbox{depth}\\
|\\
\fbox{Circuits of superpolynomial depth and size:  Obstructions}\\
\uparrow\\
|\\
\fbox{Circuits of depth two: expanders} \\
\end{array}\]
\caption{The relationship between obstructions and expanders} 
\label{fobsvsexpintro}
\end{figure}

An initial  motivation for  going for
explicit construction of obstructions as in Figure~\ref{fgct6} 
was provided by explicit construction of expanders 
\cite{sarnak,margulis}. 
As explained in Section~\ref{sObsvsExp}, the  obstructions in GCT are in a 
certain sense generalizations of the
expanders 
from constant depth  to 
superpolynomial depth circuits. Specifically,
obstructions are to superpolynomial depth circuits what 
expanders are to constant depth, in fact, depth two circuits;
 cf. Figure~\ref{fobsvsexpintro}.
In view of this  relationship, explicit construction of obstructions as
in Figure~\ref{fgct6}
would be  in the setting of superpolynomial depth circuits what 
explicit construction of expanders is in the setting of constant depth 
circuits. 
As we remarked earlier, the standard quantum group also corresponds to
circuits of depth two. That is, expanders and the standard quantum group 
both 
correspond to  the class of depth-two circuits. Hence it does not 
seem to be a coincidence
that the Riemann hypothesis over finite fields, which enters in the 
theory of the standard quantum group, also enters in the theory of 
expanders \cite{lubot,sarnakbook}.

Existence of expanders can be proved by a simple probabilistic method. 
In contrast, existence of expanders  may not be provable by 
 a probabilistic  method. Indeed, this is roughly the main content
of \cite{rudich},
which says that a nonconstructive method, such as a probabilistic
method,  should not  work 
in the context of the $P$ vs. $NP$ problem under reasonable assumptions.
This is why in  GCT we go for explicit construction of obstructions
in the spirit of explicit construction of expanders. The 
$P/poly$-naturalizability barrier in \cite{rudich} should  not be
applicable to such
explicit, constructive proof techniques. 
This issue is addressed 
in more detail in Section~\ref{snatural}.

\subsection{Is there a simpler proof technique?}
Finally, one may ask
if the $P\not = NP$ conjecture may be proved 
by a substantially simpler proof technique. This seems unlikely for
the following reasons.

The results in complexity theory such as \cite{agrawal,reingold} suggest
that explicit constructions may be  more or less essential  for 
derandomization. In conjunction with 
the hardness vs. randomness principle \cite{kabanets,nisan},
 this suggests that 
explicit constructions may also  be more or less essential for 
(the difficult) lower bound problems as well. Hence, 
the difficulty in any viable proof technique for the
$P\not = NP$ conjecture may  be  intimately linked to the difficulty
(complexity) 
of  the explicit construction of obstructions, i.e., 
``proofs of hardness'' as per that technique. This may  be so
regardless of whether
the technique actually constructs such obstructions explicitly or not.
Because, as per the existence-vs-construction principle \cite{kwig},
the difficulty of deciding  existence may  be
more or less the  same as that of construction in natural problems.
These and other considerations  naturally lead
to a notion of {\em explicit construction complexity} of
an {\em easy-to-verify}
proof technique towards the $P\not = NP$ conjecture, where easy-to-verify
formally means $P$-verifiable; cf. Section~\ref{spveri}.

The explicit construction (depth) complexity of expanders is O(1),
in fact, two, since they
can be constructed by (nonuniform) depth-two algebraic circuits (over a
ring of integers modulo $k$ for some $k$)
\cite{sarnak,margulis}. Whereas, as per Hypothesis~\ref{hphflipintro}, the 
explicit construction (depth) complexity of the obstructions 
in GCT over $\C$ is $\poly(m)$, $m=n^{\log n}$ (say) being 
the circuit size parameter in the lower bound problem; cf. 
Figure~\ref{fobsvsexpintro}.
The arguments in Section~\ref{spveri}
suggest that this may be essentially the  best explicit construction 
complexity that one can expect in any
$P$-verifiable  proof technique 
towards the $P\not = NP$ conjecture. 
In other words, the
massive $\Omega(m)$ 
gap between the explicit construction  complexity
of  obstructions and 
the $O(1)$ explicit construction complexity of
expanders,
as shown in   Figure~\ref{fobsvsexpintro}, may be inevitable
in any $P$-verifiable proof technique towards the $P\not = NP$ conjecture.
If so,  GCT may be among the 
``easiest'' $P$-verifiable  approaches to this conjecture 
as per the explicit construction complexity measure defined here, and hence,
it may be unrealistic to expect a technique 
that is  substantially simpler or easier.

\begin{figure}[!p]
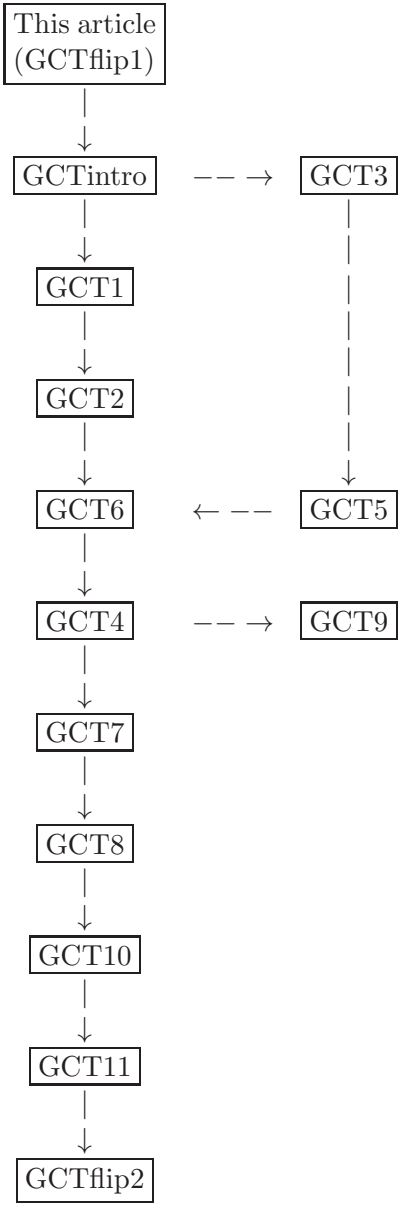

\centering
\[\begin{array} {ccc} 
\fbox{\parbox{.75in}{This article (GCTflip1)}} \\
|\\
\downarrow\\
\fbox{GCTintro} & --\rightarrow & \fbox{GCT3} \\
| & & | \\
\downarrow & & | \\
\fbox{GCT1}& & | \\
|& & | \\
\downarrow & & | \\
\fbox{GCT2} & & | \\
|& & | \\
\downarrow & & \downarrow \\
\fbox{GCT6} & \leftarrow -- & \fbox{GCT5} \\
|\\
\downarrow\\
\fbox{GCT4} & --\rightarrow & \fbox{GCT9} \\
|\\
\downarrow\\
\fbox{GCT7}\\
|\\
\downarrow\\
\fbox{GCT8}\\
|\\
\downarrow\\
\fbox{GCT10}\\
|\\
\downarrow\\
\fbox{GCT11}\\
|\\
\downarrow\\
\fbox{GCTflip2}\\
\end{array}\]
\caption{Logical dependence among the GCT papers} 
\label{ftree}
\end{figure}

In the rest of this article, we elaborate the plan 
in Figure~\ref{fgct6} further and  give a high-level description of
the results in the GCT papers. 
Logical 
dependence  among the GCT papers is shown in Figure~\ref{ftree}.

\subsection{Organization of the paper}
In Section~\ref{sbasicdefn} we recall 
a few basic facts in algebraic
geometry and representation theory which are easy to state and should be
easy to believe. The readers not familar with these fields should be
able to take these on faith. In Section~\ref{sgroupvariety} we describe
a special class of algebraic varieties, called group-theoretic varieties.
All class varieties in GCT are group-theoretic varieties. 
They are described in Section~\ref{sclass}. Obstructions are 
defined in Section~\ref{sobs}. Why they should exist is described  in
Section~\ref{swhyobsex}.
The flip is described in Section~\ref{sflip}. 
The main barrier in the implementation 
of the flip, the $P$-barrier,
is described in Section~\ref{seasybarrier}.
The main result of GCT that crosses this barrier,
assuming the mathematical positivity hypotheses PH1 and SH (PH2), 
is described in
Section~\ref{sgct6}. Why PH1 and PH2 should hold is described in
Section~\ref{swhyph1ph2}. 
Simpler analogues in representation theory 
 of the decision problems in Hypothesis~\ref{hdecisionintro} are described
in Section~\ref{sreprdec}. The $P$-barrier in this context,
its crossing subject to analogous PH1 and SH (PH2),
along with   theoretical results supporting these positivity hypotheses
 are described in
Section~\ref{spbarrepr}.
The nature of the reduction in the top arrow of Figure~\ref{fbasicintro} 
is described in Section~\ref{sreduction}. 
The basic plan in \cite{GCT6} to prove PH1 and SH
via the theory of quantum groups is described next.
The standard quantum group is intuitively 
described in Section~\ref{sstdquantum}.
The nonstandard quantum groups  are intuitively described in
Section~\ref{snonstdquantum}. Why nonstandard Riemann hypotheses should
exist and their role in the theory of 
nonstandard quantum groups  is briefly described in Section~\ref{sriemann}.
The relationship between obstructions and expanders is described  in
Section~\ref{sObsvsExp}.
Why GCT should cross the relativization and the $P/poly$-naturalizability 
barriers is described  in Section~\ref{snatural}. 
Why GCT may be among the easiest $P$-verifiable 
approaches to the $P$ vs. $NP$ problem
as per the explicit-construction-complexity measure
is described in Section~\ref{spveri}.

\section{Basics in algebraic geometry and representation theory}
 \label{sbasicdefn}
In this section we describe the
basic facts in algebraic geometry and representation
theory which are needed in this article and which should be 
easy to believe for the readers not familiar with these fields.
Their proofs  can be found in \cite{fultonrepr,mumford}. 

\subsection{Representation theory} \label{sbasicdefnrepr}
Let $G$ be a group.
We say that a vector space $V$ is a {\em representation}
 of $G$, or a {\em $G$-module},
 if there is 
 a homomorphism
\begin{equation} \label{eqreprmap}
\rho: G \rightarrow GL(V),
\end{equation} 
where $GL(V)$ is  the general linear 
group of invertible transformations of $V$. We denote $\rho(g)(v)$ 
by $g\cdot v$--the result of the action of $g$ on $v$. 
A $G$-subrepresentation $W \subseteq V$ is a subspace that 
is invariant under $G$; i.e., $g \cdot w \in W$ for every $w \in W$.
If $G$ is clear from the context, we just call it subrepresentation.
We say that $V$ is {\em irreducible} if it does not contain a proper 
nontrivial subrepresentation.
A {\em $G$-homomorphism}  from a $G$-module $U$ to a $G$-module $V$ is 
map $\psi: U \rightarrow V$ such that $\psi(g\cdot u)=g \cdot (\psi(u))$
for all $u \in U$.

We say that $G$ is {\em reductive} if 
every finite dimensional 
representation $V$ of $G$ is {\em completely reducible}.
This means it
can be expressed as a direct sum of irreducible 
representations in the form
\begin{equation} \label{eqcompletedecomp}
V=\bigoplus_\lambda {m_\lambda} V_\lambda(G)
\end{equation}
where $\lambda$ 
ranges over all indices (labels) of irreducible representations of $G$,
 $V_\lambda(G)$ denotes the irreducible representation of $G$ with label 
$\lambda$, and
${m_\lambda} V_\lambda(G)$ denotes a direct sum of $m_\lambda$ copies of
$V_\lambda(G)$. Here $m_\lambda$ is called the {\em multiplicity}
 of $V_\lambda(G)$
 in
$V$. It is a basic fact of representation theory that for reductive groups, 
the decomposition (\ref{eqcompletedecomp}) is 
unique; i.e., $m_\lambda$'s are uniquely defined. 
If $m_\lambda>0$, we say that $V_\lambda(G)$ {\em occurs} in $V$.

An example of a nonreductive group is a solvable group that is not abelian.
In this case a subrepresentation $W \subseteq V$ need not have a 
complement $W^\bot$ such that $V=W \oplus W^\bot$.

Every finite group is reductive. Thus
$S_n$, the symmetric group
on $n$ letters, is reductive.
A prime example of a continuous reductive group is the general linear 
group $GL_n(\C)=GL(\C^n)$, the  group of nonsingular $n\times n$ matrices,
and its subgroup the special linear group
$SL_n(\C)=SL(\C^n)$ of matrices with determinant one.
Any product of reductive groups is also reductive.  These are the only kinds 
of reductive groups that we need to know in this article. So whenever
we say reductive, the reader may wish to assume that the group is a
general or special linear group or a symmetric group or a product thereof.

We say that 
the  representation (\ref{eqreprmap}) of $G=GL_n(\C)$ or $SL_n(\C)$
 is polynomial if
for every $g \in G$, every entry in the matrix form  of $\rho(g)$ is
a polynomial in the entries of $g$.

Complete reducibility as in  eq.(\ref{eqcompletedecomp})
means every finite 
dimensional representation of a reductive group is composed of 
irreducible representations. These can be thought of as the building blocks
in the representation theory of reductive groups, and it is important 
to know what  these building blocks are.

\subsubsection{Irreducible representations of $GL_n(\C)$} \label{sweyl}
For $GL_n(\C)$  this was done by 
Weyl in his classic book \cite{weyl}. 
The polynomial irreducible representations of $GL_n(\C)$ are
 in one-to-one correspondence with the  tuples
$\lambda=(\lambda_1,\ldots,\lambda_k)$ of integers, where $k \le n$ and 
$\lambda_1 \ge \lambda_2 \cdots \ge \lambda_k > 0$. 
Here $\lambda$ is called a {\em partition} of length 
 $k$ and size $d=\sum_i \lambda_i$. Its bitlength
 $\bitlength{\lambda}$ is defined to be
the total bitlength of all $\lambda_i$'s.

Thus the polynomial irreducible representations of $GL_n(\C)$ are labelled by 
partitions $\lambda$ of length at most $n$, but any size. 
The irreducible representation corresponding to a partition 
$\lambda=(\lambda_1,\lambda_2,\ldots)$ 
is denoted by $V_\lambda(GL_n(\C))$, and  is
called a {\em Weyl module} of $GL_n(\C)$.
When $GL_n(\C)$  is clear from the context,
we shall denote it by simply $V_\lambda$.

Each partition $\lambda$ corresponds to a Young diagram, which consists of
$k$ rows of boxes, with $\lambda_i$ boxes in the $i$-th row. For example,
the Young diagram corresponding to $(4,2,1)$ is shown below: 

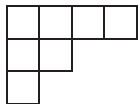
\[ 
\yng(4,2,1)
\]
 When  thinking of a partition,
it is helpful to think of the corresponding  Young diagram.
Thus each Weyl module is labelled by a Young diagram of height at most $n$.
This is a useful combinatorial tool for studying the Weyl modules.

A Weyl module $V_\lambda$ 
is explicitly constructed as follows.
This construction of Deyruts as well as  Weyl's original
construction are given in \cite{fultonrepr}. 
Let $Z$ be an $n\times n$ variable matrix. Let $\C[Z]$ be the ring of
polynomials in the entries of $Z$. 
It is a representation of $GL_n(\C)$. Action of a  matrix  
$\sigma \in GL_n(\C)$ 
on a polynomial  $f \in \C[Z]$ is given by
\begin{equation} 
(\sigma \cdot f)(Z)= f (Z \sigma).
\end{equation}

By a numbering (filling), we 
mean filling of the boxes of a Young diagram 
by numbers in $[n]$; for example:
\[ 
\young(1243,23,1)
\]
We call such a numbering a {\em (semistandard) tableau} if the numbers
are strictly increasing in each column and weakly increasing in all rows;
e.g.
\[ 
\young(1233,23,4)
\]

The partition corresponding to the Young diagram of a numbering
is called the {\em shape} of the numbering.

With every  numbering $T$, we associate a polynomial $e_T \in \C[Z]$,
which is a product of minors for each column of $T$. The $l\times l$ 
minor $e_c$ for a column $c$ of length $l$
is formed by the first $l$ rows of $Z$ and the columns indexed 
by the entries $c_j$, $1 \le j \le l$, of $c$. Thus $e_T=\prod_c e_c$,
where $c$ ranges over all columns in $T$. The Weyl module 
$V_\lambda$ is the subrepresentation of $\C[Z]$  spanned by $e_T$,
where $T$ ranges over all 
numberings of shape $\lambda$ over $[n]$. 
Its one possible basis is given by $\{e_T\}$, where $T$ ranges over 
semistandard tableau of shape 
$\lambda$ over $[n]$.

Let $B \subseteq GL_n(\C)$
be the subgroup of upper triangular matrices. It is called 
the {\em Borel subgroup} of $GL_n(\C)$.
An element  $v_\lambda \in V_\lambda$ is called a
{\em highest weight vector} if it
is an eigenvector for the action of each $b \in B$.
It is easy to show that $V_\lambda$ has a unique highest weight vector,
upto a constant multiple:  it is $e_{T_0}$, where
$T_0$ is the canonical tableau whose $i$-th row contains only $i$'s, for 
each $i$; e.g.
\[ 
\young(1111,22,3)
\]

Let $P \subseteq GL_n(\C)$ be the subgroup of upper block triangular matrices,
where the sizes of the blocks are fixed. For example: 

\[ 
\left[\begin{array} {llllll}
* & * & * & * & * & * \\
* & * & * & * & * & * \\
0 & 0 & * & * & * & * \\
0 & 0 & * & * & * & * \\
0 & 0 & 0 & 0 & * & *  \\
0 & 0 & 0 & 0 & * & *
\end{array}
\right]
\]

Such subgroups are called {\em parabolic}. 
Let $P_\lambda$ be the (projective) stabilizer of the highest weight 
vector $v_\lambda=e_{T_0}$; i.e.,
the set of all $\sigma \in GL_n(\C)$
 such that $\sigma \cdot v_\lambda = c(\sigma)
v_\sigma$, for some complex number $c(\sigma)$.
Then it is easy to show that $P_\lambda$ is parabolic, where the 
sizes of the blocks are completely determined by $\lambda$.

The irreducible representation of $GL_n(\C)$ corresponding to 
the Young diagram that consists of just one column of length $n$ is 
the determinant representation: $g \rightarrow \det(g)$. When restricted 
to the subgroup $SL_n(\C) \subseteq GL_n(\C)$ this becomes trivial. 
More generally, $V_\lambda(G)$ and $V_{\lambda'}(G)$ give the
same representation of $SL_n(\C)$ if $\lambda'$ is obtained from
$\lambda$ by removing columns of length $n$. Hence,
irreducible polynomial representations of $SL_n(\C)$  are in one to
one correspondence with partitions of length less than $n$, and 
are obtained from the ones of $GL_n(\C)$ by restriction.

\subsubsection{Irreducible representations of the symmetric group}
\label{sspecht}
Irreducible representations of $S_n$, called {\em Specht modules}, are in
one-to-one correspondence with the Young diagrams of size $n$,
as opposed to those of length $\le n$ for $GL_n(\C)$.
We  denote the Specht module corresponding to a partition 
$\lambda$ by $S_\lambda$. 
It is explicitly constructed as follows.

Let $\C[X]=\C[x_1,\cdots,x_n]$ 
be the ring polynomials in $n$ variables. 
It is a representation of $S_n$: given $\sigma \in S_n$ and $f \in \C[X]$,
\[ (\sigma \cdot f) (x_1,\cdots,x_n) = 
f(x_{\sigma(1)},\cdots, x_{\sigma(n)}).\]
Given a  numbering $T$ of $\lambda$ with distinct numbers in $[n]$, let 
$f_T$ be the polynomial formed by taking a product of discriminants for 
all columns of $T$. The discriminant for a column with entries $c_i$,
 $1\le i\le l$, is $\prod_{i < i'} (x_{c_i}-x_{c_{i'}})$. 
Then $S_\lambda$ is simply the subrepresentation of $\C[X]$  spanned
by  $f_T$, where $T$ ranges over 
all  numberings of $\lambda$ with distinct entries in $[n]$. 
Its basis is given by $\{f_T\}$,
where $T$ ranges over standard tableau of shape 
$\lambda$ with entries in $[n]$. Here a standard tableau means the 
rows as well as the columns are strictly increasing; e.g.
\[ 
\young(1236,45,7)
\]

\subsubsection{Tensor products} \label{stensor}
If $V$ and $W$ are representations of a group $G$, then their tensor 
product $V\otimes W$ is also a representation: for $\sigma \in G$,
$v\in V, w \in W$,
\[\sigma(v\otimes w)=(\sigma \cdot v) \otimes (\sigma \cdot w). \] 

Given two irreducible representations of a reductive group $G$,
a fundamental problem in representation theory is to find an 
explicit complete 
decomposition of their tensor product in terms of irreducible 
representations of $G$. The following instances of this 
problem are of central importance in GCT.

\subsubsection*{Littlewood-Richardson coefficients}
First we consider this problem when $G=GL_n(\C)$.
Given Weyl modules $V_\alpha$ and $V_\beta$, 
let 
\begin{equation} \label{elittlewood} 
V_\alpha \otimes V_\beta = 
\bigoplus_{\gamma} {c_{\alpha,\beta}^\gamma} V_\gamma,
\end{equation}
be the complete decomposition of  their tensor product into irreducible 
Weyl modules of $G$. Here
the multiplicities 
$c_{\alpha,\beta}^\gamma$ are called Littlewood-Richardson coefficients.
The Littlewood-Richardson rule gives the sought
explicit formula for these multiplicities. It is as follows.

Align the left top corners of the Young diagrams for  $\alpha$ and $\gamma$. 
If the Young diagram for $\alpha$ is not contained in the one for
$\gamma$, then $c_{\alpha,\beta}^\gamma$ is zero.
Otherwise, form a skew shape $\gamma \setminus \alpha$ by removing 
the boxes in $\gamma$ belonging to $\alpha$. A skew tableau with 
content $\beta$ and shape $\gamma \setminus \alpha$ 
is a filling of this skew diagram with $\beta_1$ ones,
$\beta_2$ twos  and so on, such that all columns are strictly increasing
and all rows are weakly increasing. For example, the following is a 
skew tableau of skew shape $(4,3,3,2) \setminus (2,2,1)$: 

\begin{equation} 
\young(\hfil\hfil 11,\hfil\hfil 2,\hfil 23,13)
\end{equation}

We say that a skew tableau is a {\em Littlewood-Richardson tableau} if, when
its entries are read from right to left, top to bottom, the number
of $i$'s read upto any point is at most the number of $(i-1)$'s
read up to that point, for any  $i$. For example, the skew tableau above is a 
Littlewood-Richardson skew tableau. Let $C_{\alpha,\beta}^\gamma$
be the set of
 Littlewood-Richardson tableau of shape $\gamma \setminus \alpha$ 
with content $\beta$. 
The Littlewood-Richardson coefficient $c_{\alpha,\beta}^\gamma$ is
simply the cardinality of  $C_{\alpha,\beta}^\gamma$: i.e., 
\begin{equation} 
  c_{\alpha,\beta}^\gamma = |C_{\alpha,\beta}^\gamma|=
\sum_{T \in  C_{\alpha,\beta}^\gamma} 1.
\end{equation}
Such a formula is called {\em positive}, because it is like the formula for
the permanent which involves only positive signs.
In contrast, there 
are many formulae for such multiplicities, based on the theory of characters
of group representations \cite{fultonrepr}, which involve
alternating signs, like the usual formula for
the determinant.  Positivity here is a deep issue; cf.
 \cite{stanleypos}.

Formally, the Littlewood-Richardson rule 
implies that the  Littlewood-Richardson coefficient belongs to
the complexity class $\#P$, just like the permanent. 
This is the real significance of the Littlewood-Richardson rule from
the complexity-theoretic perspective.
Furthermore, just like the permanent, the Littlewood-Richardson coefficient
is $\#P$-complete \cite{hari}.

\subsubsection*{Kronecker coefficients}
Now we turn to the symmetric group. Since it is reductive, the
tensor product of two Specht modules $S_\alpha$ and $S_\beta$ 
decomposes as:
by: 
\begin{equation} \label{espechtdecomp} 
S_\alpha \otimes S_\beta = 
\bigoplus_{\gamma} {k_{\alpha,\beta}^\gamma} S_\gamma,
\end{equation}
where the multiplicities 
$k_{\alpha,\beta}^\gamma$ are called Kronecker  coefficients.

No  positive rule akin to the Littlewood-Richardson rule
is known for the Kronecker coefficients.
In fact, this is a fundamental open problem in the representation theory
of symmetric groups,
which arose almost with the birth of representation theory in the work
of Frobenius, Schur, Weyl and others in the beginning of the twentieth century;
cf. \cite{macdonald,stanleypos} for its history and significance.
In the  language of complexity theory, the problem is:

\begin{question} 
Does the Kronecker coefficient belong to $\#P$?
\end{question}
Though this is not how it was stated in representation theory.
The answer is conjecturally yes \cite{GCT4}. Indeed,  this is
the main focus of the work in \cite{GCT4,GCT8}: roughly,
\cite{GCT8} says that such a rule exists assuming a  conjecture 
regarding the  nonstandard quantum group defined in \cite{GCT4}. 
This is the first entry point of nonstandard quantum groups in GCT.

\subsection{Algebraic geometry} \label{sbasicdefnalg}
Let $V=\C^n$. 
Let $X=(x_1,\ldots,x_n)$ be the  variable $n$-vector whose 
entries stand for the  coordinates of $V$.
An {\em affine algebraic set}  $Z \subseteq V$  is the
set of zeroes of a collection of polynomials in $\C[X]=\C[x_1,\ldots,x_n]$.
An affine algebraic 
set is called {\em irreducible} if it cannot be expressed as the union of two 
proper affine algebraic subsets. An irreducible affine algebraic subset 
$Z$ of $V$ is 
called an  {\em affine variety}.  
Its {\em ideal} $I(Z) \subseteq \C[X]$ is 
the set of all polynomials that vanish on $Z$, and its  coordinate 
ring $\C[Z]$ is defined to be $\C[X]/I(Z)$. The elements of 
$\C[Z]$ are polynomial functions on $Z$.

Let  $P^{n-1}=P(V)$ be the projective space of lines in 
$V$ through the origin. We  say that $V$ is the {\em affine cone}
 of $P(V)$.
Given a nonzero $v\in V$, we  also denote by $v$
the point in $P(V)$ that corresponds to the line in $V$ passing through $v$ 
and the origin; the meaning should be clear from the context.
The {\em homogeneous coordinate ring} of $P(V)$
is defined to be $\C[X]$. Its elements
are homogeneous functions on $V$, the affine cone of $P(V)$.
A {\em projective algebraic set}  $Y$  in $P(V)$ is the
set of zeroes of a collection of homogeneous forms (polynomials).
The {\em affine cone}
  $\hat Y \subseteq V$ of $Y\subseteq P(V)$ is defined to be 
the union of the  
lines in $V$ corresponding to the points in $Y$. A projective  algebraic 
set is called {\em irreducible} if it cannot be expressed as the union of two 
proper projective 
algebraic subsets. An irreducible projective algebraic subset $Y$ of $P(V)$ is 
called a {\em projective variety}.  
Its {\em ideal} $I(Y)\subseteq \C[X]$ is 
the set  of all homogeneous 
forms that vanish on $Y$, and its {\em homogeneous coordinate 
ring} $R(Y)$ is $\C[X]/I(Y)$. The elements of 
$R(Y)$ are homogeneous functions on the affine cone $\hat Y$ of $Y$.
The degree $d$-component  $R(Y)_d$ of  $R(Y)$ is
the subspace of homogeneous forms of degree $d$.
The {\em Hilbert function}  $h_Y(d)$
of $Y$ is defined to be the dimension of $R(Y)_d$.

By a {\em (Zariski)-open subset}
  of $Y$ we mean the complement of an algebraic 
subset of $Y$. An open subset of a projective variety is also called 
a {\em quasi-projective variety}.

Now suppose  $V$ is a representation of  a reductive group $G$. Then $G$ also 
acts on $P(V)$, since it takes line to a line. Furthermore, $\C[X]$ is 
also a representation  of $G$: given $\sigma \in G$ and
$f \in \C[X]$, we define 
\begin{equation} \label{eqbasicaction}
(\sigma \cdot f)(X)=f(\sigma^{-1} X).
\end{equation}
The variety $Y$ is called a {\em $G$-variety}
 if its ideal $I(Y) \subseteq \C[X]$
is a $G$-subrepresentation of $G$; i.e.,
  $\sigma \cdot f \in I(Y)$ for all $\sigma \in G$,
$f \in I(Y)$.  In this case, the homogeneous coordinate 
ring $R(Y)$  of $Y$ is also a representation of $G$.
Furthermore, given a point $p \in Y$, the point $\sigma(p)$
also belongs to $Y$. In other words, {\em $G$ acts on the variety}
 $Y$ by moving
its points around. 

If $Z$ is a projective subvariety of $Y$, then it is a basic fact
that there exists a 
degree preserving surjection from $R(Y)$ to $R(Z)$; i.e., from 
$R(Y)_d$ to $R(Z)_d$ for every $d$. This surjection is obtained by simply
restricting a polynomial function 
on the affine cone $\hat Y$ to the
subcone $\hat Z \subseteq \hat Y$.
If both $Y$ and $Z$ are $G$-varieties,
then this surjection is a $G$-homomorphism. 
By complete reducibility, it then follows that 
$R(Z)_d$ is a $G$-submodule of $R(Y)_d$ for every $d$. 
Pictorially, 
\begin{equation}  \label{eqgmodulesub}
R(Z)_d \hookrightarrow R(Y)_d,
\end{equation}
for every $d$.

Given a point $v \in P(V)$, let $G v$ denote its $G$-orbit. 
It can be shown that $G v$ is a quasi-projective variety.
Let $G_v=\{ \sigma \ | \ \sigma \cdot v = v\}$ be its stabilizer. 
Then $G v$ as a set is isomorphic to the coset set  $G/H$, $H=G_v$. 
Quasiprojective varieties of the form $G/H$ are called homogeneous
spaces. These have been intensively studied in algebraic geometry.

Let $\Delta_V[v]=\overline{G v} \subseteq P(V)$
denote the  closure of the $G$-orbit of $v$  in the usual complex topology
\footnote{This coincides with the closure in the Zariski-topology
 \cite{mumford}.}. 
We  call such a  variety {\em an orbit closure}.
It can be shown that $\Delta_V[v]$ is a projective $G$-variety. 
One can think of $\Delta_V[v]$ as a closure  of 
the homogeneous space $G/G_v$. Such spaces are called 
almost-homogeneous spaces \cite{akhiezer}. 
These have also been intensively studied.
Let $R_V[v]$ be the homogeneous coordinate ring of $\Delta_V[v]$, and
$R_V[v]_d$ its degree $d$-component.
Since $G$ acts on $\Delta_V[v]$, each $R_V[v]_d$ is a finite dimensional
representation of $G$.

The simplest example of $\Delta_V[v]$ arises as follows.
Let $V_\lambda$ be a Weyl module of $G=GL_n(\C)$. 
Let $v_\lambda \in P(V_\lambda)$ be the point corresponding to the
highest weight vector in $V_\lambda$; we call it the {\em highest weight
point}. Then it can be shown that the orbit $G v_\lambda \cong G/P_\lambda$,
where $P_\lambda$ is the stabilizer of $v_\lambda$, is already
closed. This is called a {\em flag variety}. It has 
been intensively studied in algebraic geometry for over a
century, and its algebraic geometry is now more or less completely 
understood; e.g. see \cite{smt}.

The flag varieties by their very definition are smooth. But
the algebraic geometry of general orbit closures 
can be extremely complicated, and essentially, intractable. 
Because, even if the orbit $G v$ is smooth, its closure 
can be highly singular, and the singularities can be pathological.
Indeed, the moral of the story that can be gained from \cite{lunavust}
 is that 
the algebraic geometry of a general orbit closure is 
{\em essentially hopeless}.

\section{Group-theoretic varieties} \label{sgroupvariety}
Fortunately,  the class varieties that arise
in  GCT are all exceptional kinds of orbit closures,
which we call {\em group-theoretic orbit closures} or 
{\em group-theoretic varieties}. 
The articles  \cite{GCT2,GCT10} together roughly say that 
problems regarding the
algebraic geometry of these group-theoretic class varieties
can be ``reduced'' to problems in
(quantum) group theory. This is what makes them tractable, and this
is how the theory of quantum groups enters in GCT.
In this section, we shall briefly describe a group-theoretic 
variety in an abstract form.

Let $V$ and $G$ be as in Section~\ref{sbasicdefnalg}.
We say that  $v \in P(V)$
 is {\em characterized by its stabilizer} $H =G_v \subseteq G$,
if it is the only point in $P(V)$ stabilized (left invariant) by $H$.
Stabilized by $H$ means,  for every $\sigma \in H$ ,
$\sigma \cdot v =v$.

For example, the highest weight point $v_\lambda \in P(V_\lambda)$ 
(Section~\ref{sbasicdefnrepr}) is characterized by 
its stabilizer $P_\lambda$. This indicates that the points that are 
characterized by their stabilizers are  very special. 

Suppose $v$ is characterized by its stabilizer. Then $v$, and hence,
its orbit closure $\Delta_V[v]$ is 
completely determined  by the group triple: 

\begin{equation} \label{eqtriple}
H=G_v \hookrightarrow G \stackrel{\rho}{\rightarrow} K=GL(V),
\end{equation}
where $\rho$ represents the representation map; cf. (\ref{eqreprmap}). 
This leads to the following:

\begin{defn}  \label{dcharstab}
Assume that $v$ is characterized by its stabilizer, and that the
associated group triple $H\hookrightarrow G \hookrightarrow K$ is explicitly
known. Then we say that the orbit closure $\Delta_V[v]$ is a 
{\em group-theoretic variety}. It is completely determined by the preceding
group triple. We  call $H\hookrightarrow G$ the {\em primary couple}
associated with the orbit closure $\Delta_V[v]$, and 
$G \hookrightarrow K$, the {\em secondary couple}. We also say that
$v$ is the {\em characteristic point} of the group triple 
$H\hookrightarrow G \hookrightarrow K$ and the primary couple 
$H \hookrightarrow G$.

If every point in $V$ is a function on some space
then we say that $v$ is the
{\em characteristic function}
 of the triple $H\hookrightarrow G \hookrightarrow K$ and
the primary couple $H\hookrightarrow G$. (This happens 
if, for example, $V$ is the space of polynomial
functions on an affine $G$-variety $X$, with the  action given by 
(\ref{eqbasicaction})).
\end{defn}

Here explicitly means
the composition factors of $H$ as well as the connecting 
homomorphisms  in (\ref{eqtriple}) 
are specified explicitly; for the details regarding how,
see \cite{GCT6,GCTflip2}. 
All varieties that arise in GCT are either group-theoretic orbit-closures
in the above sense, or their  generalizations, which are 
again essentially determined by  group triples as above, and hence,
will also be called group-theoretic varieties. 
The simplest example of a group-theoretic variety is a flag variety
(Section~\ref{sbasicdefnalg}).

If a variety is group theoretic, then, in principle, we ought to be
able to understand its algebraic geometry if we understand the structure 
of the associated group triple, along with the connecting homomorphisms, 
in depth. We shall elaborate on what in depth means later 
in Section~\ref{snonstdquantum}. Briefly,
 it means understanding the structure of
the group triple at the quantum level.

\section{Class varieties} \label{sclass}
Now we turn to the class varieties associated with 
the complexity classes $NC,P, \#P$ and $NP$ in \cite{GCT1} on which 
the obstructions in GCT  live.
All the class varieties that arise in GCT will be orbit closures of the
following special form.

Let $Y=[y_0,\cdots,y_{l-1}]$ denote a variable  $l$-vector.
For $n<l$, let $X=[y_1,\cdots,y_n]$, and $\bar X=[y_0,\cdots,y_n]$ be its 
 subvectors of size $n$ and $n+1$. We also denote $y_i$, $1\le i \le n$,
by $x_i$.
Let $V=\sym^s(Y)$ be 
the  space of homogeneous forms of degree $s$ in the $l$ variable-entries of
$Y$. It has 
a natural action of 
$G=SL(Y)=SL_l(\C )$ and $\hat G=GL(Y)=GL_l(\C)$, just as in
(\ref{eqbasicaction}).

Similarly, 
let  $W=\sym^{r}(X)$, $r < s$, be the representation of $GL(X)=GL_n(\C)$.
We have a natural embedding $\phi: W \rightarrow V$, which maps 
\begin{equation}\label{eqphibasic} 
w\in W \rightarrow  y^{s-r}w \in V,
\end{equation} 
where $y=y_0$ is used as the
homogenizing variable.  The image $\phi(W)$  is contained in 
$\bar W=\sym^s(\bar X)$, a representation of $GL(\bar X)=GL_{n+1}(\C)$.

The basic recipe for constructing  class varieties is as follows.
Say we want to separate a complexity class $C_1$ from a complexity class 
$C_2 \supseteq C_1$. 

We pick a form $g=g(Y)=g(y_0,\ldots,y_{l-1}) \in P(V)$
 which is a complete function for
the complexity class $C_1$. Then the orbit closure $\Delta_V[g;l]=\Delta_V[g]$
is called the 
{\em class variety} associated with $C_1$, or simply
the $C_1$-variety based on the complete function $g$.
In principle, we can let $g$ be any complete function for the class $C_1$.
But for the algebraic geometry of $\Delta_V[g]$ to be tractable,
we have to choose $g$ so that $\Delta_V[g]$ is group-theoretic; i.e.,
so that
$g$ is characterized by its stabilizer as in Definition~\ref{dcharstab}, or 
in a slightly relaxed sense (cf. Section 7 in \cite{GCT1}), which is 
good enough for our purposes. 

Similarly, we choose a form $h=h(X)=h(x_1,\ldots,x_n) \in P(W)$
which is complete for
the class $C_2$. Then the orbit closure  $\Delta_W[h;n]=\Delta_W[h]
\subseteq P(W)$ is called the {\em base
class variety} associated with $C_2$, or simply the base $C_2$-variety
based on $h$. 
Let $f=\phi(h)$, with $\phi$ as in (\ref{eqphibasic}).
We call the orbit closure 
$\Delta_V[f;n,l]=\Delta_V[f] \subseteq P(V)$ the {\em 
extended class variety} associated with $C_2$, or the extended $C_2$-variety
based on $h$. This extension is necessary so that the 
$C_2$-variety $\Delta_V[f]$ and the $C_1$-variety $\Delta_V[g]$ 
live in the same ambient space $P(V)$. Again, $h$ has to be 
chosen so that it is characterized by its stabilizer (almost) so that
the varieties $\Delta_W[h]$ and $\Delta_V[f]$ are group-theoretic.

Let us suppose  to the contrary that $C_2 \subseteq C_1$.
Then it would turn out
that $f \in \Delta_V[g;l]$, and hence, 
$\Delta_V[f;n,l]$ is a $G$-subvariety of $\Delta_V[g;l]$: 
\[ \Delta_V[f;n,l] \hookrightarrow \Delta_V[g;l].\] 
The goal is to show that such an embedding does not exist when $l$ is
small enough, say, $l=n^{\log n}$, $n\rightarrow \infty$. This will
show that $C_1 \not = C_2$. Here $l$ will be a parameter in the
lower bound problem that depends on the depth and/or the size of the
circuit.

We  now demonstrate  this recipe  in two basic separation problems
in complexity theory.

\subsection{$NC$ vs. $P^{\#P}$}\label{spermvsdet}
Let  $NC$ be the standard class of functions that can be computed 
by circuits of polylogarithmic depth, and $\#P$  the counting
class associated with $NP$. 
The determinant is complete for the class $NC$ and the permanent 
 for the class $\#P$ \cite{valiant}. 
The $P^{\#P} \not = NC$ conjecture over $\C$ \cite{valiant} says that
the permanent cannot be computed by a circuit over $\C$
of polylogarithmic depth.
Using  the determinant and the permanent, we  now construct 
class varieties for $NC$ and $\#P$.
The $P^{\#P} \not = NC$ conjecture over $\C$ 
will then be reduced to 
showing that the extended class variety for $\#P$ 
is not contained in the one for $NC$.

Let $Y$ be an $m\times m$ variable matrix,
which can also be thought of as a variable $l$-vector, $l=m^2$, by linearly 
ordering its entries in any order. Let
$X$ be its, say, the principal bottom-right 
$n\times n$ submatrix, $n<m$, which can also be thought of as a 
variable $k$-vector, $k=n^2$. Let $V=\sym^m(Y)$ be the space of 
homogeneous forms of degree $m$ in the variable entries of $Y$,
and $W = \sym^n(X)$,
the space of 
homogeneous forms of degree $n$ in the variable entries of $X$.
We have a natural action of $G=SL(Y)=SL_l(\C)$ on $V$ and the 
projective space $P(V)$: namely, $\sigma \in G$ maps 
a form $q(Y) \in P(V)$ to $q(\sigma^{-1}Y)$, where we think of 
$Y$ as a variable $l$-vector.
Similarly, we have an action of $H=SL(X)=SL_k(\C)$ on $P(W)$. 

Using   any entry $y$  of $Y$ not in 
$X$ as a 
 homogenizing variable we get an
 embedding 
$\phi: W \rightarrow V$, which maps 
 any $w\in W$ to $y^{m-n}w \in V$. 

Let $g=\det(Y) \in P(V)$ be the determinant form.
Let $\Delta_V[g;l]=\Delta_V[g] \subseteq P(V)$ be its orbit closure.

It is shown in \cite{GCT1} that if  a form $h(X)\in P(W)$ can be 
computed by a  circuit of depth less than  $\log^c n$, then 
 $f=\phi(h)$ lies in 
$\Delta_V[g;l]$ for  $m=2^{\log ^c n}$. Conversely, if $f$ lies in
 $\Delta_V[g;l]$ then $h[X]$ 
can be approximated infinitesimally closely\footnote{This means,
 for every $\epsilon >0$, there exists 
a form $\tilde f \in V$, which has a circuit of depth $O(\log^{2 c} m)$,
such that $||f-\tilde f||< \epsilon$, in the usual norm on $V$.}
 by a circuit of
depth $O(\log^{2 c} m)$. 
 Since, the permanent is $\#P$-complete \cite{valiant},
this is not expected to happen if $h=\perm(X)$ and 
 $m=2^{O(\polylog (n))}$.
 This leads to:

\begin{conj} \cite{GCT1} \label{corbit1}
Let $h=\perm(X)\in P(W)$ and $f=\phi(h)$.
Then $f \not \in \Delta_V[g]=\Delta_V[g;l]$, if $m=2^{O(\polylog (n))}$, as
 $n \rightarrow \infty$.
Since $\Delta_V[g]$ is a $G$-variety, this is 
equivalent to saying that  $\Delta_V[f] \not \subseteq \Delta_V[g]$.
Pictorially:
\[ 
\Delta_V[f] \not \hookrightarrow \Delta_V[g].
\]
\end{conj}

Here $\Delta_V[g]=\Delta_V[g;l]$
is called the {\em class variety} associated with
the class $NC$,
or simply the {\em $NC$-variety} based on the complete determinant function.
We  also denote it by $X_{NC}(g;l)$ or simply $X_{NC}(l)$.
We call
$\Delta_W[h]$ the {\em base class variety} and
$\Delta_V[f]$  the {\em extended class variety}
associated with
the class $\#P$, or simply the {\em base $\#P$-variety} and the
{\em extended $\#P$-variety}, respectively.
We also denote them by $X_{\#P}(h;n)$ and $X_{\#P}(f;n,l)$, or simply,
$X_{\#P}(n)$ and $X_{\#P}(n,l)$.  
The goal (Conjecture~\ref{corbit1}) 
is to show that the   extended class variety for $\#P$ cannot be
contained in the  class variety for $NC$, when $m=2^{\polylog(n)}$: i.e.,
\[ X_{\#P}(n,l) \not \hookrightarrow X_{NC}(l).\] 
This will show that the
permanent cannot be computed by circuits of polylogarithmic depth.

Next we describe why these class varieties are group-theoretic.
For this, we need to show that the determinant and the permanent 
are characterized by their stabilizers.

The stabilizer  of $\det(Y) \in P(V)$ in 
$G = SL(Y)= SL_{m^2}(\C)$ is
 known  to be a reductive subgroup $G_{det}$  which  consists of  
linear transformations in $G$ 
 of the form (thinking of $Y$ as an $m\times m$ matrix):
\begin{equation} \label{eqrepr1}
  Y \rightarrow A Y^{*} B,
\end{equation} 
where $Y^*$ is either $Y$ or $Y^T$, $A,B \in GL_{m}(\C)$.
That the determinant is 
characterized by its stabilizer follows from classical invariant theory
\cite{fultonrepr}.
Hence the $NC$-variety defined here is group-theoretic.
The associated  group triple  is 
\begin{equation}\label{eqtripleNC}
 G_{det} \hookrightarrow G \hookrightarrow GL(V),
\end{equation}
and $G_{det} \hookrightarrow G$  the primary couple.
The embedding $G_{det} \rightarrow G$ almost looks like the natural embedding

\begin{equation} \label{eqdetstab}
GL(\C^m) \times GL(\C^m) \rightarrow  GL(\C^m \otimes \C^m),
\end{equation}
given by: $(g,h)\rightarrow g \otimes h$, where $g\otimes h$ denotes the
Kronecker product. That is, 
\begin{equation} \label{eqkroneckerproduct}
(g \otimes h)\cdot (x\otimes y)=(g \cdot x)
\otimes (h \cdot y). 
\end{equation}

The stabilizer   of $\perm(X) \in P(W)$   in
 $SL(X)=SL_{n^2}(\C)$ 
is a reductive subgroup generated  by linear transformations in $SL(X)$ of the 
form (thinking of $X$ as an $n \times n$ matrix):
\begin{equation}  \label{eqrepr2}
 X \rightarrow \lambda X^* \mu,
\end{equation}
where  $X^*$ is either $X$ or $X^T$, $\lambda$ and $\mu$ are either diagonal or permutation
matrices, and  $n\ge 3$.
It is easy to show that the permanent is also
characterized by its stabilizer.
Hence the base $\#P$-variety  defined in this section is 
group theoretic; the extended $\#P$-variety is also group-theoretic.

\subsection{$P$ vs. $NP$ problem over $\C$} \label{spvsnp}
The class varieties associated with 
the classes $P$ and $NP$ can be constructed in principle  using 
any $P$-complete and  $NP$-complete 
functions. But again it is necessary to choose these functions 
in a special way so that the resulting class  varieties turn out to be
group-theoretic (Section~\ref{sgroupvariety}). 
Such $P$-complete 
and (co)-$NP$-complete functions, called $H(Y)=H(y_1,\ldots,y_l)$
 and $E(X)=E(x_1,\ldots,x_n)$ respectively,
have been constructed in \cite{GCT1}. We do not need to know their 
definitions here.

Let $W=\sym^r(X)$ be the space of forms of degree $r=\deg(E(X))$ in
the entries of $X$. Thus $E(X)\in P(W)$. 
Let $V=\sym^s(Y)$ be the space of forms of degree $s=\deg(H(Y))$ in
the entries of $Y$. Thus $H(Y) \in P(V)$.
We identify $X$ with a suitable subset of $Y$, and define
a map $\phi: P(W) \rightarrow P(V)$ as in (\ref{eqphibasic}) 
by choosing a variable $y$ in
$Y\setminus X$ as a homogenizing variable. 

Now, using the recipe above,
we can  associate with $E(X)$, for every $n$ and
$l \ge n$,
a group-theoretic  variety (orbit closure) 
$\Delta_V[f;n,l]=\Delta_V[f] \subseteq P(V)$,
where $f=\phi(h)$ and $h=E(X)$. It is a $G$-variety, for 
$G=SL_l(\C)$.
It will be called the {\em (extended) class variety} for $NP$ or simply
the  {\em $NP$-variety} based on the form $E(X)$,
and will be denoted by $X_{NP}(E;n,l)$ or simply $X_{NP}(n,l)$.
Similarly, we can associate 
with $H(Y)$ a group-theoretic 
$G$-variety $\Delta_V[g;l]=\Delta_V[g] \subseteq P(V)$, where $g=H(Y)$. It is
called the {\em class variety for $P$} or simply the
{\em $P$-variety} based on the form $H(Y)$, and is denoted by
$X_P(H;l)$ or simply $X_P(l)$.

\begin{remark} \label{ractualvariety}
The actual $P$-variety   $X_P(H;l)$  in the $P$ vs. $NP$ problem 
is not meant to be $\Delta_V[g,l]$, as defined here,
but rather  the variety $\hat \Delta[H(Y)]$
defined in Section 7 of \cite{GCT1}. But we shall ignore that difference
here.
\end{remark}

It can be shown  \cite{GCT1}
that if $E(X)$ is computable by a circuit of 
size $m$ then $X_{NP}(E;n,l)$ can be embedded within $X_{P}(H;l)$
for $l=O(m^2)$:
\begin{equation}\label{eqembed}
X_{NP}(n,l)=X_{NP}(E;n,l) \hookrightarrow X_P(l)=X_{P}(H;l).
\end{equation}

In this context:

\begin{conj}  \cite{GCT1}
This embedding cannot exist if $m=n^{\log n}$, or more generally,
$m=2^{n^a}$, for a small enough $a>0$, 
as $n\rightarrow \infty$.
\end{conj}
This will show that $P\not = NP$ over $\C$.
This transforms the $P$ vs. $NP$ problem over $\C$ into a problem in
geometric invariant theory.

Again, these class varieties are group-theoretic, in a slightly relaxed
sense than defined in Section~\ref{sgroupvariety},
but which is good enough for 
the  purposes of GCT \cite{GCT1}.

\section{Obstructions} \label{sobs}
An obstruction in the $P$ vs. $NP$ problem (characteristic
zero) is defined to be a representation  that lives 
on the extended 
class variety associated with $NP$ but not on the class variety 
associated with $P$. We now elaborate   what this means.

Let $R(n,l)=R(E;n,l)$ and $S(l)=S(H;l)$
denote the homogeneous coordinate rings of $X_{NP}(n,l)=X_{NP}(E;n,l)$
and $X_P(l)=X_P(H;l)$, respectively. 
We  call them the {\em class rings} associated with the complexity
classes $NP$ and $P$. Let $R(n,l)_d$ and $S(l)_d$ denote 
their degree $d$-components, consisting of  homogeneous polynomial
functions of degree $d$. Since $G$ acts on the class varieties, it also
acts on the class rings (see Section~\ref{sbasicdefnalg}).
That is, each $R(n,l)_d$ or
$S(l)_d$ is a finite dimensional representation of $G$.

If the embedding (\ref{eqembed}) exists, then 
$R(n,l)_d$ can be embedded as a $G$-submodule  of $S(l)_d$, for each
$d$; cf. (\ref{eqgmodulesub}):
\begin{equation} \label{eqRSembed}
R(n,l)_d \hookrightarrow S(l)_d.
\end{equation}
In particular, every irreducible representation (Weyl module) 
$V_\lambda=V_\lambda(G)$
 of $G$ that occurs
within $R(n,l)_d$ as a subrepresentation 
also occurs within $S(l)_d$ as a subrepresentation.

\begin{defn} \label{dobs}
We say that $S=V_\lambda$ is an {\em obstruction}, 
for $n,l$ and the pair $(E,H)=(E(X),H(Y))$,
if it occurs in $R(n,l)_d$ but not in $S(l)_d$, for some $d$.

In this case we say that $V_\lambda$ is an obstruction of degree $d$.
We also refer to $\lambda$ as an obstruction of degree $d$.

Obstruction in the setting of the $NC$ vs. $P^{\#P}$ problem over 
$\C$ is defined similarly.
\end{defn} 
This notion of obstruction in \cite{GCT2}
is a refinement of the earlier notion in \cite{GCT1}.

The specification of  an obstruction is given in the form of its
label $\lambda$.
The existence of such an obstruction for  given $n$
and $l$
is a ``proof'' that 
the embedding in  (\ref{eqRSembed}), and hence, the one
in  (\ref{eqembed}) cannot exist.

In this context:

\begin{conj} \cite{GCT2,GCT10} \label{cobsex}
An obstruction for $n,l$ and the pair $(E,H)$
exists  if $m=n^{\log n}$,
or more generally, $m=2^{n^a}$, for
a small enough $a>0$, as $n\rightarrow \infty$; recall that
$l=O(m^2)$.
Furthermore, there exists such an obstruction of a small degree
$d(n,m)=2^{m^b}$, $b>0$ a large enough constant.

Similar conjecture can be made in the context of the $NC$ vs. $P^{\#P}$
problem. In this case, the degree $d(n,m)$ can be $m^b$, $b>0$ a large
enough constant.
\end{conj} 

If such an obstruction $V_{\lambda(n)}$ 
exists for every  $n\rightarrow \infty$, with $m$ 
as above, then it follows that $P\not = NP$ over $\C$.
We   say that $\{V_{\lambda(n)}\}$ or $\{\lambda(n)\}$ is
an {\em obstruction} family for the $P$ vs. $NP$ problem over $\C$.
The goal is to prove existence of such a family.

\section{Why should  obstructions exist?} \label{swhyobsex}
A priori, it is not at all clear why such obstructions should even
exist. In this section, we explain why they should.

An intuitive reason for existence of obstructions is as follows. 
The article \cite{deligne1} roughly says that (algebraic) groups are 
completely determined by their representations. On the other hand,
the group-theoretic class 
varieties are essentially determined by the associated
group triples, and hence, as per the philosophy in \cite{deligne1}, the 
representation-theoretic information associated with these group triples.
Hence,  a ``witness''
for nonexistence of the embedding as in (\ref{eqembed}) ought
to be present in the representation-theoretic information associated 
with the group triples, assuming that $P\not = NP$--which we
take on faith. This is intuitively why a representation-theoretic
obstruction ought  to exist. 
Specifically,
there should exist  a representation-theoretic witness (obstruction)
that explains why one group-theoretic class variety, with associated 
group triple $H_1\hookrightarrow G \rightarrow K$, cannot be
embedded in another group theoretic variety  with  associated group triple
$H_2 \hookrightarrow G \rightarrow K$; in our problem 
$G$ and $K$ in both triples would be the  same.

But why should such a representation-theoretic 
obstruction be specifically of the type as defined here?

To see this, let us first consider a simpler example.
Instead of triples, let us consider couples. Let us say we are given
two couples $\rho_1: H_1 \hookrightarrow G$, and 
$\rho_2: H_2 \hookrightarrow G$, where
$G=GL_l(\C)=GL(W)$, $W=\C^l$.
This means $W$ is a representation of 
$H_1$ and $H_2$. Let us assume that it is an irreducible representation
of $H_1$ and $H_2$, and furthermore, that both $H_1$ and $H_2$ 
are reductive, and that $H_2$ is not a  conjugate of $H_1$.
Now  the coset sets $G/H_1$ and
$G/H_2$ can be given the structure of affine algebraic varieties 
\cite{mumford2}. Since $H_2$ is not a conjugate of $H_1$, 
$G/H_1$ cannot be embedded in $G/H_2$ (and vice versa). 
The goal is to find a representation theoretic obstruction for the
nonexistence of such an  embedding.
We  say that $V_\lambda(G)$ is an {\em obstruction} for this pair of
couples $(\rho_1,\rho_2)$ if it
occurs as a $G$-submodule in the 
coordinate ring of $G/H_1$ but not in the coordinate ring of $G/H_2$.
This is equivalent to saying that
 $V_\lambda(G)$ contains an $H_1$-invariant,
when considered as an $H_1$-module via $\rho_1$, but not an $H_2$-invariant,
when considered as an $H_2$-module via $\rho_2$; this is a consequence
of the Peter-Weyl theorem \cite{springer}. 
This then is an obstruction very similar to the one in 
Definition~\ref{dobs}. Its existence implies that $G/H_1$ cannot be
embedded in $G/H_2$. 
The work \cite{larsen} implies that
such as an obstruction always exists when $H_1$ and $H_2$ are as above.

Conjecture~\ref{cobsex} is a natural generalization of this
well characterized situation. It says that there exists a similar obstruction 
for the embedding among the group-theoretic varieties under consideration.
This, as expected, is a much harder issue. 
The existence of such an obstruction depends crucially on 
the following conjecture concerning 
the algebraic geometry of the class varieties under consideration.

\begin{conj} \label{csftclass}

\noindent (a) (cf. \cite{GCT2}) 
The algebraic geometry of the class variety for $NC$ 
is completely determined by the representation theory of the
associated group triple. Specifically, let 
$\Pi$ be the  set of $G$-submodules of
$\C[V]$ whose duals do not contain a $G_{det}$-invariant; i.e., the
trivial $G_{det}$-module; cf. (\ref{eqtripleNC}). Let $X(\Pi) \subseteq P(V)$
be the zero set of the forms
in the $G$-modules in $\Pi$. Then $X_{NC}=X(\Pi)$.

\noindent (b) (cf. \cite{GCT10}) Analogous, but more complex, statements  hold
for the class varieties associated with the complexity classes $P,NP$ and
$\#P$.
\end{conj}

For  precise statements see \cite{GCT2,GCT10}.

\begin{remark} [Erratum]
In \cite{GCT2} it is conjectured that $X_{NC}=X(\Pi)$ as a scheme 
\cite{hartshorne}. This stronger conjecture may not hold as it is.
Rather, its variant, as would be described in \cite{GCT10}, is expected
to hold.
\end{remark}

Concrete support for this conjecture  is provided by the following 
two results.
The first result is the second fundamental theorem of invariant theory.
It  says
that the analogue of Conjecture~\ref{csftclass} holds for flag varieties and
their generalizations \cite{smt}. Thus Conjecture~\ref{csftclass}
may be thought of as
a natural 
generalization of the second fundamental theorem of invariant theory to 
the group-theoretic class varieties under consideration.
The second result, specific to the 
setting under consideration,  is the following.

\begin{theorem} 
(Theorem 2.11 in \cite{GCT2}) \label{tweaksft}

A weaker form of Conjecture~\ref{csftclass}  holds for the $NC$-variety.
Specifically, there is a dense open neighbourhood $U \subseteq P(V)$ of the
orbit $G g$ of the determinant $g=\det(Y)$ such that 
$X_{NC}\cap U= X(\Pi) \cap U$, assuming a reasonable 
technical condition.
\end{theorem}

The article  \cite{GCT10} gives justifications for and a plan 
to prove Conjecture~\ref{csftclass}.
It is shown in \cite{GCT2} that obstructions as
in Definition~\ref{dobs} indeed exist in the 
context of $NC$ vs. $P^{\#P}$ problem,
for all $n\rightarrow \infty$,
assuming 
\begin{enumerate} 
\item  Conjecture~\ref{csftclass} (a), and
\item that the permanent  cannot be
approximated infinitesimally closely 
by circuits of polylogarithmic depth.
\end{enumerate}
The argument for existence of obstructions in the context of the $P$ vs. 
$NP$ problem based 
Conjecture~\ref{csftclass} (b) is similar \cite{GCT10}.

The first statement here crucially depends on the group-theoretic
nature of the class variety for $NC$.
If in place of  the determinant 
we substitute  other function, this need not hold.
The second statement is 
a slightly strengthened form of the statement that we are 
finally trying to prove: namely, that the permanent  cannot be computed by
 circuits of small depth.
This circular reasoning  tells us why obstructions should exist.
But it gives no help in showing that 
they exist unconditionally.

We turn to this task in the next section.
A remark before we do so.
The existence of obstructions here crucially depends on the exceptional 
nature of $H(Y)$. But we have made no use  so far of the exceptional 
nature of $E(X)$. 
In fact,
obstructions of such kind should exist for any hard (co-NP-complete) function
$h(X)$ in place of $E(X)$.
But the approach for constructing  obstructions
described in the next section  crucially depends on the exceptional
nature of $E(X)$--i.e., on the group-theoretic nature of the 
class variety $X_{NP}(E;n,l)$ for $NP$  based on $E(X)$.

\section{The flip} \label{sflip}
Now we come to the  real problem: how to prove 
the existence of obstructions for the specific $E(X)$ under consideration.
One may wish 
to try a probabilistic strategy for proving existence of 
obstructions: just 
choose a label $\lambda(n)$ of high enough degree randomly, and show that
$V_{\lambda(n)}$  is an obstruction with a good probability. 
But this technique
would not work in the context of the $P$ vs. $NP$ problem because it
is $P/poly$-naturalizable \cite{rudich}. Hence
we  shall go for {\em explicit construction}
of obstructions in the spirit of explicit construction
of expanders \cite{sarnak,margulis,reingold2}. The 
$P/poly$-naturalizability barrier
in \cite{rudich} would  not apply to an approach based on explicit
constructions (Section\ref{snatural}).
This approach is based on 
the following  hypothesis governing   the flip:

\begin{hypo} {\bf (PHflip1)} \label{hphflipformalnew}

The following problems 
belong to $P$. Specifically:

\noindent (a) (Verification):
There exists  a
$\poly(l,n,\bitlength{d},\bitlength{\lambda})$-time
algorithm for deciding, given $l,n,d$ and $\lambda$, if
$V_\lambda$ is an obstruction of degree $d$ for $n,l$ and the 
pair $(E,H)$ (Definition~\ref{dobs}). Here
$\bitlength{d}$ and $\bitlength{\lambda}$ denote the bitlengths 
of $d$ and $\lambda$, respectively.

\noindent (b) (Explicit construction of obstructions): 
Suppose $l=n^{\log n}$, or $2^{n^a}$, for a small enough constant $a>0$.
Then, for every $n\rightarrow \infty$, a label  $\lambda(n)$ of an 
obstruction for $n$ and  $l$ 
can be constructed explicitly  in 
$\poly(n,l)$ time, thereby proving existence of an obstruction 
for every such $n$ and $l$.

\noindent (c) (Discovery of obstructions in general):
There exists a $\poly(l,n)$-time algorithm for deciding 
if there exists  an obstruction 
for $n,l$ and the pair $(E,H)$, and
for  constructing the label of one, if it exists.

Similar hypothesis holds for the $NC$ vs. $P^{\#P}$ problem.
\end{hypo} 

In view of the definition of obstruction (Definition~\ref{dobs}),
The statement (a) for verification follows from the 
following:

\begin{hypo}  {\bf (PHflip2)} \label{hdecision}
\noindent (a) There exists  a
$\poly(l,n,\bitlength{d},\bitlength{\lambda})$-time
algorithm for deciding, given $l,n,d$ and $\lambda$, if
$V_\lambda(G)$ occurs in $R(n,l)_d$.

\noindent (b) 
There exists  a
$\poly(l,\bitlength{d},\bitlength{\lambda})$-time
algorithm for deciding, given $l,d$ and $\lambda$, if
$V_\lambda(G)$ occurs in $S(l)_d$.

Similar hypothesis holds for the $NC$ vs. $P^{\#P}$ problem.
\end{hypo}

As mentioned in Section~\ref{sreductionintro}, once Hypothesis~\ref{hdecision} is proved,
the polynomial time algorithms for the decision problems therein have
to be transformed into a polynomial time algorithm for 
explicit construction of obstructions as in Hypothesis~\ref{hphflipformalnew}
 (b), thereby proving 
Conjecture~\ref{cobsex}, and hence
the lower bound under consideration.
This issue will be addressed in detail in Section~\ref{sreduction} later.

\ignore{Assume that $l=n^{\log n}$ or $2^{n^a}$, for a small enough $a>0$.
Once   Hypothesis~\ref{hphflipformal}  proven, 
an easy algorithm for discovering an obstruction therein
has to be transformed, as described in Section~\ref{sstdintro},
into an easy, i.e., feasible algorithmic proof of existence of
an obstruction for every $n\rightarrow \infty$,
as in Conjecture~\ref{cobsex}, thereby proving that
$P\not = NP$ over $\C$. 
This method would also 
construct such an obstruction $\lambda(n)$ for every $n$
explicitly, if we wish, though that 
is strictly speaking not necessary.
This is what is meant by
the flip, i.e.,  ``reduction'' from hard nonexistence ($P\not = NP$ over
$\C$)  to the 
easy existence problems (Hypothesis~\ref{hphflipformal}).
}

\begin{figure}[!h]
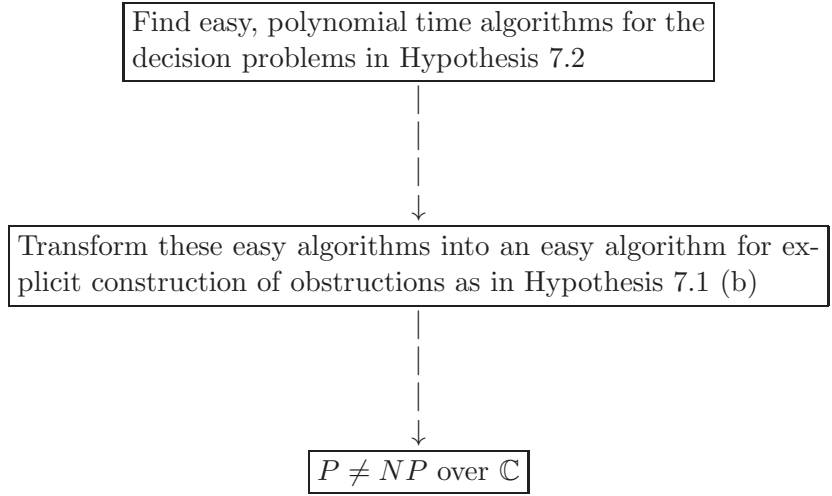

\centering
\[\begin{array} {c} 
\fbox{\parbox{3in}{Find  easy, polynomial time algorithms for 
the decision problems in Hypothesis~\ref{hdecision}}} \\
|\\
| \\
|\\
\downarrow \\
\fbox{\parbox{4.2in}{
Transform these easy algorithms into an easy 
algorithm for explicit construction of obstructions
as in Hypothesis~\ref{hphflipformalnew} (b)}} \\
|\\
|\\
|\\
\downarrow \\
\fbox{$P \not = NP$ over $\C$} \\
\end{array}\]
\caption{The  flip} 
\label{ftheflip}
\end{figure}

The  whole discussion in this section is summarized in Figure~\ref{ftheflip}.

\section{Why should the flip work?: the  $P$-barrier} \label{seasybarrier}
But why should there exist 
easy algorithms as in Hypotheses~\ref{hphflipformalnew} and \ref{hdecision}?
This  turns out to be, paradoxically, the hardest aspect of the
flip: just to prove easiness.
In this section, we elaborate   its nature further.

Clearly,
the function $E(X)$ has to be extremely 
special for Hypotheses~\ref{hphflipformalnew} and \ref{hdecision} to hold.
If, instead of $E(X)$, we  consider a general co-NP-complete function
$h(X)$ then, obstructions  can still be 
expected to exist (cf. Section~\ref{swhyobsex}), but 
Hypotheses~\ref{hphflipformalnew} and \ref{hdecision}
would fail severely, as we  now  explain.

So fix a general integral function $h(X)=h(x_1,\ldots,x_n)$,
which is co-NP-complete, when
considered over $F_2$ by reduction modulo $2$. 
Let $X_{NP}(h;n,l) \subseteq P(V)$
be the class variety associated with it by following
the recipe in  Section~\ref{spvsnp} with $h(X)$ in place of $E(X)$.
Here $V=\sym^s(Y)$ is the space of forms of degree $s=\deg(H(Y))$ 
in $l=O(m^2)$ variable entries of $Y$.
The dimension $M$ of the ambient projective space $P(V)$ here
is exponential in $l=O(m^2)$, $m$ being the  the circuit size.
Using the  currently best  available algorithms for constructing 
a Gr\"obner basis \cite{mayr2}, and for  various problems in invariant
theory \cite{sturmfels}, 
analogues of the decision problems in Hypotheses~\ref{hphflipformalnew} and 
\ref{hdecision} for
$h(X)$ can be solved in at best $O(\dim(\C[V]_d)=O(d^M)=O(d^{2^{\poly(m)}})$
space, where $\C[V]_d$ denotes the degree $d$ component of $\C[V]$,
the homogeneous coordinate ring of $P(V)$. This 
is so even for the decision problems in Hypothesis~\ref{hdecision} and
hence for the verification  problem in Hypothesis~\ref{hphflipformalnew} (a).
This is the
best that we can expect for general $h(X)$ in view of the 
lower bound \cite{mayr} 
for the construction of Gr\"obner bases. In other words, 
for a general $h(X)$ the time taken by a  best procedure to even 
verify if $V_\lambda(G)$, for a given $\lambda$, is an obstruction 
would take space that is double 
exponential in $m$, and hence, time that is triple exponential in $m$.

As we shall argue in  Section~\ref{spveri}, for any approach towards
the $P\not =NP$ conjecture to be viable, at least the problem of
verifying an obstruction (i.e., a ``proof''or ``witness'' 
of hardness as per that approach)
should be easy; i.e., belong to $P$. Intuitively, because
however hard it may be to discover a proof, its verification, once
found, should be easy.
The main $P$-barrier in the course of GCT is  this huge
gap between the triple exponential bound given by
the  currently best  techniques for
a general $h(X)$ and the polynomial bound stipulated for verification in
Hypothesis~\ref{hphflipformalnew} (a) and in
Hypothesis~\ref{hdecision}.

\section{On crossing the $P$-barrier} \label{sgct6}
We now come to the main result of \cite{GCT6}
which crosses this  $P$-barrier under reasonable assumptions.
It gives polynomial-time  algorithms for the decision problems in
Hypothesis~\ref{hdecisionintro}, and hence,  for verifying 
an obstruction (Hypothesis~\ref{hphflipintro} (a)),
assuming the mathematical  positivity hypotheses PH1 and SH
(Hypotheses~\ref{hph1genintro}-\ref{hshgenintro}).

\subsection{A basic prototype  with constant depth complexity}
 \label{slittlegct6}

To motivate these positivity hypotheses, we first 
consider a
basic prototype  of the decision problems in Hypotheses~\ref{hdecision} in
a simplified setting:

\begin{problem} (Littlewood-Richardson problem) \label{plittle}
Given $\alpha,\beta$ and $\lambda$, 
decide 
if the Littlewood-Richardson coefficient $c_{\alpha,\beta}^\lambda$
(cf. Section~\ref{stensor}) is positive
(nonzero). 

Equivalently, consider the diagonal homomorphism:
\begin{equation} \label{eqlittlerho}
  \rho: H=GL_n(\C) \rightarrow G=H\times H.
\end{equation}
Given an irreducible $G$-module $V_{\alpha}(H)\otimes V_\beta(H)$,
decide if an irreducible $H$-module $V_\lambda(H)$ occur in it,
when considered as an $H$-module via the diagonal homomorphism.
\end{problem} 

This  problem 
corresponds  to circuits of depth two in the following sense.
Let $X$ be an $n\times n$ variable matrix.
 Let $V=\sym^1(X)$ 
be the space of linear forms in the entries of $X$. 
We have the action of $G$ on $P(V)$ given by:
\[ ((h1,h2) \cdot f) (X) = f(h_1^{-1} X h_2),\] 
for any $h1,h2 \in H$ and $f \in P(V)$. Let $f(X)=\mbox{\trace}(X)$.
Then the stabilizer of $f$ in $G$ is precisely $H$, and $f$ is 
characterized by its stabilizer. Hence, $f(X)=\trace(X)$
is the characteristic function (Definition~\ref{dcharstab})
of the couple (\ref{eqlittlerho}). It
can be computed by a circuit of depth two. Hence, 
the characteristic class of the couple (\ref{eqlittlerho}) can be
defined to the class of circuits of depth two.
In this sense, the setting of the 
Littlewood-Richardson problem is
roughly dual to the setting of expander graphs (Section~\ref{sobsvsexpintro}),
which too correspond to circuits of depth two. 

In \cite{GCT3,loera,knutson2} it is shown that this problem indeed belongs to
$P$, thereby establishing the analogue of Hypothesis~\ref{hdecision} in this 
 setting.
Two main ingradients in this proof, in addition to linear programming,
are PH1 and SH for Littlewood-Richardson coefficients 
(Hypotheses~\ref{htlittleph1} and \ref{htlittlesh}).
In \cite{GCT5}, it  is shown that the problem of deciding nonvanishing of
a  generalized  Littlewood-Richardson coefficient 
for the classical connected reductive groups other than $GL_n(\C)$,
namely the simplectic and the orthogonal groups, 
also belongs to $P$, assuming the following generalized form of SH
in this context.

Let 
$\tilde c_{\alpha,\beta}^\lambda(k)=c_{k \alpha,k\beta}^{k \lambda}$
be the  stretching function 
for a generalized
Littlewood-Richardson coefficient $c_{\alpha,\beta}^\lambda$,
where $\alpha,\beta$ and $\lambda$ are no longer partitions, but rather
their generalizations \cite{fultonrepr}. It is known to be 
a quasi-polynomial \cite{berenstein,loera}. 

\begin{hypo} \label{htlittlegenph2} 
 {\bf (PH2):} 
The quasi-polynomial $\tilde c_{\alpha,\beta}^\lambda (k)$ is positive.
\end{hypo}

This was  conjectured in \cite{loera}  on the basis of considerable experimental 
evidence. Its weaker form is:

\begin{hypo}  \label{htlittlegensh}
{\bf (SH):} The quasi-polynomial $\tilde c_{\alpha,\beta}^\lambda(k)$
 is saturated.
\end{hypo}

In \cite{GCT5} it is shown that the problem of deciding if a generalized 
Littlewood-Richardson coefficient is nonzero also belongs to $P$ assuming
PH2, or its weaker form, SH.

\subsection{From constant to superpolynomial depth} \label{sgct6superpoly}
The goal now is to lift the polynomial time algorithms and
the mathematical positivity hypotheses PH1 and PH2 above from 
the simplified constant-depth setting to the superpolynomial-depth
setting of Hypotheses~\ref{hphflipformalnew} (a)  and \ref{hdecision}.
This is done in 
\cite{GCT6} in two steps. 
We  only consider the $P$ vs. $NP$ problem,
considerations for the $NC$ vs. $P^{\#P}$ problem being similar.
We use the same notation as in Section~\ref{sobs}. 

The first step is the following mathematical result which
allows formulation of the mathematical hypotheses PH1,PH2, and SH.
Let $s_d^\lambda(H;l)$ and $s_d^\lambda(E;n,l)$
 denote the multiplicities  of the Weyl module $V_\lambda(G)$ 
in $S(H;l)_d$ and $R(E;n,l)_d$, respectively. 
Let us associate with them the following stretching functions:

\begin{equation} 
\tilde s_d^\lambda(H;l)(k) = s_{k d}^{k \lambda}(H;l),
\end{equation}
and 
\begin{equation} \label{eqstretching}
\tilde s_d^\lambda(E;n,l)(k) = s_{k d}^{k \lambda}(E;n,l).
\end{equation}

Then:

\begin{theorem} (cf. Theorem 3.4.11 in \cite{GCT6}) \label{tmainquasipoly}

\noindent (Rationality Hypothesis):
Assume that the singularities of the class varieties 
$X_P(H;m)$ and $X_NP(E;n,l)$ are  rational.

Then  the stretching functions $\tilde s_d^\lambda(H;l)(k)$ and 
$\tilde s_d^\lambda(E;n,l)(k)$ 
are quasi-polynomials.

Similar result also holds in the context of $NC$ vs. $P^{\#P}$ problem.
\end{theorem} 

Rationality (niceness) \cite{kempf} of singularities here
is supported by the algebro-geometric results and arguments
in \cite{GCT2,GCT10}.

The second step is the following complexity-theoretic result:

\begin{theorem} 
 \label{tmaincomplexity}
(cf. Theorems 3.4.11 and  3.4.13 in \cite{GCT6})
The decision problems in Hypothesis~\ref{hdecision},  and hence,
the problem of verifying an obstruction (Hypothesis~\ref{hphflipformalnew}
(a)) are indeed in $P$
assuming the rationality hypothesis above, and
PH1 and PH2 (or  weaker  SH)  in the introduction 
(Hypotheses~\ref{hph1genintro}-\ref{hshgenintro}). 

Similar result also holds in the context of the $NC$ vs. $P^{\#P}$ 
problem, assuming analogous hypotheses PH1, PH2 (or weaker SH) in this
setting.
\end{theorem}

Theorem~\ref{tmaincomplexity}  reduces the complexity-theoretic 
positive hypotheses in Hypothesis~\ref{hdecision}
to the
mathematical positivity hypotheses PH1 and SH (PH2), and
the rationality hypothesis, unconditionally.
Furthermore, \cite{GCT6} also gives theoretical and experimental 
results in support of these positivity hypotheses, and suggests a plan for 
proving them via the theory of quantum groups. We shall discuss this plan
later in Sections~\ref{snonstdquantum}-\ref{sriemann}.

\begin{figure}[!h]
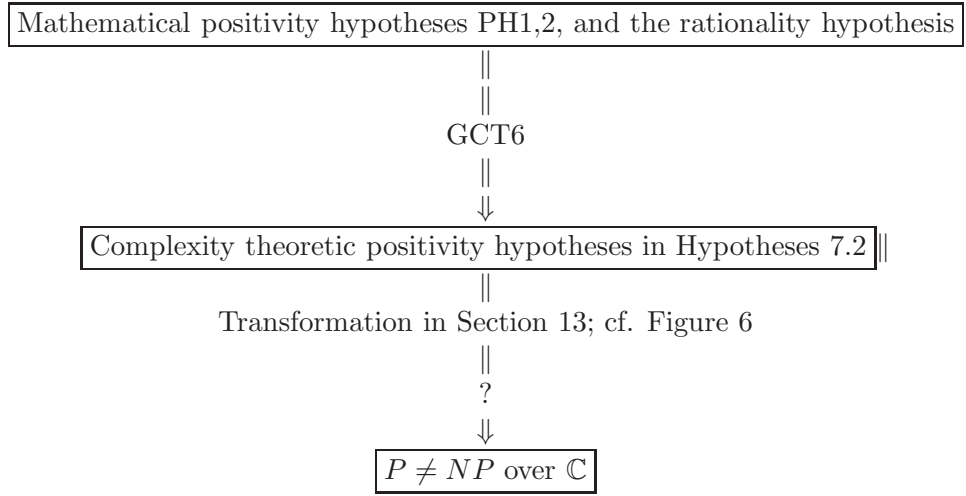

\centering
\[\begin{array} {c} 
\fbox{Mathematical positivity hypotheses PH1,2, and
the rationality hypothesis} \\
\|\\
\| \\
\mbox{GCT6}\\
\|\\
\Downarrow \\
\fbox{Complexity theoretic positivity hypotheses in
Hypotheses~\ref{hdecision}}
\|\\
\|\\
\mbox{Transformation in Section~\ref{sreduction}; cf. Figure~\ref{ftheflip}}\\
\|\\
?\\
\Downarrow\\
\fbox{$P \not = NP$ over $\C$} \\
\end{array}\]
\caption{The  main result of GCT6} 
\label{fGCT6plan}
\end{figure}

The whole discussion of this section is summarized in Figure~\ref{fGCT6plan}.
The top double arrow is unconditional, the bottom arrow is conjectural.

\subsection{Saturated and positive integer programming} \label{sgct6satpgm}
The algorithm in Theorem~\ref{tmaincomplexity} is based on
a polynomial time  algorithm in \cite{GCT6} 
for a restricted form of integer programming, called 
saturated (positive) integer programming. We  briefly 
explain it in this section.

Let $A$ be an $m\times n$ integer matrix, and $b$ an integral $m$-vector.
An integer programming problem asks if the polytope $P: A x \le b$ 
contains an integer point. In general, it is NP-complete.
So let us begin with the well known special case of 
 integer programming which
belongs to $P$. This is the unimodular integer programming problem,
wherein the constraint matrix $A$ is unimodular. This means
the polytope $P$ is 
integral. In this case, $P$ has an integer point iff $P$ is nonempty. The
latter can be checked in polynomial time by standard linear programming 
methods. 

Saturated (positive) integer programming 
is a generalization of unimodular integer programming, wherein 
a variant of linear programming still works, even when $P$ is nonintegral,
provided $P$ satisfies certain saturation or positivity hypothesis,
which make up for the loss of unimodularity.

It is defined as follows.
Let $f_P(n)$ be the 
Ehrhart quasi-polynomial of $P$ \cite{stanleyenu}. 
An integer programming problem 
is called {\em saturated} 
if the Ehrhart quasi-polynomial $f_P(n)$
is guaranteed to be saturated (cf. Section~\ref{spbarrierintro}), if $P$ is nonempty.
It is called {\em positive}
if $f_P(n)$
is guaranteed to be {\em positive} (cf. Section~\ref{spbarrierintro}),
 if $P$ is nonempty.
We allow $m$, the number of constraints, to be exponential in $n$. Hence, 
we cannot assume that $A$ and $b$ are explicitly specified. Rather, 
it is assumed that the polytope $P$ is specified  in the
 form of a (polynomial-time) 
separation oracle as in
\cite{lovasz}.
Given a point $x \in \R^n$, the separation oracle tells if
$x \in P$, and if not, gives a hyperplane that separates $x$ from $P$.

The following is the main complexity-theoretic result in \cite{GCT6}.

\begin{theorem} \label{tgct6sat} 
A saturated, and hence positive, 
integer programming problem  has an oracle-polynomial-time algorithm.
\end{theorem} 

Furthermore, this polynomial time algorithm is conceptually extremely simple.
It is essentially a variant
of linear programming: it uses a generalization of the  ellipsoid method
\cite{khachian}
for  linear programming  in \cite{lovasz},
and a polynomial time algorithm for computing Smith normal forms in
\cite{kannan}. 
Thus the saturated and positive integer programming 
paradigm, in essence, says that
linear programming works for integer programming provided 
the  saturation or the positivity property  holds.

Theorem~\ref{tmaincomplexity}
follows from Theorem~\ref{tmainquasipoly} because 
PH1 and SH (PH2) (cf. Hypotheses~\ref{hph1genintro}-\ref{hshgenintro}) imply that 
the decision problems in Hypothesis~\ref{hdecision} can be transformed in
polynomial time into saturated (positive) integer programming problems.
Thus, in essence,  a variant of linear programming 
works for the decision problems in Hypothesis~\ref{hdecision}, provided 
PH1 and SH (PH2) hold.

But these saturation and positivity hypotheses (PH1 and SH) 
are nontrivial, and,
as we shall see in Sections~\ref{sstdquantum} to \ref{sriemann},
their validity  intimately seems to depend on deep phenomena in
algebraic geometry and the theory of quantum groups. 
We can already see an indication of this here.
For example, even to state PH1, SH or PH2,
we need to show that
the stretching functions used in their statements are quasi-polynomials,
as shown in Theorem~\ref{tmainquasipoly}. Without it,
PH1, SH  and PH2 are meaningless.
But the proof of Theorem~\ref{tmainquasipoly} already depends on 
nontrivial machinery
in algebraic geometry; e.g. the cohomology vanishing result in \cite{kempf},
and
the result in \cite{boutot}, which, in turn,  needs 
resolution of singularities in characteristic zero \cite{hironaka} and
other  cohomology vanishing  results. Hence it should not be 
surprising if proving  these positivity hypotheses 
needs  far more. We shall describe the basic 
plan  in \cite{GCT6} for proving them later
(Sections~\ref{snonstdquantum}-\ref{sriemann}). 

\section{Why should  PH1 and PH2 hold?} \label{swhyph1ph2}
But, first, we have to explain 
why PH1 and PH2 should hold in the
first place.
This depends, as mentioned earlier, on the exceptional nature of $H(Y)$ and
$E(X)$. Specifically, on the fact that the associated class varieties
$X_P(H;l)$ and $X_{NP}(E;n,l)$ are
group-theoretic. We  now elaborate on this.

First, let us consider the  analogue of the decision problem 
in Hypothesis~\ref{hdecision}
for the simplest group-theoretic variety,
namely, a flag variety (Section~\ref{sbasicdefnalg}).
Given a flag variety $Z = G v_\mu\subseteq P(V_\mu)$,
where $V_\mu$ is a Weyl module of $G=SL_l(\C)$, the decision problem 
is to decide if $V_\lambda(G)$ occurs in $R(Z)_d$, the degree $d$ 
component of the homogeneous coordinate ring of $Z$. By the Borel-Weil
theorem \cite{fultonrepr},
$R(Z)_d=V_{d \mu}^*$, the dual of $V_{d\mu}$.
Hence, $V_\lambda$ occurs in
$R(Z)_d$ iff $V_\lambda=V_{d \mu}^*$. It is easy to show that this is so
 iff the Young diagram 
for $\lambda$ is obtained by flipping the  complement
of  the Young diagram for
$d \mu$ in the smallest rectangle containing it. This  can be  decided in
$\poly(\bitlength{d},\bitlength{\lambda},\bitlength{\mu})$ time. 
The analogues of PH1 and PH2 in this setting  clearly hold,
since the multiplicity of $V_\lambda$ in $R(Z)_d$ is just $0$ or $1$.

Now let us move to 
a general group-theoretic class variety. Let 
$(H\hookrightarrow G \hookrightarrow K)$ be the associated group triple. 
Since the class variety in question is (essentially) determined by this
triple, all questions concerning the  variety should, in principle, be
reducible to representation-theoretic questions regarding this triple;
cf. \cite{GCT10}, and  Sections~\ref{sgroupvariety} and
\ref{snonstdquantum}.

In \cite{GCT6} and \cite{GCT10}  analogues of the
decision problems in Hypothesis~\ref{hdecision}  for the couples 
$H\hookrightarrow G$ and $G \hookrightarrow K$ are formulated.
Furthermore,
theoretical and experimental evidence for PH1 and PH2 for the
decision problems associated with these couples is provided. Since the
triples are qualitatively similar to the couples,  though much harder,
this provides the main evidence in support of PH1 and PH2 for the
class varieties under consideration. We shall turn to this evidence 
in the next section.

\section{Decision problems in representation theory} \label{sreprdec}
We now describe   the decision problems associated with the couple
$H\hookrightarrow G$, the couple $G \hookrightarrow K$ being similar. 
A general decision problem is as follows:

\begin{problem} {\bf (The subgroup restriction problem)} \label{pgct6subgroup}

Let  $\rho:H \rightarrow G$  be as above, with $G$ connected
(and some mild  technical restrictions
on $\rho$ as described in \cite{GCT6}).
Assume that both $H$ and
$G$ are reductive. Let $V_\pi(H)$ 
be an irreducible representation 
of $H$, and 
$V_\lambda(G)$ an irreducible representation  of $G$, where
$\pi$ and $\lambda$ denote the classifying labels of these representations.
Let  $m_\lambda^\pi$ be the multiplicity   of $V_\pi(H)$
in $V_\lambda(G)$, considered as an $H$-module via $\rho$.
Given specifications of the embedding $\rho$ and the labels 
$\lambda,\pi$, decide nonvanishing of the multiplicity $m_\lambda^\pi$.
\end{problem} 
The general decision problems in Hypotheses~\ref{hdecision}
 can be thought of as
harder variants of this problem obtained by going from couples
to triples. 
All couples that arise in GCT are either of the  type in this decision
problem,  or of a hybrid type obtained by combining 
this type with the type considered earlier in
connection with the flag variety, when $H=P_\mu$ is parabolic;
cf. \cite{GCT10} for a discussion of the hybrid types.

Problem~\ref{pgct6subgroup}
is a  fundamental decision problem of representation theory.
Indeed, one of the main motivations in the classical
works of representation theory, e.g. \cite{weyl}, for classifying of all
irreducible representations of reductive groups was to be able to solve this 
problem  satisfactorily. But despite all  progress in
representation theory in the last century, this problem at its 
very heart remained open. 
PHflip in \cite{GCT6} 
says that this fundamental decision  problem of representation theory
has an  easy polynomial time algorithm.

Here we shall describe PHflip in  only  the following three
special cases of the above  decision problem, referring the reader to
\cite{GCT6} for a full discussion and results for 
the general decision problem.

\subsubsection{Littlewood-Richardson problem}
Let $H=GL_n(\C)$, $G=H\times H$, the embedding 
\[ \rho:H\rightarrow H\times H=G\]
being diagonal. Then the multiplicity in
Problem~\ref{pgct6subgroup} is just the Littlewood-Richardson coefficient,
because every irreducible representation of $G$ is of the
form $V_{\alpha}\otimes V_\beta$, where $V_\alpha$ and $V_\beta$ 
are irreducible representations of $H=GL_n(\C)$ for partitions
$\alpha$, and $\beta$, and the multiplicity of an $H$-module
$V_\lambda$ in $V_{\alpha}\otimes V_\beta$, considered as 
an $H$-module via the diagonal map $\rho$, is precisely the
Littlewood-Richardson coefficient $c_{\alpha,\beta}^\lambda$.
We have already noted that its nonvanishing can be decided in
polynomial time (Section~\ref{slittlegct6}).

\subsubsection{Kronecker problem}
Let $H=GL_n(\C)\times GL_n(\C)$ and 
\[\rho: H \rightarrow G=GL(\C^n \otimes \C^n)=GL_{n^2}(\C) \] 
the natural embedding given by: 
$\rho(h_1,h_2)=h_1\otimes h_2$, for any $h_1,h_2 \in H$. Here 
$h_1\otimes h_2$ is the Kronecker product as defined
in (\ref{eqkroneckerproduct}). 
Let  $k_{\lambda,\mu}^\pi$ be  the multiplicity of the 
$H$-module $V_\lambda(GL_n(\C)) \otimes V_\mu(GL_n(\C))$ in the $G$-module
$V_\pi(G)$, considered as an $H$-module via the embedding $\rho$.
Then  it can be shown \cite{fultonrepr} 
that the Kronecker coefficient as defined in Section~\ref{sspecht}
 is a special (dual)
case of this when  $\lambda,\mu$ and $\pi$ there coincide with the 
$\lambda,\mu$ and $\pi$ here. For this reason, we  call
$k_{\lambda,\mu}^\pi$ a Kronecker coefficient.

\begin{problem} (The  Kronecker problem) \label{pgct6kronecker} 
Given partitions $\lambda,\mu$ and $\pi$, decide nonvanishing of the Kronecker 
coefficient  $k_{\lambda,\mu}^\pi$. 
\end{problem} 

The following is an analogue of Hypothesis~\ref{hdecision} in this context:

\begin{hypo} \cite{GCT6} (PHflip-kronecker) 
Given partitions $\lambda,\mu$ and $\pi$, nonvanishing of the Kronecker 
coefficient  $k_{\lambda,\mu}^\pi$ can be decided in
$\poly(\bitlength{\lambda},\bitlength{\mu},\bitlength{\pi})$ time.
\end{hypo}

\subsubsection{The plethysm problem} \label{splethysmproblem}
The Kronecker coefficient  is known  \cite{kirillov}  to be a special
case of the plethysm coefficient in 
the following more general problem.

\begin{problem} (The  plethysm problem) \label{pgct6plethysm}
Given partitions $\lambda,\mu$ and $\pi$, decide nonvanishing of the plethysm
constant $a_{\lambda,\mu}^\pi$. This is the multiplicity  of 
the irreducible representation 
$V_\pi(H)$ of $H=GL_n(\C)$ 
in the irreducible representation 
$V_\lambda(G)$ of $G=GL(V_\mu)$,
where $V_\mu=V_\mu(H)$ is an irreducible representation $H$.
Here $V_\lambda(G)$ is considered an $H$-module via 
the representation map 
\[ \rho:H\rightarrow G=GL(V_\mu).\] 
\end{problem} 

The following is an analogue of Hypothesis~\ref{hdecision} in this context:

\begin{hypo} \cite{GCT6} (PHflip-plethysm) \label{hphflipplethysm} 
Given partitions $\lambda,\mu$ and $\pi$, nonvanishing of the plethysm
constant  $a_{\lambda,\mu}^\pi$ can be decided in
$\poly(\bitlength{\lambda},\bitlength{\mu},\bitlength{\pi})$ time.
\end{hypo}

\section{The $P$-barrier in representation theory} \label{spbarrepr}
At the surface, this hypothesis too
seems impossible  because the dimension of $G$  here
can be exponential in the dimension of $H$. 
This happens when 
the dimension of the representation $V_\mu(H)$ is exponential in 
$\dim(H)$. But the total bitlength of $\lambda,\mu$ and $\pi$ 
can be polynomial in $\dim(H)$. Hypothesis~\ref{hphflipplethysm}
in this case says that
nonvanishing of the plethysm constant can still be decided in
$\poly(\bitlength{\lambda},\bitlength{\mu},\bitlength{\pi})$ time. 
A priori it is not even clear 
that the plethysm constant can be evaluated in PSPACE in this case.
Since the usual character-theory-based algorithms in representation 
theory for its evaluation \cite{fultonrepr,macdonald}
take space that is  polynomial in the dimension of $G$, and hence,
exponential in the dimension of $H$.

The main $P$-barrier in representation theory is this huge gap between
the exponential space bound for the plethysm or the general 
decision Problem~\ref{pgct6subgroup}  given by the usual methods
of representation theory and the polynomial time bound stipulated 
in Hypothesis~\ref{hphflipplethysm} 
for the plethysm constant and the  hypothesis
in \cite{GCT6} for the general decision Problem~\ref{pgct6subgroup}.

\subsection{Crossing the $P$-barrier}
We now describe the main results of \cite{GCT6} which together
cross this $P$-barrier in representation theory subject to
the analogous mathematical positivity hypotheses PH1 and SH (PH2).
We shall only concentrate on the plethysm problem, since it is the
crux of the matter.

Associate
with a plethysm constant $a_{\lambda,\mu}^\pi$ the
stretching function 
\begin{equation} 
\tilde a_{\lambda,\mu}^\pi (k)=a_{k \lambda,\mu}^{k \pi}.
\end{equation} 

Note that $\mu$ is not stretched here.

Then the following is an (unconditional) analogue of 
Theorem~\ref{tmainquasipoly} 
in this context:

\begin{theorem} \label{tquasiplethysm} (cf. Theorem 1.6.1 in \cite{GCT6})
The stretching function
$\tilde a_{\lambda,\mu}^\pi(k)$ is a quasi-polynomial function of $k$.
\end{theorem}

The following are the analogues of PH1 and PH2 in this context:

\begin{hypo} {\bf (PH1)} \label{htph1plethysm}

For every $(\lambda,\mu,\pi)$ there exists a polytope 
$P=P_{\lambda,\mu}^\pi \subseteq \R^m$ with 
$m=\poly(\bitlength{\lambda},\bitlength{\mu},\bitlength{\pi})$ such that:
\begin{equation} \label{eqph1int}
a_{\lambda,\mu}^\pi=\phi(P),
\end{equation}
where $\phi(P)$   is equal to the number of integer points in $P$,
and the Ehrhart quasi-polynomial of $P$ coincides with the 
stretching quasi-polynomial $\tilde a_{\lambda,\mu}^\pi(k)$ in
Theorem~\ref{tquasiplethysm}.
(And some additional technical constraints)
\end{hypo}

\begin{hypo} \label{htph2plethysm}
{\bf (PH2)}

 The stretching  quasi-polynomial 
$\tilde a_{\lambda,\mu}^\pi(k)$ is positive (cf. Section~\ref{spbarrierintro}).
\end{hypo}

PH2  implies the following saturation hypothesis:

\begin{hypo} {\bf (SH)} \label{htshplethysm}

The quasi-polynomial 
$\tilde a_{\lambda,\mu}^\pi(k)$ is  saturated
 (cf. Section~\ref{spbarrierintro}).
\end{hypo} 

The following is an analogue of Theorem~\ref{tmaincomplexity} in 
this context:

\begin{theorem} \label{tplethysmcomplexity} \cite{GCT6}
Assuming PH1 and SH (or, more strongly, PH2), 
nonvanishing of a plethysm constant $a_{\lambda,\mu}^\pi$ can be decided
in $\poly(\bitlength{\lambda},\bitlength{\mu}, \bitlength{\pi})$ time; i.e.
the problem of deciding nonvanishing of a plethysm constant 
belongs to $P$, as per Hypothesis~\ref{hphflipplethysm}.
\end{theorem}

PH1 above implies that $a_{\lambda,\mu}^\pi$ belongs to $\#P$ 
just like the Littlewood-Richardson coefficient.
Its weaker form is:

\begin{theorem} \label{tpspace}
The plethysm constant $a_{\lambda,\mu}^\pi$ can be computed in 
PSPACE, i.e., in 
$\poly(\bitlength{\lambda},\bitlength{\mu}, \bitlength{\pi})$ space.
\end{theorem} 

That  this holds  even if the
dimension of $G=GL(V_\mu)$ is exponential in $n$
is crucial in the context of GCT. Because the dimension of 
$K=GL(V)$ in the triples $H\hookrightarrow G \hookrightarrow K=GL(V)$ 
associated with the class varieties (Section~\ref{sclass}) is 
exponential in the circuit size $m$. Hence, without this result, it is 
not at all clear why the structural constants $s_d^\lambda(H,l)$ 
and $s_d^\lambda(E;n,l)$ in Hypothesis~\ref{hph1genintro} should even belong 
$PSPACE \supseteq \#P$, as implied by it.

Theorem~\ref{tpspace} and  Theorem~\ref{tquasiplethysm} together provide  
good theoretical evidence for PH1 (Hypothesis~\ref{htph1plethysm}). 
Indeed, Theorem~\ref{tquasiplethysm}, together with other evidence
in \cite{GCT6}, suggests that 
$\tilde a_{\lambda,\mu}^\pi(k)$ is the Ehrhart quasi-polynomial
of some polytope $P=P_{\lambda,\mu}^\pi$.
Furthermore, Theorem~\ref{tpspace}
says that the dimension $m$ of the ambient space $\R^m$ containing $P$ 
should be polynomial in the bitlengths $\bitlength{\lambda},
\bitlength{\mu}$ and $\bitlength{\pi}$. If not, it would not 
be possible to count the number of integer points in $P$ in PSPACE,
since even the bitlength  of any integer point in $\R^m$ 
would not be polynomial. For further
theoretical and experimental results in support of PH1 and PH2 in this
context,  see  \cite{GCT6}. 
These  constitute the main evidence in support of
PH1 and PH2 for the group-theoretic class varieties 
(Hypotheses~\ref{hph1genintro}-\ref{hph2genintro}),
because mathematical  positivity is a very abstract 
property, which should remain invariant when we go from couples to triples.

\section{Reduction} \label{sreduction}
Now we turn to the 
reduction in the top arrow in Figure~\ref{fbasicintro}. 
For this, we have to describe:
\begin{enumerate} 
\item  How to transform the easy algorithms 
in Theorem~\ref{tmaincomplexity} into an easy algorithm for discovering an
obstruction as in Hypothesis~\ref{hphflipformalnew} (c),
and 
\item How to transform this
easy algorithm for discovery into a
constructive proof of existence of 
obstructions--as expected (Section~\ref{swhyobsex})--for every $n$ and
$l=n^{\log n}$,
by showing how such an obstruction-label
can be easily constructed in this case  {\em explicitly}.
\end{enumerate} 
This would imply that  $P\not = NP$ over $\C$.

These transformations cannot  be carried out at present, since
we do not know the polytopes $P_d^\lambda(H;l)$ and $P_d^\lambda(E;n,l)$
explicitly. We only know that they should exist as per PH1
(Hypothesis~\ref{hph1genintro}). But once their explicit descriptions become
available, it should be possible to carry out the above two transformations
along the lines that we now suggest.

\subsection{Towards easy discovery} \label{stowardseasy}
First, let us 
describe why it should be possible to extend and transform the polynomial-time 
algorithms in Theorem~\ref{tmaincomplexity}
 to obtain a polynomial time algorithm 
for discovering an obstruction (Hypothesis~\ref{hphflipformalnew} (c))
once explicit descriptions
of the polytopes $P_d^\lambda(H;l)$ and $P_d^\lambda(E;n,l)$ 
become available.

For the sake of simplicity, let us assume that 
the quasi-polynomials in Theorem~\ref{tmainquasipoly}
are actually polynomials;
i.e., their periods are one, though this is not expected.
In that case, it can be shown that \cite{GCT6}
PH1 and SH imply  that there exist
polytopes $P(n,l)=P(E;n,l)$ and $Q(l)=P(H;l)$ of $\poly(n,l)$ dimensions
such that 
an obstruction for $n,l$ and the pair $(E,H)$ 
 exists iff the relative difference 
$T(n,l)=P(n,l)\setminus Q(l)$ is nonempty, and furthermore, 
an explicit obstruction can also be constructed in polynomial time
once we are given a rational point in $T(n,l)$.
So it suffices to  check if $T(n,l)$ is nonempty, and if so, find a rational
point in it. This can be
done in polynomial time using the convex (linear) programming algorithm in
\cite{lovasz,vaidya} if $Q(l)$ has only $\poly(l)$ explicitly 
described facets. This is so 
even if  $P(n,l)$ has exponentially many facets.
But if $Q(l)$ has exponentially many facets--as happens even in
the context of the simpler Littlewood-Richardson problem 
(Problem~\ref{plittle})--then 
an  oracle-based algorithm as  in \cite{lovasz} cannot be used to
get a polynomial time algorithm for this problem \cite{babai}.

But this does not appear to be  a serious problem. Indeed, a general principle 
in combinatorial optimization, as illustrated in \cite{lovasz}, is that
complexity-theoretic properties of 
polytopes with exponentially many facets are similar to the ones with
polynomially many facets if these facets have a well-behaved regular 
structure. For example, if $Q(l)$ and $P(n,l)$ were perfect matching
polytopes for non-bipartite graphs--which can have 
exponentially many facets--nonemptiness of $T(n,l)$ can be easily
decided in polynomial time  \cite{vempala}
 using the polynomial time algorithm \cite{edmonds}
for finding a perfect matching in a nonbipartite graph.
The facets of the analogues of $P(n,l)$ and $Q(l)$ in the Littlewood-Richardson
problem, called Littlewood-Richardson cones \cite{zelevinsky},
have an explicit
description with very nice algebro-geometric and representation-theoretic
properties \cite{kly}. The same is expected to be the case in our setting.

This is why we expect that nonemptiness of $T(n,l)$ and computation
of a rational point in it, if it is nonempty, can be done
in polynomial time, 
once  explicit descriptions of $P(n,l)$ and $Q(l)$ become known. 
This would  give a polynomial time algorithm for discovering an obstruction,
if it exists, as per Hypothesis~\ref{hphflipformalnew} (c),
assuming that the quasi-polynomials 
in Theorem~\ref{tmainquasipoly} are polynomials.

Furthermore, it is expected, for the mathematical reasons given in
\cite{GCT6},
that there exist genuinely simple, i.e.,  purely
combinatorial greedy-type  algorithms for the  problems under consideration
 that do not even need 
linear programming. That is, the story is expected to be the same as 
for the min-cost flow problem in combinatorial optimization,
for which a linear-programming-based polynomial-time algorithm was found
first \cite{tardos} to be followed by several genuinely simple and
purely combinatorial polynomial time algorithms; e.g. see \cite{orlin}.
Similarly, 
it is reasonable to expect 
that the algorithms in Theorem~\ref{tmaincomplexity} and the subsequent 
algorithm for discovery of obstructions  can be simplified further
to eventually get   simple  greedy algorithms for these problems
akin to the Hungarian method, once explicit descriptions of 
$P(n,l)$ and $Q(l)$  become known.

So far we are assuming that the quasi-polynomials in 
Theorem~\ref{tmainquasipoly} 
are polynomials. This need  not be so. In fact, this is 
not so even in the simplified setting of plethysm constants \cite{GCT6}.
When the quasi-polynomials in Theorem~\ref{tmainquasipoly}
have nontrivial periods,
the obstructions can be classified in two types: {\em geometric} and
{\em modular} \cite{GCT6}. Geometric obstructions are similar 
to the ones that would arise if these quasi-polynomials were polynomials.
A polynomial time algorithm for their existence and construction 
may be designed  along the lines we just described. 

Let us next describe briefly what needs to be done in the case of modular
obstructions. Theorem~\ref{tmaincomplexity} says  that
for the decision problems
therein linear programming in conjunction
with  modular techniques (computation of Smith
normal forms \cite{kannan}) works even in the modular setting, i.e., when
the quasi-polynomials have nontrivial periods.  Hence, once we have 
a polynomial time algorithm for discovering  a geometric construction,
it should be possible to extend it to a polynomial time algorithm for
discovering a modular obstruction in conjunction with appropriate
modular techniques; cf. \cite{GCT6} for the problems that need to
be addressed in this extension. 

\subsection{From easy algorithm for discovery to easy proof of existence}
\label{sfromto}
Assuming that we have an easy polynomial time algorithm for 
discovering an obstruction as per Hypothesis~\ref{hphflipformalnew} (c),
let us now describe why it should be possible to prove using this algorithm,
or rather the underlying structure and techniques, that there always exists an 
obstruction, as expected (Section~\ref{swhyobsex}),
for every $n\rightarrow \infty$, assuming $l=n^{\log n}$ (say).

For the sake of simplicity, let us again assume that that the 
quasi-polynomials in Theorem~\ref{tmainquasipoly}
are polynomials, and that 
we have an easy Hungarian-type greedy 
algorithm as discussed above for deciding nonemptiness of 
$T(n,l)$, and for computing a  point it it, if it is nonempty.
Then we have to show, using the techniques and the structure underlying
this algorithm, that $T(n,l)$ is always nonempty when 
$l=n^{\log n}$, $n\rightarrow \infty$.
Such a proof would also give us a polynomial time  procedure for 
explicit construction of an obstruction $\lambda(n)$, for every $n$.
Hence we shall call it a {\em $P$-constructive proof}.

To see how to get such a $P$-constructive proof, let us consider an analogy.
Let us imagine  that
$Q(l)$ is empty, so that $T(n,l)=P(n,l)$ is a polytope, and that it
is the perfect-matching polytope of a bipartite graph $G(n,l)$.
Then $T(n,l)$ is nonempty iff $G(n,l)$ has a perfect matching, 
which  can be thought of as an obstruction in this analogy. 
The analogous goal then is to show using the techniques and structure 
underlying  the Hungarian method
that $G(n,l)$  always has a perfect matching, as expected, when
$l=n^{\log n}$, and $n \rightarrow \infty$. 
In other words, we have to
give a $P$-constructive proof for existence of a perfect
matching in every such  $G(n,l)$.
In this analogy, the technique underneath the
Hungarian  method  can be easily
used to give  a constructive proof of  Hall's marriage theorem--namely,
that every bipartite  graph $H$
in which every subset, on any side of the graph, has at least as many 
neighbours as the size of that subset has
a perfect matching--which  then has to  be
used to show that $G(n,l)$ always 
has a perfect matching whenever $l=n^{\log n}$,
$n\rightarrow \infty$. 

Now in our setting $T(n,l)$ is not a perfect matching polytope. 
But if it has a nice structure like the perfect matching polytope 
then it should be possible to prove structure theorems in the spirit of
Hall's marriage theorem for $T(n,l)$ using the structure of the 
Hungarian-type greedy algorithm 
for deciding nonemptyness of $T(n,l)$ and then
use it to prove nonemptyness of $T(n,l)$, for every $n\rightarrow \infty$,
when $l=n^{\log n}$.

For such a  transformation of a polynomial time algorithm for discovery into 
a $P$-constructive proof of existence to work, it is crucial 
that:
\begin{enumerate} 
\item The polyhedral set $T(n,l)$ has a nice, regular structure like
the perfect matching polytope. Fortunately, 
the polytopes $P(n,l)$ and $Q(n,l)$  that would arise in our setting
should be even nicer than the perfect matching polytope.
For example, in the simpler setting of the Littlewood-Richardson problem
(Problem~\ref{plittle}), $P(n,l)$ and $Q(l)$  become
Littlewood-Richardson cones \cite{zelevinsky},
which have extremely regular structure with
remarkable representation-theoretic and algebro-geometric properties
\cite{fultonkly,kly}. The same is expected to be the case for the 
actual $P(n,l)$ and $Q(l)$.

\item The algorithm for discovery not only works 
in polynomial time, but also has a 
simple structure like the Hungarian method. 
The  Hungarian-type greedy algorithms that we expect for the problems under 
consideration should have such structures.
\end{enumerate}

Hence, it is reasonable to expect  that an easy Hungarian-type 
algorithm for deciding
nonemptyness of $T(n,l)$ can 
be   transformed into the sought  $P$-constructive 
proof of obstructions.
The story is expected to be similar, albeit much harder, when the
quasi-polynomials in Theorem~\ref{tmainquasipoly} have nontrivial periods;
cf \cite{GCT6}.

The above scheme for  the  transformation of an
algorithm for discovery  into a   constructive proof of existence 
banks on the fact that  the  algorithm to be
transformed is easy, i.e., works in polynomial time, besides having a 
simple structure.
The underlying informal principle, which cannot be
proved, is that the mathematical complexity of an
algorithmic (constructive)   proof is intimately linked to the computational
complexity of the algorithm on which it is based; see  
 Section~\ref{spveri} for a detailed treatment of this issue.
This is
why there is no nontrivial result, comparable to Hall's theorem, for
Hamiltonian paths. Because
the problem of finding such a path is NP-complete.

The reader may wonder why we are talking about 
explicit construction of obstructions,
when, strictly speaking, we only need to
know their existence. This is because the nature of obstructions in our 
case is such that  their explicit construction, if they
exist, can be done with only a little additional cost over the cost 
of deciding existence. 
To see this, let us again assume, for the sake of simplicity,
that the quasi-polynomials in
Theorem~\ref{tmaincomplexity} are polynomials. Then 
a technique that can decide nonemptiness of $T(n,l)$
should also be able to compute
a  point in it, as a proof of nonemptiness, at only
a little additional cost, just as in linear programming.
In other words, the complexity of deciding existence of an obstruction
should be more or less the same as that of constructing it, if it exists.
This is why we mainly talk of explicit construction of obstructions
though, in principle, just their existence would suffice.

Our discussion so far says  that  PH1 and SH (PH2) are the crux 
of the matter. If they can be  proved, and explicit descriptions of the
polytopes therein become available, it should be possible to 
transform the easy algorithms in Theorem~\ref{tmaincomplexity} 
into an easy algorithm for explicit construction of obstructions
as per Hypothesis~\ref{hphflipformalnew} (b).

\section{Standard quantum group}
\label{sstdquantum}
Now we proceed to the  basic plan in \cite{GCT6} for proving PH1 and
SH. This is motivated by a story in the theory of standard quantum groups
in the context of the Littlewood-Richardson
problem (Problem~\ref{plittle}).
We describe that story in this section.

For this we need the notion of a standard quantum group, by which we 
mean the quantum group in  \cite{drinfeld,jimbo,rtf}.
We can not formally define here  this object, but 
we can at least give an intuitive idea. Let $GL(\C^n)$ be the group of
nonsingular $n\times n$ matrices. It can be thought of as the group of
nonsingular transformations of $\C^n$. Let $x_i$'s denote the coordinates 
of $\C^n$. These commute. That is:
\[ x_i x_j = x_j x_i.\] 
Let us now see what happens if the coordinates  become noncommuting.
This is precisely what happened  in quantum physics. We discovered
 that the position and the momentum, which for centuries 
we thought were 
commuting observables, do not actually commute. Quantum 
groups were invented precisely to investigate the related 
phenomena in theoretical
physics. Let $\C^n_q$ denote the
quantum space whose coordinates $x_i$'s are noncommuting, and satisfy the
following relation: 

\[ x_i x_j = q x_j x_i, \quad i<j\] 
where $q \in \C$ is some fixed number.
The {\em standard} 
quantum group $GL_q(\C^n)$ is
the ``group'' of invertible linear transformations of this quantum space.
This is not a ``group'' in any ordinary sense.
Its precise description is given in \cite{drinfeld,jimbo,rtf}.
We do not need that here.
Let us just think of a quantum group as what a group becomes when the
coordinates become noncommuting.

Let us now  explain how  quantum groups enter in the story of
Littlewood-Richardson coefficients.
This is because the most transparent proof of the Littlewood-Richardson
rule  came via the theory of
quantum groups \cite{kashiwara1,littelmann,lusztigbook}.
The earlier proofs, though elementary and
combinatorial, were highly mysterious. Moreover, the theory 
of quantum groups gave the first proof of 
the generalized  Littlewood-Richardson
rule \cite{} 
for general (connected) reductive groups, instead of just $GL_n(\C)$. 

Let us now elaborate the nature of this proof. We begin by observing that
the Littlewood-Richardson problem (Problem~\ref{plittle}) is an
instance of the general  decision
Problem~\ref{pgct6subgroup} associated with the diagonal group homomorphism 
\begin{equation}\label{eqdiaglittle}
 \rho: H= GL(\C^n) \rightarrow H \times H = GL(\C^n) \times GL(\C^n).
\end{equation}
If we understood the structure of this homomorphism in depth,
we ought to understand why PH1 and SH (and also PH2) hold for
the Littlewood-Richardson coefficients. As we mentioned earlier, in depth
means at the  quantum level. To understand the
homomorphism (\ref{eqdiaglittle}) at the quantum level, we need to quantize it.
Ideally,  one would want its quantization in the form of 
a homomorphism 

\begin{equation} \label{eqdiagq}
 \rho_q: H_q= GL_q(\C^n) \rightarrow H_q \times H_q = GL_q(\C^n) \times GL_q(\C^n).
\end{equation}
where $H_q$ is the standard quantum group associated with $H$.
This does not hold as it is; i.e., $H_q$ is not a quantum subgroup
of $H_q \times H_q$. But this is essentially so. That is,
it holds in a certain dual setting--this is the main result in 
\cite{drinfeld,jimbo,rtf}.
Thus the theory of quantum group can be regarded as the
theory of the quantization $\rho_q$.

Once this theory is developed sufficiently, the
Littlewood-Richardson rule as well as PH1 for Littlewood-Richardson
coefficients (Hypothesis~\ref{htlittleph1})
turn  out to be
a consequence, in a nontrivial way, of a deep positivity result in the
theory of the standard quantum groups 
\cite{kashiwara2,kashiwaraglobal,lusztigcanonical,lusztigbook}:
namely, their 
representations and coordinate rings have {\em canonical} bases,
also called global crystal bases, 
whose structural constants, which determine their multiplicative and
representation theoretic structure, are all nonnegative. 
For this reason, we say that the canonical bases are {\em positive},
and refer to existence of a canonical basis as a positivity property
(hypothesis) PH0.

We now give a brief intuitive description of
the canonical basis.
Let $X$ be an $n \times n$ variable matrix. The coordinate
algebra $R={\cal O}(G)$ of the group $G=GL(\C^n)$ is defined to 
be the $\C$-algebra 
generated by the entries $x_{ij}$ of $X$ and $\det(X)^{-1}$,
where $\det(X)$ denotes the determinant of $X$. Its elements are
regular functions on $G$, considered as an affine variety. 
There is a natural left action of $G$ on $R$ given by 
$f(X) \rightarrow f(\sigma^{-1} X)$, for any $\sigma \in G$, and
a similar right action.

These notions can now be quantized.
It is possible to associate a coordinate ring 
$R_q={\cal O}(G_q)$ with the standard quantum group $G_q=GL_q(\C^n)$,
whose elements can be intuitively thought of as functions on $G_q$.
Unlike $R$, $R_q$ is not commutative. Its precise definition can be
found in \cite{rtf}. There are natural left and right actions of
$G_q$ on $R_q$.

A canonical basis $B$  of $R_q$ is a very special basis with the
following properties: 

\noindent (1) It is representation-theoretically well behaved.
This means there is a filtration 
\[ B_0 = \emptyset \subset B_1 \subset B_2 \subset \cdots \subset B \]
 with $\cup_i B_i = B$, such that 
$\langle B_i \rangle / \langle {B_{i-1}} \rangle$
is an irreducible $G_q$-module. Here
$\langle {B_i} \rangle$ denotes the span of the basis elements in $B_i$.

\noindent (2) Positivity property of the multiplicative structure constants:

Given two elements $b,b' \in B$, let 
\[ b b'= \sum_{b'' \in B} f_{b,b'}^{b''} b'',\] 
be the expansion of the product in terms of the basis $B$.
Then each $f_{b,b'}^{b''}$ is an explicit  polynomial in $q$ and $q^{-1}$ with
nonnegative coefficients. Here $f_{b,b'}^{b''}$ are called 
multiplicative structure constants. What this says is that each
multiplicative structure constant has an   {\em explicit positive} formula,
akin to that of the permanent.
Here explicit means that each nonnegative coefficient of 
$f_{b,b'}^{b''}$ has an interpretation in terms of
a nonnegative  topological invariant (akin to Betti numbers) of an algebraic 
variety.

\noindent (3) Positivity property of the representation-structure constants:

Given any element $b \in B$ and a generator $e$ of a certain algebra 
defined in \cite{drinfeld,jimbo}, which is ``dual'' to $R_q$, let

\[ e \cdot  b = \sum_{b'} g_{e,b}^{b'} b'\] \
be the expansion of $e \cdot b$, the result of applying $e$ to $b$,
in terms of the basis $B$. Then each $g_{e,b}^{b'}$ is also an explicit
polynomial in $q$ and $q^{-1}$ with nonnegative coefficients.
That is, each representation-structure constant also has an explicit  positive 
formula.

These positivity properties do not actually hold
as stated--that is still a conjecture \cite{lusztigbook}--but their slightly
weaker form holds unconditionally \cite{lusztigbook}. 
We shall ignore that difference here.

If we specialize the canonical basis at $q=1$, we get a 
canonical basis of $R$, the coordinate ring of $G$, with analogous
positivity property.
But, as of now,  the only way to 
prove existence of  such a canonical basis of $R$ is 
via the theory of quantum groups as above. This shows the power of this
theory. 

One can easily imagine that there ought  be a connection between 
existence of bases whose  structural constants have explicit positive formulae 
(PH0) and existence of an explicit  positive (polyhedral) formula for
Littlewood-Richardson coefficients (PH1). That is indeed so, as we mentioned
earlier,
but in a quite nontrivial way; cf. \cite{kashiwara1,littelmann,lusztigbook}.
We shall simply take this
connection on faith here. Pictorially:

\begin{equation} \label{eqph0ph1}
  PH0 \rightarrow PH1.
\end{equation}

One does not really need the full power of PH0 to deduce PH1.
Just  existence of a local crystal basis \cite{kashiwara1},
which is the limit (crystalization)
of a canonical basis as $q \rightarrow 0$, is
sufficient. But when we move to the nonstandard setting in GCT,
even the full power of PH0 is needed for some other reasons;
\cite{GCT8,GCT10}.

The implication (\ref{eqph0ph1}) 
provides arguably the most satisfactory proof of PH1 for
 Littlewood-Richardson coefficients, which, in addition, also provides
deep additional information (existence of canonical bases)
which the  combinatorial proofs \cite{fultonyoung} cannot  provide.
Such canonical bases are central to the
approach in \cite{GCT6,GCT10}
towards  PH1 and PH2 for the group-theoretic class varieties 
(Section~\ref{snonstdquantum}).
Hence, as far as GCT is concerned, quantum groups 
are a must. 

SH  for the
usual Littlewood-Richardson coefficients is the saturation theorem
in \cite{knutson}. It 
comes from a reformulation of
PH1 in terms of special polytopes (called Hive polytopes) and their 
subsequent detailed study. Thus pictorially:

\begin{equation} \label{eqph1sh}
 PH1 \rightarrow SH,
\end{equation}
again in a nontrivial way.

But how is PH0  proved?
The only known 
proof of PH0 \cite{lusztigcanonical,lusztigbook}  is based on a deep positivity
property in mathematics: the Riemann Hypothesis over finite fields 
\cite{weil2}, 
and related results \cite{beilinson}. In other words, nonnegativity of the
structural constants associated with $H_q$ is connected at a profound
level with the lining up of the zeros of the zeta functions of some
algebraic varieties on one axis. We shall denote the Riemann hypothesis
over finite fields by $PH+$. Then
pictorially: 

\begin{equation}\label{eqphplus}
PH+ \rightarrow PH0,
\end{equation}
in a highly nontrivial way. 

\begin{figure}[!h]
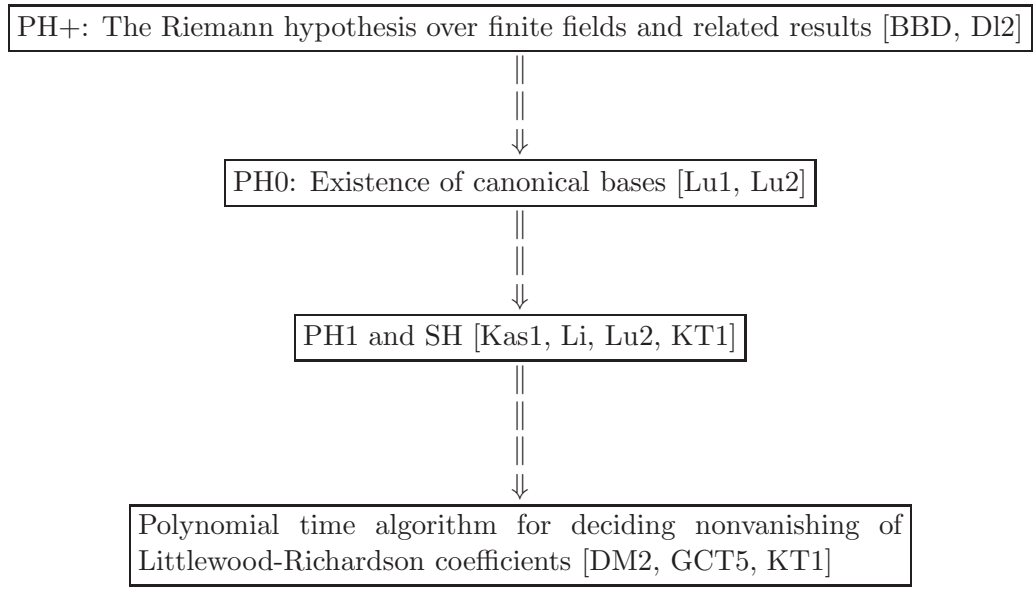
 
\centering
\[\begin{array} {c} 
\fbox{PH+: The
Riemann hypothesis over finite fields and related results 
\cite{beilinson,weil2}} \\
\|\\
\|\\
\Downarrow\\
\fbox{PH0: Existence of canonical bases \cite{lusztigcanonical,lusztigbook}} \\
\|\\
\| \\
\Downarrow \\
\fbox{PH1 and SH \cite{kashiwara1,littelmann,lusztigbook,knutson}} \\
\|\\
\|\\
\|\\
\Downarrow \\
\fbox{\parbox{4in}{Polynomial time algorithm for deciding nonvanishing of 
Littlewood-Richardson coefficients \cite{loera,GCT5,knutson}}}
\end{array}\]
\caption{A story in the theory of standard quantum groups} 
\label{fgctlittle}
\end{figure}

Putting implications (\ref{eqph0ph1})-(\ref{eqphplus}) together with the story
in Section~\ref{slittlegct6},
we arrive at 
Figure~\ref{fgctlittle} which summarizes the story in this section.

\section{Nonstandard quantum groups} \label{snonstdquantum}
Now we turn to the problem of proving PH1 and SH 
that actually arise in GCT
(Hypotheses~\ref{hph1genintro}-\ref{hshgenintro}, and 
\ref{htph1plethysm}-\ref{htshplethysm}). 
The basic plan in \cite{GCT6} for this 
is simply to lift the story in Figure~\ref{fgctlittle} 
from  height two to superpolynomial height--i.e., from the circuits of
height two that the Littlewood-Richardson problem corresponds to to
the circuits of superpolynomial height that the decision problems in
Hypothesis~\ref{hdecision} correspond to.
Roughly, it goes as follows:

\noindent (1) {\bf Quantization:} Quantize the couples 
\[ H \hookrightarrow G, \quad G \hookrightarrow K\] 
and the triples 
\[ H \hookrightarrow G \hookrightarrow K,\]
associated  with the class varieties in a manner akin to the
quantization (\ref{eqdiagq})
of (\ref{eqdiaglittle}) via standard quantum groups.

\noindent (2) {\bf PH0 for couples and triples:}
Prove that the coordinate rings and
representations of the quantum groups that arise in this quantization
have canonical bases akin to the canonical bases for the standard 
quantum groups whose structure constants, which determine their 
multiplicative and representation theoretic structure, are all nonnegative.

\noindent (3) {\bf PH0 for class varieties:}
Use the canonical bases for  the 
(quantized) triples associated with the class varieties 
to construct analogous canonical bases for the coordinate rings 
for appropriate quantizations of the  class varieties with 
 nonnegative structure constants.

\noindent (4) {\bf PH1, SH:} Deduce PH1 and  SH from PH0 
in the spirit of the middle arrow in Figure~\ref{fgctlittle}.

\begin{figure}[!h]
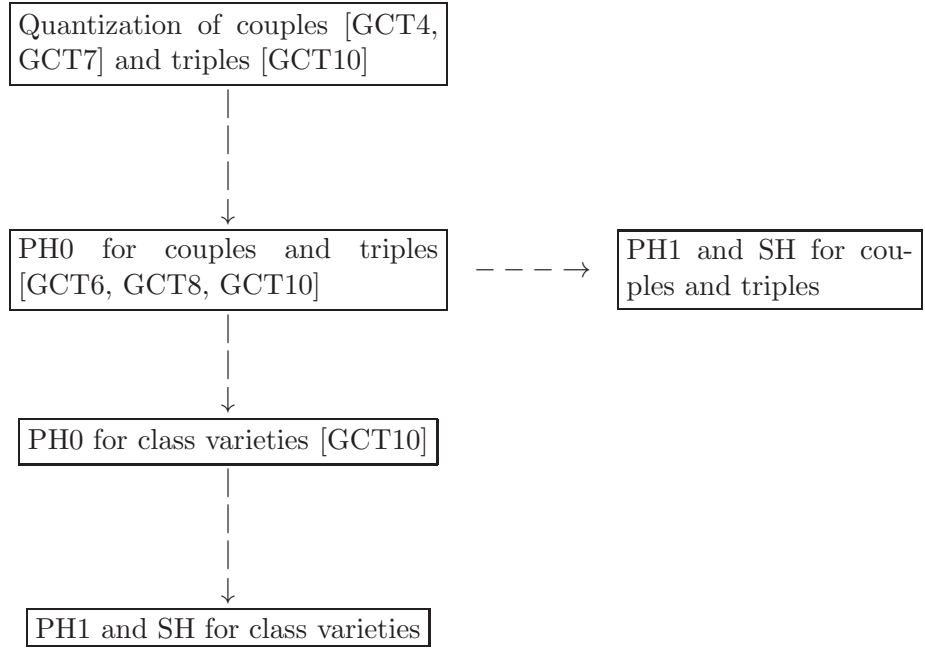

\centering
\[\begin{array} {ccc} 
\fbox{\parbox{2.2in}{Quantization of couples \cite{GCT4,GCT7} and triples
\cite{GCT10}}} \\
|\\
| \\
|\\
\downarrow \\
\fbox{\parbox{2.2in}{PH0 for couples and triples \cite{GCT6,GCT8,GCT10}}}
&---\rightarrow& \fbox{\parbox{1.5in}{PH1 and SH for couples and triples}} \\
|\\
|\\
\downarrow\\
\fbox{PH0 for class varieties \cite{GCT10}} \\
|\\
|\\
|\\
\downarrow \\
\fbox{PH1 and SH for class varieties} \\
\end{array}\]
\caption{The basic plan for proving PH1 and SH in \cite{GCT6}} 
\label{fph1sh}
\end{figure}

Figure~\ref{fph1sh} shows this pictorially.

We shall now elaborate Figure~\ref{fph1sh}.

\subsection{Quantization} 
Let us begin with the first step of quantization. We shall only
worry about the couples. To be concrete, let
\begin{equation} \label{eqprimary} 
 H\hookrightarrow G=GL(\C^k),
\end{equation} 
be as in Problem~\ref{pgct6subgroup},
where $H$ is connected, reductive subgroup of $G$.
Quantization of this couple is the crux  of the problem. All other
quantizations that are needed are hyped up versions of this, so we shall
only concentrate on this.

The standard theory of quantum groups can not be used for 
quantizing this couple, as expected. Specifically,
let $G_q=GL_q(\C^n)$ be the standard quantum group associated with $G$.
In a similar fashion,  one can associate \cite{drinfeld,jimbo,rtf} 
a standard quantum group $H_q$ with $H$. Then, $H_q$ cannot be embedded
as a quantum subgroup of $G_q$ (where the notion of subgroup in the
quantum setting is akin to the usual notion of a subgroup). Hence
the goal is to associate a quantization $\hat G_q$ with $G$ 
akin to the standard quantum group $G_q$  so that 
the standard quantum group $H_q$ 
is a quantum subgroup of $\hat G_q$. In that case:
\begin{equation} \label{eqprimaryq}
H_q \hookrightarrow \hat G_q,
\end{equation}
can be considered to be a quantization of (\ref{eqprimary}).

This quantization  step is
addressed in the following result for the 
couples in Problems~\ref{pgct6kronecker}-\ref{pgct6plethysm},
which are the main prototypes of the 
couples that arise in GCT.

\begin{theorem} \label{tnonstdquantum}

\noindent (1) (cf. \cite{GCT4}) The couple
\[ H=GL(\C^n) \times GL(\C^n) \rightarrow GL(\C^n \otimes \C^n)=G,\]
associated with the Kronecker problem (Problem~\ref{pgct6kronecker})
can be quantized in the form:
\[ H_q \rightarrow \hat G_q,\] 
where $H_q$ is the standard quantum group associated with $H$ and
$\hat G_q$ is the new {\em nonstandard} quantum group associated with
$G$. Furthermore, $\hat G_q$ has a quantum unitary subgroup $\hat U_q$
in the sense of \cite{wor1}, which is a quantization of the unitary 
subgroup $U=U_{n^2}(\C)\subseteq G=GL_{n^2}(\C)$.

\noindent (2) (cf. \cite{GCT7}) More generally, the couple
\begin{equation} \label{eqquantumpl}
 H=GL_n(\C) \rightarrow G=GL(V_\mu(H)),
\end{equation}
associated with the plethysm  problem (Problem~\ref{pgct6plethysm})
can also be quantized in the form:
\[ H_q \rightarrow \hat G_q,\] 
where $H_q$ is the standard quantum group associated with $H$ and
$\hat G_q$ is the new {\em nonstandard} (possibly singular) 
quantum group associated with $G$. 
Here  $H$ can even be any connected classical reductive group.
\end{theorem}

The nonstandard  quantum group in \cite{GCT4} 
is  qualitatively similar to the standard 
quantum group in \cite{drinfeld,jimbo,rtf} in the sense
that it has a maximal  quantum unitary subgroup just as in the 
the standard case. This, in conjunction with work in
\cite{wor1},  allows the mathematical  machinery  related to
unitariness--such as harmonic analysis, existence of orthonormal bases--to
be transported to its theory. This is important
in the context of PH0. Indeed, PH0 in the theory of standard quantum 
groups is intimately related to existence of  unitary quantum subgroups.
Because the local crystal bases for representations of the standard quantum
group \cite{kashiwara1}, which were later 
globalized to canonical (global crystal) bases
in \cite{kashiwara2}, arose in the study of special orthonormal Gelfand-Tsetlin
bases for representations of the standard quantum group. 
This is first main reason why PH0 is expected to hold for the nonstandard
quantum group in \cite{GCT4}.

The general nonstandard quantum group \cite{GCT7} can be singular, i.e., its 
quantum determinant can vanish. Hence, we cannot define its quantum unitary 
subgroup  in the sense of \cite{wor1}. Fortunately, this is not matter,
because analogues of the main required
results in \cite{wor1} still hold; cf. \cite{GCT7} for a precise statement.
Hence PH0 is  expected to hold for the general nonstandard
quantum groups in \cite{GCT7} as well.

But at the same time these 
nonstandard quantum groups are fundamentally different
from the standard quantum groups.
Hence the terminology nonstandard.
For a detailed description of the differences
between the standard and nonstandard quantum groups, see
\cite{GCT4,GCT7,GCT8}. 
Here we only give a brief description from the complexity-theoretic
perspective.
Towards this end, we  associate a complexity level 
with each of these quantum groups.
This is briefly done as follows.

Suppose $H\hookrightarrow G$ is a primary couple associated with 
a group-theoretic 
class variety for some complexity class $C$ (Definition~\ref{dcharstab}).
Then the 
complexity class of  this primary couple as well as its quantization,
if it exists, is defined to be just $C$.

As we have already noted, the theory of  the standard quantum group is
the theory of quantization of the couple (cf. (\ref{eqdiaglittle}))

\[ GL_n(\C) \rightarrow GL_n(\C) \times GL_n(\C).\] 
This is a primary couple associated with orbit-closure of the trace of an
$n\times n$ matrix (Section~\ref{slittlegct6}), 
which can be computed 
by a  circuit of depth two  using only additions or multiplications by 
constants.
Hence, the standard quantum group corresponds to the complexity class of 
problems that can be solved by circuits of depth two using only 
additions or multiplications by constants, just like expanders 
(Section~\ref{sObsvsExp}). There is no
lower bound problem here to speak of. That is why the standard quantum 
group cannot be used for deriving any lower bound, again like 
expanders.

The couple associated with the Kronecker problem coincides with
the primary  couple 
\begin{equation} \label{eqdetstab2}
GL(\C^m) \times GL(\C^m) \rightarrow  GL(\C^m \otimes \C^m).
\end{equation}
 associated with the $NC$-class variety;
cf. (\ref{eqdetstab}). Not exactly. The primary couple associated 
with the $NC$-variety is slightly different from this, 
but the difference is trivial, and can be ignored.
Theory of the nonstandard quantum group in Theorem~\ref{tnonstdquantum} (a) 
is the theory of quantization of this couple. Hence, 
the complexity class of this nonstandard quantum group can be defined
to be $NC$.

The couple (\ref{eqquantumpl}) that is quantized in \cite{GCT7} 
is not a primary couple of any class variety. But it is qualitatively
similar to the primary couple associated with the $NP$-class-variety 
(Section~\ref{spvsnp}). For this reason, the nonstandard 
quantum group in \cite{GCT7} can be roughly taken to be of superpolynomial 
complexity.

\subsection{PH0 for couples and triples} 
The article \cite{GCT8} gives  a conjecturally correct algorithm 
to  construct  canonical bases of
the coordinate rings of the nonstandard quantum groups
in  \cite{GCT4,GCT7}. These are  natural generalizations of the 
canonical basis in  \cite{kashiwaraglobal,lusztigbook} for the coordinate
ring of the standard quantum group. Further
theoretical and experimental evidence in support of  PH0 for the nonstandard 
quantum group in \cite{GCT4}  is also  given.
For the  problems that have to be addressed in the context of 
the  triples associated with
the class varieties under consideration, see \cite{GCT10}.

\subsection{PH0 for class varieties}
Since the group-theoretic class varieties are essentially determined 
by the associated group triples, once PH0 is proved for the triples,
it should, in priciple,  be possible 
to ``transport'' this knowledge 
from group theory to algebraic geometry, thereby proving PH0 for the
class varieties. 
In \cite{GCT10} is basic plan for this ``transport'' is suggested, with
a  description of the various mathematical problems 
that need to be resolved.

A crucial bridge between group theory and
algebraic geometry for this transport is provided
by Conjecture~\ref{csftclass}, which has to be proved first.
It may be remarked that quantum groups were indeed brought into GCT
precisely for the purpose of proving this conjecture, thereby extending
the proof in \cite{GCT2} for its weaker form 
 (Theorem~\ref{tweaksft}).
A basic plan for this extension via nonstandard quantum groups
is also suggested in \cite{GCT10}.

\subsection{PH1 and SH} 
The journey from PH0 to PH1 in the nonstandard setting should be
akin to the one in the standard setting; cf. \cite{GCT6}.

In summary, the nonstandard quantum groups 
have to be used as a rope, as it were, to pull the  proofs of
the various mathematical positivity hypotheses from the 
constant depth (of the standard quantum groups) to superpolynomial 
depth.

\section{Ultimate mystery: nonstandard Riemann hypotheses?}\label{sriemann}
Now we come to the final chapter of this story: How to prove
PH0, and specifically, correctness of the algorithm in 
\cite{GCT8} for  constructing canonical bases of the coordinate
rings of the nonstandard quantum groups in \cite{GCT4,GCT7} and 
their required  conjectural properties.

For the standard quantum group, as we mentioned in Section~\ref{sintro},
the topological
proof in \cite{lusztigcanonical,lusztigbook} depends on the
Riemann hypothesis over finite fields \cite{weil2} and the related work
\cite{beilinson}. 
The main open problem at the heart of GCT is to extend this work, 
and use it to prove nonstandard PH0. But
the standard Riemann hypothesis over finite fields is not expected to
work in the nonstandard setting; cf. \cite{GCT8}.
Briefly this is because the relevant 
quantized noncommutative algebraic varieties in the nonstandard setting
simply ``disappear'' when specialized at $q=1$. Specifically, unlike
in the standard case, the Hilbert
function of these varieties at $q\not = 1$ is different from the
Hilbert function of the corresponding classical varieties at $q=1$. 
Hence, they look very different from the classical algebraic varieties.
This is why the Riemann hypothesis over finite fields may not  be used
as in the standard case for proving PH0. 

Thus we seem to need nonstandard extensions of the Riemann hypothesis over 
finite fields in the quantized noncommutative
setting to prove the PH0's under consideration. We cannot 
even formulate such extensions. But we believe such
nonstandard extensions  exist. We now briefly explain why.

For this, we need to indicate the nature of the experimental evidence
\cite{GCT8}
in support of PH0 for the most basic nonstandard quantum group for 
the Kronecker problem in \cite{GCT4}. Specifically, around a  thousand 
 structural constants associated  with a canonical basis 
for a certain dual  of this quantum group were computed, each
structural constant being a polynomial in $q$ of degree more than ten.
All the coefficients of these structural polynomials turned out be 
nonnegative. 
In the standard case, the cause for such nonnegativity 
was  the Riemann hypothesis over finite fields.
There ought to be a similar theoretical  cause for
  nonnegativity in the nonstandard setting.
For,  without a cause, the probability of 
over ten thousand coefficients being nonnegative would be
absurdly small--naively $1/2^{10000}$. This estimate, being naive,
should not be taken literally. But  it does suggest that the experimental
evidence for positivity should  only be a shadow of the ultimate 
cause--nonstandard analogues of the Riemann hypotheses over finite fields.

This leads us to believe  that nonstandard 
extensions of the Riemann hypothesis over finite fields for the
various nonstandard quantum groups that arise in GCT exist, and
now, having seen the shadow, we have to search for the ultimate cause
whose shadow it is. 
If this search succeeds, then 
we can expect to pull the 
proofs of PH0 from the standard to the nonstandard setting,
using the rope provided by the nonstandard quantum groups, and the 
power provided by nonstandard Riemann hypotheses,
thereby leading to the
proof of $P \not = NP$ conjecture in characteristic zero; cf. 
Figure~\ref{fgct6}.

Eventually, this whole story in characteristic zero,
along with the nonstandard Riemann
hypotheses and the accompanying positivity hypotheses, may be
lifted, as suggested in \cite{GCT11},
to algebraically closed fields of positive characteristic, and
finally, finite fields, thereby proving the $P\not = NP$ conjecture
in its usual form. 
This would then constitute the ultimate flip in Figure~\ref{fgctultimate}.

\section{Obstructions vs. expanders}  \label{sObsvsExp}
We now explain the relationship between explicit construction of
obstructions and explicit construction of expanders as shown in
Figure~\ref{fobsvsexpintro}. 

\begin{figure}[!h]
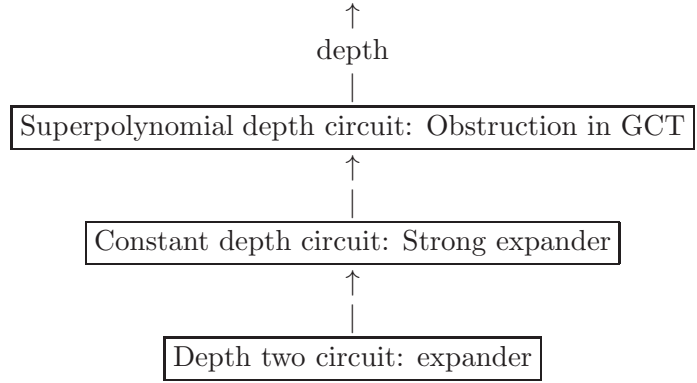

\centering
\[\begin{array} {c} 
\uparrow \\
\mbox{depth}\\
|\\
\fbox{Superpolynomial depth circuit:  Obstruction in GCT}\\
\uparrow\\
|\\
\fbox{Constant depth circuit: Strong expander} \\
\uparrow\\
|\\
\fbox{Depth two circuit: expander} \\
\end{array}\]
\caption{The relationship between obstructions and expanders} 
\label{fobsvsexp}
\end{figure}

As per the hardness-vs-randomness principle 
\cite{agrawal,wigderson,kabanets,nisan}, derandomization is
intimately linked to lower bound problems. In particular, 
restricted  kinds of lower bounds follow from existence of
efficient pseudo-random generators. 
At present,  we do not have pseudo-random generators based on expander-like 
structures that  can  yield  a lower bound result for
constant depth circuits. But for the sake of discussion, let us imagine 
that the expander in, say,   \cite{sarnak,margulis}
can be generalized 
further to obtain a hypothetical structure, which we shall call a
{\em strong expander}, using which we can obtain 
an efficient  pseudo-random generator, whose existence, in turn, implies 
separation of the class $NC^1$ from $AC^0$.
Here $NC^1$ is the class of
problems that can be solved by circuits of  logarithmic depth,
and  $AC^0$ the class of problems that can be solved by circuits of
constant depth. Furthermore, let us also assume that the problem of 
constructing such a strong 
expander belongs to (nonuniform, algebraic) $AC^0$, as it does 
for  the expander in \cite{sarnak,margulis}.
Now the existence of such a family $\{E_n\}$ of strong expanders
would imply that an  explicit function in $NC^1$, depending
on the pseudo-random generator, cannnot 
be computed by a circuit of constant depth. 
Hence  such  strong expanders can  be regarded as
obstructions, i.e. proofs of hardness,
for computation of an explicit $NC^1$-function by 
constant-depth circuits. In this sense
GCT obstructions are to superpolynomial-depth circuits are what 
strong expanders are to constant-depth circuits. 
This is pictorially depicted in Figure~\ref{fobsvsexp}.

The  expander in \cite{sarnak,margulis} can actually be constructed 
by a nonuniform algebraic circuit of depth two (a basic ring operation
is taken as unit cost).
Hence, it can be expected to 
serve as an obstruction for computation by a circuit of
depth at most two--really, just one. Because the depth of a
circuit for computing  an explicit structure whose
existence separates $NC$ from (nonuniform) $AC^k$, the class of
circuits of depth $k$, should be at least   $k$--really higher than $k$.
So an expander, as against 
the hypothetical strong expander, actually belongs to depth-two circuits.
But there is no nontrivial lower bound
problem for circuits of depth one.
This is  why the 
expanders that we have at present cannot be used in lower bound problems.

Now let us compare explicit construction of expanders
with the suggested  method for explicit construction of obstructions 
in   Figure~\ref{fgct6}.

First,  let us observe that, though 
the explicit construction of expanders \cite{sarnak,margulis}
is  ``extremely easy'' 
(nonuniform $AC^0$), its  correctness is based on a  nontrivial 
{mathematical positivity hypothesis}:

\noindent {\bf PHspectral:} 
The spectral gap of an expander  is bounded below by a positive constant. 

The mathematical positivity hypotheses PH1 and PH2
(Hypothesis~\ref{hph1genintro}-\ref{hshgenintro})
can  be regarded as nonspectral analogues 
of PHspectral in the setting of superpolynomial depth circuits.

Second, the proof of PHspectral in \cite{sarnak} for expanders depends on 
the Riemann hypothesis over finite fields (for curves) \cite{weil2}.
It should not be a surprise then that what is
needed to prove the positivity hypotheses PH1, SH (PH2) mentioned above
is, in essence, an
extension of the Riemann hypothesis over finite fields and the
results surrounding it.
But given the big gap between constant depth  and superpolynomial
depth circuits it would have been a great surprise if the 
existing {standard}  Riemann Hypothesis over finite field were 
to suffice. %And it does not seem to, as expected.
Instead, what seems to be needed 
are {nonstandard extensions} of the Riemann hypothesis over finite fields,
and the related results; cf. Section~\ref{sriemann}.
In the case of expanders, the Riemann hypothesis over finite
fields is not indispensible, since there are  alternative constructions
of expanders with  proofs of correctness based on linear algebra
  \cite{reingold2}. 
But, again given a big gap between constant depth and superpolynomial depth,
it should not be surprising if  nonstandard extensions of the Riemann 
hypothesis  turn out
to be indispensible in the context of the $P$ vs. $NP$ problem.

\section{On relativization and  $P/poly$-naturalization  barriers} 
\label{snatural}
In this section we point out why the flip 
should  be nonrelativizable and non-$P/poly$-naturalizable.

We  already mentioned one reason for why  the flip should be 
nonrelativizable:
namely, the ``reduction'' from hard nonexistence to easy existence
is not a formal Turing machine reduction.
There is also another reason.
For this, let us examine why
the proof of $IP=PSPACE$ result \cite{shamir}  does not seem  relativizable. 
Mainly because it is based on  the construction
of an explicit low-degree polynomial.
This seems already enough to make it nonrelativizable, though
the proof technique is not fully explicit. (Because
it makes use of  estimates on the number of
roots of a low degree polynomial. Any technique based on counting or
estimates is, by definition, not fully explicit).
In contrast, the flip is to be implemented using
explicit algebro-geometric and representation-theoretic constructions.
This is why it should be nonrelativizable.

Now we turn to the $P/poly$-naturalizability barrier \cite{rudich}.
Intuitively, 
this too 
should be crossed simply because everything is to be done explicitly and
constructively. Recall that explicit construction of
obstructions  is for superpolynomial depth
circuits what 
explicit construction of expanders is for depth-two circuits (Section~\ref{sObsvsExp}).
The usual probabilistic (nonconstructive) 
proof for existence of expanders may be considered to be 
$P/poly$-naturalizable--as the probabilistic proofs \cite{sipser}
of lower bounds for constant depth
circuits--whereas the proof via explicit construction in 
\cite{sarnak,margulis,reingold2} may be considered non-$P/poly$-naturalizable.
This is only an analogy. Strictly speaking,
there is no notion of $P/poly$-naturalization
for constant depth circuits. Rather,
this barrier lies between
the circuits of constant depth to which the expanders correspond and 
the circuits of superpolynomial depth to which the obstructions correspond.
But this analogy should intuitively explain why the flip should cross this
barrier.

Now we  turn to a more  formal argument.
We begin by  recalling  the notion of a
$P/poly$-naturalizable proof \cite{rudich}. We use the formal
term $P/poly$-naturalizable proof  instead of the informal term natural proof,
because otherwise GCT, and hence, the algebro-geometric and 
quantum-group-theoretic techniques that enter into it would have to be called 
unnatural. That may seem paradoxical, especially since quantum groups
arose in the study of natural phenomena in theoretical physics.

Let $F_n$ be the set of $n$-variable boolean functions.
By a {\em property} of boolean functions, we mean a family of 
subsets  $C_n\subseteq F_n$ for every $n$.
It is called {\em useful} if the circuit size of any function 
$h(X)=h(x_1,\ldots,x_n) \in C_n$ 
is super-polynomial. It is called {\em $P/poly$-natural}
 if it contains a subset
$C_n^*$ satisfying the following two constraints:

\noindent {\bf Constructivity}: Whether a given 
$h(X)$ belongs to $C_n^*$ can be
decided in time polynomial in the size $N=2^n$ of the truth table of 
$h(X)$.

\noindent {\bf Largeness}: 
\begin{equation} \label{eqlarge1}
|C_n^*|/|F_n|\ge 1/N^k,
\end{equation}
for some fixed $k$.

A proof technique based on a useful $P/poly$-natural property is called 
$P/poly$-naturalizable. The article \cite{rudich} says that
the $P\not = NP$ conjecture would  not have a $P/poly$-naturalizable 
proof under reasonable assumptions.

Next, we translate this notion to the setting wherein  the base field $K$ of
computation is algebraically closed, as in this article.
We assume that $K=\C$ or 
$K=\bar F_p$, the algebraic closure of a finite field $F_p$. 
Let $F_n$ be the set of $n$-variable 
polynomials  of degree $d(n)$ for some fixed function
$d(n)=2^{\poly(n)}$.
If $K=\C$, we assume that each polynomial in
$F_n$ is  an integral polynomial
whose coefficients have $\poly(n)$ bitlength.
If $K=\bar F_p$, we assume that all coefficients  belong to $F_p$,
and that the bitlength $\bitlength{p}=\poly(n)$.
Let $N$ denote the total number of coefficients of $h(X)$.
The total bitlength of the  specification  all coefficients of $h(X)$ is
$N$, ignoring a $\poly(n)$ factor.
Hence we let it play the role of the  truth-table-size in what follows.
This leads to the following  straightforward generalization of the notion
of a $P/poly$-naturalizable proof over $\C$ or $\bar F_p$.

By a property, we now mean   a subset $C_n \subseteq F_n$, for each $n$.
It is called useful if the circuit size over $K$ of any function 
$h(X) \in C_n$ 
is super-polynomial. It is called $P/poly$-natural if it contains a subset
$C_n^*$ satisfying the following two constraints:

\noindent {\bf Constructivity}: Whether a given 
$h(X)$ belongs to $C_n^*$ can be
decided in $\poly(N)$ time, where each operation over $K$ is considered
to be of unit cost.

\noindent {\bf Largeness}:
\begin{equation} \label{eqlarge2}
|C_n^*|/|F_n|\ge 1/N^k,
\end{equation}
for some fixed $k$.

A proof technique based on a useful $P/poly$-natural property is called 
$P/poly$-naturalizable. 
The results in \cite{rudich} are proved only over a finite field. 
But the constructivity and largeness constraints over algebraically closed
fields here are natural extensions of the ones over finite field. 
Hence, we shall assume in what follows that they are meaningful even
over algebraically closed fields. It would be interesting to know 
if the  techniques in \cite{rudich} can be lifted in some form
to such fields.

In the context of the flip,
we  next   formulate 
a property which is conjecturally useful and 
which should violate both the largeness and the constructivity constraints.
This should be enough to cross the $P/poly$-naturalizability barrier.

Let us 
follow the notation as in Section~\ref{sclass}. Let $K=\C$.
Let $h(X)\in P(W)$ be an integral 
homogeneous form in $F_n$ 
that belongs to co-NP (i.e., the problem of deciding if it is nonzero for
given $x_i$'s belongs co-NP).

Let UP (useful property)  be the conjunction of the following two  properties.

\noindent {\bf UP1:}  The form $h=h(X)$ is co-NP-complete.

\noindent {\bf UP2: (Characterization by stabilizers)}

The form $h$, as a point in $P(W)$, 
is characterized by its stabilizer $G_h \subseteq GL(W)$, not exactly
as in Definition~\ref{dcharstab},
but in a relaxed manner as described in Section 7 in \cite{GCT1}.
So also the form $f=\phi(h)$ as a point in $P(V)$.
This means the associated class varieties $\Delta_W[h;n]=\Delta_W[h]$,
and $\Delta_V[f;n,l]=\Delta_V[f]$, 
as defined in Section~\ref{spvsnp} with $h(X)$ playing the role of $E(X)$,
are group-theoretic. Let 
$(H_1\hookrightarrow G_1 \hookrightarrow K_1)$ and 
$(H_2\hookrightarrow G_2 \hookrightarrow K_1)$ be the group-triples 
associated with the varieties $\Delta_W[h;n]$ and 
$\Delta_V[f;n,l]$, respectively. We assume that 
$H_1$ is reductive,
that its simple composition factors are explicitly known, and
that it is built from  these composition factors by simple operations:
to keep the matters simple, we only allow direct or wreath products,
which suffice in GCT. We also assume that all the simple composition factors
are either classical connected groups, tori or  alternating groups,
as in GCT, though, again, this is  strictly not necessary.
We also assume that 
all  homomorphisms in these triples  are explicit  as  defined
in  \cite{GCT6}--this is necessary.

Here UP1  is stipulated only so that 
obstructions to efficient computation of $h(X)$ should exist 
(Section~\ref{swhyobsex}).
Otherwise, it is never explicitly used in GCT. The approach may
work for other hard, though not co-NP-complete functions.
UP2  is the main property  that GCT needs for proving
existence of obstructions. Hence we shall only concentrate on it in
what follows.

It is shown in \cite{GCT1} that $E(X)$ has property UP2; whereas
UP1 over $\C$ is shown in \cite{gurvits}.
The permanent has an analogous property,
where co-NP-completeness is replaced by $\#P$-completeness.
The function $H(Y)$ also has an analogous property with $P$-completeness
replacing co-NP-completeness. But in the case of $H(Y)$ the
class variety is not the usual orbit closure $\Delta[H(Y)]$,
but rather $\hat \Delta[H(Y)]$ as defined in \cite{GCT1}; cf.
Remark~\ref{ractualvariety}.

In these definitions,  we can also let the base field $K$ be a
finite field $F_p$, or its algebraically closure 
$\bar F_p$, since 
characterization by stabilizers is a well-defined notion over any field.

\subsection{From usefulness to superpolynomial lower bounds}
Though GCT strives to prove superpolynomial lower bounds for the particular 
functions $E(X)$ and $\perm(X)$, its  main techniques should,
in principle, extend to any $h$ satisfying UP.
We now briefly indicate  how. 
This should justify the name  UP.

Define an obstruction for the pair $(h(x),H(Y))$ as in
Definition~\ref{dobs}, with $h(X)$ playing the role of $E(X)$.
Such obstructions should exist for every 
$n\rightarrow \infty$, $l=n^{\log n}$, as long as $h(X)$ is
co-NP-complete (cf. Section~\ref{swhyobsex}). 
Associate with the class variety $\Delta_V[f;n,l]$ a stretching
function $\tilde s_d^\lambda(h;n,l)(k)$ as in (\ref{eqstretching})
with $h(X)$ playing the role of $E(X)$.

The results in \cite{GCT6} now imply
the following  analogue of Theorem~\ref{tmainquasipoly}
for $h(X)$:

\begin{theorem} \cite{GCT6} \label{tgct6h}
Assuming that the singularities of the class variety 
$\Delta_V[f;n,l]$ and $\Delta_W[h;l]$ are rational,
the stretching function  $\tilde s_d^\lambda(h;n,l)(k)$ 
associated with the class variety 
$\Delta_V[f;n,l]$ is a quasi-polynomial
\end{theorem} 
It is may be conjectured that the singularities will be rational,
as needed here, as long as $h$ satisfies UP2.

Using this theorem, we can formulate PH1, PH2, and SH for $h(X)$
just as for $E(X)$ (cf. Hypotheses~\ref{hph1genintro}-\ref{hshgenintro}).

\begin{remark} 
The statements of PH1 and PH2 given in this paper are assuming that
all simple composition factors of the reductive groups under consideration
are either classical connected 
groups or tori or alternating groups. In the presence of
composition factors of other types, some variations are necessary
\cite{GCT6}.
\end{remark}

The following is an analogue of Theorem~\ref{tmaincomplexity} in this context.

\begin{theorem} \cite{GCT6} \label{tgct6hx}
Assuming the rationality hypothesis (cf. Theorem~\ref{tgct6h}),
PH1 and SH,  analogues of the decision problems in
Hypothesis~\ref{hdecision}
for $h(X)$  belong to $P$. In particular, the problem of 
verifying  an obstruction for the pair $(h(X),H(Y))$
belongs to $P$.
\end{theorem}

These results suggest, just as for $E(X)$, the following 
strategy for proving a superpolynomial lower bound for $h(X)$:

\noindent {\bf (1):} Let 
$H_i \hookrightarrow G_i \hookrightarrow K_i$ be the triples 
that occur in the definition of UP2.
Quantize the couples $H_i \hookrightarrow G_i$ and
$G_i \hookrightarrow K_i$.  That is, prove 
analogues of Theorem~\ref{tnonstdquantum} for these. Also quantize the triples
along the scheme suggested in \cite{GCT10}.

\noindent {\bf (2):} Prove existence of canonical bases (PH0) for the 
coordinate rings and representations of the quantum groups that 
arise in this quantization along the lines of the basic scheme 
in \cite{GCT8}. For formal statements of PH0 see 
\cite{GCT6,GCT10}.

\noindent {\bf (3):} Use these canonical bases to prove existence of 
canonical bases (PH0) for the coordinate rings of the class varieties
$\Delta_V[f;n,l]$ and $\Delta_W[h,l]$ along the lines suggested in
\cite{GCT10}.

\noindent {\bf (4):}  Use PH0 to deduce PH1 and SH, 
as suggested in \cite{GCT6}. The polytope in PH1
should be  more or less determined once
PH0 holds, just as in the standard case; cf. Section~\ref{snonstdintro}.

\noindent {\bf (5):} 
Theorem~\ref{tgct6hx}, in conjunction with PH1 and SH for  $h(X)$,
then  implies polynomial 
time algorithm for the analogue of the decision problem in 
Hypothesis~\ref{hdecision} (a) for h(X) in place of $E(X)$.

\noindent {\bf (6):}  Carry out the steps (1)-(5) 
for the $P$-complete function $H(Y)$ as well.
It will imply a
polynomial time algorithm for the decision problem in 
Hypothesis~\ref{hdecision} (b)
for  $H(Y)$.
This step is  the same as for $E(X)$.

\noindent {\bf (7):} Transform the easy, polynomial time algorithms 
in steps (5) and (6),
along the lines suggested in Section~\ref{sreduction} and \cite{GCT6},
into a $P$-constructive  proof of existence of an obstruction $\lambda(n)$
for every $n\rightarrow \infty$, assuming that $l=n^{\log n}$.
As pointed out in \cite{GCT6}, this transformation may need additional
positivity hypotheses in the spirit of PH1 and SH. But these 
can be expected to hold, assuming  $h(X)$ satisfies UP2.
The polytope in PH1 for $h(X)$ in the step (4) can
also be expected to have 
a regular well-behaved 
structure, as needed for this step, assuming $h(X)$ satisfies UP2.

\noindent {\bf (8):} Existence of an obstruction family $\{\lambda(n)\}$
would imply  a superpolynomial size circuit 
lower bound for $h(X)$, and hence, that $P\not = NP$ over $\C$.

For the problems that need to be addressed over a finite field,
or an algebraically closed field of positive characteristic,
see \cite{GCT11}.

\subsection{On violation of the largeness  constraint}
Now  let us see why UP2 should imply violation of the
largeness constraint. We cannot prove this
formally over $\C$. But this
can be proved formally over a finite field $F_p$ or 
an algebraically closed field  $\bar F_p$  of positive characteristic
\cite{GCT10}. In fact, it turns out that  violation of the largeness
constraint is far more  severe  than what is formally required.
Namely, when $K$ is  a finite field, it can be shown that
\[ |C_n|/|F_n|\ge 1/2^{\Omega(N)},\] 
where $C_n$ is the set of $h(X)$ which satisfy UP2. 
This may be compared with (\ref{eqlarge1}).

The proof of violation of the largeness constraint over $\bar F_p$
does not carry over to $\C$ for technical reasons.
Specifically, the key ingradient in this proof
is the Riemann hypothesis over finite
fields, or rather its extension as proved in \cite{weil2}. 
To transport this to the case when  $h(x)$ is integral would presumably
require an analogous statement in arithmetic algebraic geometry.

It may be remarked that, in contrast, the proof of violation of the
largeness constraint over a finite field is elementary. 
Thus the difficulty of proving the violation of the largeness constraint
over $K$
seems inversely related to the difficulty of proving the $P\not = NP$
conjecture  over $K$. When $K=F_p$, the conjecture is hardest to prove,
and hence, the proof of violation is easy. When $K=\C^*$, the conjecture
should be  easier than over $\bar F_p$ or $F_p$. Accordingly,
proving violation of the largeness constraint formally  turns out
to be the hardest.

\subsection{On violation of the constructivity constraint}
Next  let us see why UP2  should also imply violation of the 
constructivity constraint. We cannot hope to show this formally,
since this is a lower bound statement in itself. But rather we can
give  good evidence. First of all, to compute the stabilizer of
$h(X)$, we have to solve a system of polynomial equations.
Determining feasibility of  a general system of polynomial equations 
in $k$ variables  is $NP$-complete and
is conjectured to take $\bitlength{p}^{\Omega(k)}$ time, when $K$ is the
finite field $F_p$.
Analogous conjecture may be made for the specific 
system of polynomial equations that arises in the computation of the
stabilizer.
Assuming this, it follows \cite{GCT10} that
deciding if $h(X)$ has a nontrivial stabilizer 
would take time  that is superpolynomial in $N$--this is the truth-table size
when $K=F_p$. 

\section{$P$-verifiable and $P$-constructible proof techniques
and their explicit construction complexity}\label{spveri}
In this section we suggest why GCT may be among the ``easiest'' 
``easy-to-verify'' approaches 
to the $P\not =NP$ conjecture as per a certain measure of proof-complexity,
called the explicit construction complexity.
For this, we have to introduce the notion of an easy-to-verify
(i.e. $P$-verifiable) proof technique
and then define its explicit construction complexity (class).

\subsection{$P$-verifiable proof technique} 
Suppose we are given a proof technique (approach) towards to 
the $P$ vs. $NP$ problem 
that  seeks to prove a superpolynomial 
lower bound for a specific hard function $h(X)$ under consideration.
We assume that the approach seeks to
prove, explicitly or implicitly,  existence of a specific 
cause for the hardness of $h(X)$,
which we shall refer to as  an {\em obstruction}.
Thus an obstruction is, roughly, a ``cause'', a ``witness'' or a ``proof'' of 
hardness. 

But what do we mean by a proof?
The final proof of the $P\not = NP$ conjecture, if true, would constitute
the ultimate obstruction to efficient computation of every (co)-NP-complete
$h(X)$.
The size of this proof would be just $O(1)$, and so also
the cost its verification.
By obstruction, we do not mean this final proof of hardness, but rather
an intermediate proof of hardness whose existence the approach strives 
to demonstrate for every $n\rightarrow \infty$, when 
the circuit size  $m=n^{\log n}$, say.

The nature of such an obstruction will depend on the proof technique.
We cannot define it formally.
Hence we will only give an intuitive idea with an example.
Suppose there is  an efficient pseudo-random generator
whose existence implies a restricted type of lower bound result
in the spirit of \cite{nisan}. Then the explicit computational circuit
for this pseudo-random generator, i.e., for  all its output bits together 
would be an obstruction in this context. Because existence of this
pseudo-random generator serves as a witness for hardness.
If the pseudo-random generator is based on an explicit structure in 
the spirit of an expander, then  this structure too can  be considered
to be  an obstruction.  
More generally, the hardness-vs-randomness principle \cite{kabanets,nisan}
suggests that  proof techniques for difficult lower bounds may need 
more or less explicit constructions of some structures. These structures,
which serve as witnesses for hardness, can then be taken as obstructions.

In the rest of this section, we confine ourselves only to
those  techniques towards the $P\not = NP$ conjecture which contain,
explicitly or implicitly, the 
notion of an obstruction in this spirit--a witness 
for hardness--which admits a 
well-defined description that can be assigned bit length.
The arguments henceforth are subject to this assumption.

Next we try to  formalize  the notion of a ``viable'' proof technique
towards the $P$ vs. $NP$ problem. 
For this, let us begin with a technique that should certainly 
not be considered viable--the trivial brute-force proof technique.
This is defined 
as follows.
Assume that the base field $K$ is finite. Fix 
any co-NP-complete function $h(X)=h(x_1,\ldots,x_n)$.
Then this proof technique 
strives to prove, for every $n$,  existence of 
the  trivial proof of hardness (obstruction), which consists of just the 
enumeration of  all circuits of size $m=n^{\log n}$, with a specific 
value of $X$ for each circuit on which the function evaluated by the
circuit differs from $h(X)$.
The size of this  trivial obstruction is exponential in $m$, and the
time taken to verify it is also exponential in $m$.
Any viable proof technique for the $P$ vs. $NP$ problem ought  to
be at least 
better than this trivial proof technique in some well defined sense.
One obvious  sense in which it could be better is that
there  exists an obstruction  whose size is not exponential in $m$,
but rather polynomial in $m$, and the time taken to verify an obstruction
is not exponential in $m$, but rather polynomial in $m$.

This leads to:

\begin{defn} \label{dpveri1}
We say that a proof technique for   the $P\not = NP$ conjecture  is 
{\em $P$-verifiable} if 
\begin{enumerate} 
\item There is a well-defined notion of obstruction, either implicit or
explicit in the technique,
\item There exists a  {\em short} obstruction to  
computation of the
specific function $h(X)=h(x_1,\ldots,x_n)$ under consideration
by a circuit of size $m=n^{\log n}$ (say), for every $n\rightarrow \infty$,
though the technique may only strive to prove existence of any obstruction,
not necessarily short.
By short, we mean  the obstruction
has a label (combinatorial specification) of bit length $\poly(m)$. 
\item The problem of verifying an obstruction is easy; i.e.,
belongs to $P$. Specifically, takes
time that is polynomial in $n,m$ and the bit length of the obstruction.
\end{enumerate} 
\end{defn}
The meaning of easy here
is the most obvious and natural definition  in the 
context of the $P$ vs. $NP$ problem.
Thus, intuitively a $P$-verifiable proof technique
is an {\em easy-to-verify} proof technique.
That is, the problem of discovering a proof of hardness (obstruction)
in the technique belongs to $NP$.
The definition above makes sense over any base field $K$ of computation,
with obvious modifications in the spirit of the ones 
in Section~\ref{snatural}.

The following  naive arguments suggest that for a technique towards
the $P\not = NP$ conjecture to be viable it out to be 
$P$-verifiable.

First, the usual experience in mathematics suggests that 
however hard the discovery of a proof may be its verification, once
found, should be easy, and furthermore, the proofs that are found
are usually  reasonably short.
In the definition of $P$-verifiability, short and 
easy are given the most obvious and natural interpretations in the 
context of the $P$ vs. $NP$ problem: description of polynomial size (short),
and can be done in polynomial time (easy).

Second, given a technique, it seems necessary to 
justify why it  is better than the trivial brute-force technique.
A $P$-verifiable proof technique is better than it
as per the most obvious complexity measures:
(1) space (short), and  (2) time (cost of verification).

Third, the  article \cite{rudich}
roughly says that a nonspecific approach
that is applicable to  a large fraction of hard functions should not
work in the context of the $P$ vs. $NP$ problem.  Thus  approaches
based on probabilistic methods or estimates of various kinds--such as 
Bezout-type estimates in algebraic geometry, or  estimates for
discrepancies and deviations in analysis or number theory--should 
not work. Proof of hardness  as per any such approach--namely, the
value  of the measure or the estimate which is the cause of hardness--should 
be hard to verify. Since to verify the value, we may  have to compute 
it and see that it really tallies with what is given, and such
computations should typically take time that is exponential in the bitlength
of the value. Thus a $P/poly$-naturalizable proof should also be 
non-$P$-verifiable, and hence,
the definition of $P$-verifiability here 
seems  consistent with the arguments in \cite{rudich}.

Admittedly, these are only  naive arguments.
One can ask if there exists  a viable  proof technique for
the $P\not = NP$ conjecture 
that is better than the trivial brute-force 
technique as per some  measure
of complexity other than the obvious ones--space and time.
But since we cannot  think of  any such 
nonobvious complexity measures which are 
also natural in the context of the $P$ vs. $NP$ problem, 
we shall confine ourselves to only $P$-verifiable proof techniques
in what follows.

By Hypothesis~\ref{hphflipformalnew} (a), which is supported by the
results that we described in this article, GCT is a $P$-verifiable proof
technique over $\C$; the story over a finite field should be similar
\cite{GCT11}.

\subsection{$P$-barrier for verification}
Every $P$-verifiable technique for 
the  $P\not = NP$-conjecture has to cross the $P$-barrier for verification;
i.e., surmount the difficulty of showing that verification of an obstruction 
is easy.  The magnitude and
difficulty of the $P$-barrier should   be of  the same order
regardless of which
$P$-verifiable  approach to the $P\not = NP$ conjecture is taken.

This is easy to see when 
the base field  $K$ is  finite. Then  the length of 
any  obstruction in the trivial brute-force technique  mentioned in the 
beginning of this section  is
exponential. The main task here is to come with a 
proof technique that admits  short obstructions which can be verified
easily. The magnitude of this $P$-barrier--the difference between
the exponential and
the polynomial--is the same regardless of which
approach to the $P\not = NP$ conjecture is taken.

Next, let us assume that  $K=\C$, as in this paper.
Let $n$ be the number of input parameters. Let  $m(n)=n^{log n}$ (say) be
the circuit-size parameter,  $h(n)=n^{log n}$ the  height-parameter, 
and $d(n) \le 2^{h(n)}$ the degree parameter in the lower bound problem under
consideration.
For a given $n$, the set of functions over $\C$ computable by circuits 
of size at most $m=m(n)$, height at most $h=h(n)$ and  degree at most $d=d(n)$
is an algebraically  {\em constructible} \cite{mumford} 
subset $S$ of the space $V$ of all forms in $m$ variables
of degree $d(n)$.  A constructible subset means it 
is in the boolean algebra generated by
closed algebraic subsets of $V$; this is a  generalization of an
affine variety. 

The goal in the lower bound problem under consideration is to show
that $h(X)$ does not belong to $S$, when $m=n^{\log n}$.
Let $\bar S$ be the closure of $S$. It is an affine variety.
If $h(X)$ is co-(NP)-complete,
it is reasonable to assume
that it does  not belong to $\bar S$ as well; 
i.e., roughly speaking,
it  cannot be approximated infinitesimally closely by a circuit of
size $m(n)$ and height $h(n)$.
So it would suffice to show this.

An obvious obstruction here would be a polynomial in the ideal of $\bar S$ 
which does not vanish on $h(X)$. To decide if a given polynomial
belongs to the ideal of $\bar S$, an obvious method is  to compute 
a good basis of this ideal, such as Gr\"obner basis, and then
use it for this decision.
But the problem of Gr\"obner basis computation is EXPSPACE complete 
\cite{mayr}. This means 
computation of the Gr\"obner basis of $\bar S$ 
can take space that is exponential in the dimension of the ambient 
space $V$, which in turn is exponential in $m$. In other words,
space  that is  double exponential in $m$, and hence, time that is
triple exponential in $m$. 
Given that the  $S$ in our problem  is really bad, this is the best 
that we can expect  from any  general purpose technique for verifying 
an obstruction that reasons about $S$ directly in this fashion.

For  the technique to be  $P$-verifiable, the huge gap  between
this triple exponential bound for  a general purpose direct 
technique and the polynomial bound 
in Definition~\ref{dpveri1} has to be bridged.
The magnitude and order of this
gap--the $P$-barrier--is exactly the same that 
we encountered in Section~\ref{seasybarrier}.

\begin{remark} 
The triple exponential size  of  this gap when $K=\C$
as against the   exponential size over  $K=F_p$   does not mean that
the $P\not = NP$ conjecture 
is easier when $K=F_p$. In fact, it is the other way around.
Since the (nonuniform) $P\not = NP$ conjecture in characteristic zero 
(over $Z$) is a weaker implication of the conjecture over finite field
(the usual case) \cite{GCT1}.
Hence, the exponential gap over $F_p$ would be much  harder to bridge than 
the triple exponential gap over $\C$. See  \cite{GCT6} for the problems
that need to be addressed over finite fields.
\end{remark}

The  class variety  $X_P(l)$ for $P$ (Section~\ref{spvsnp})
is constructed in \cite{GCT1} precisely to cope up with the
triple exponential gap over $\C$.
It is a
nice algebraic variety that contains $S$, or rather its projectivization.
So instead of trying to show that a given $h(X)$ is not in $S$,
one strives to show that it is not in $X_P(l)$. Since the algebraic 
geometry of $X_P(l)$ is exceptional, this problem becomes easier--especially
when  $h(X)$ is also exceptional, like $E(X)$. 

The quantum-group and algebro-geometric machinery is needed in GCT just 
to cross the $P$-barrier  for verification (over $\C$).
This suggests that  mathematics required for
any  $P$-verifiable  approach towards the $P\not = NP$ conjecture
may not  be substantially simpler, or easier.

\subsection{$P$-constructible proof technique}
In fact, it may  be much harder unless it is also $P$-constructible
in the following sense. 

\begin{defn} \label{dpconstr}
We say that a  $P$-verifiable 
proof technique for the $P\not = NP$ conjecture is
{\em $P$-constructible} if the discovery of an obstruction in this technique
is also
easy. That is, there exists an algorithm, which, 
given $n$ and $m$, can decide whether there exists an obstruction
in $\poly(m)$ time,
and if so, also construct  a short  obstruction 
in $\poly(m)$ time.
\end{defn}

Thus the problem of discovering an obstruction
in a $P$-constructible proof technique.
belongs to $P$. But 
the  proof  technique itself   need  not 
give a polynomial time algorithm for discovering 
an obstruction explicitly. That is, this may only be implicit in the proof,
or it may be left to posterity.
We call the technique $P$-constructive if it gives such an algorithm 
more or less explicitly:

\begin{defn} 
A $P$-constructible proof technique is called $P$-constructive,
if it also yields a procedure 
to construct an obstruction explicitly in $\poly(n,m)$
time, if one exists.
\end{defn}

The relationship between $P$-constructible (constructive) and 
$P$-verifiable proof strategies is
akin to the relationship between $P$ and $NP$.
The $P\not =NP$ conjecture says that the discovery of a proof is, in general,
harder than its verification. Hence,  just as $P$ denotes the 
class of easy problems within $NP$, the $P$-constructible and 
$P$-constructive  proof strategies
are in a
sense the ``easy'' ones  among the $P$-verifiable proof strategies, wherein
discovery is also easy like 
verification.

By Hypothesis~\ref{hphflipformalnew} (c),
as supported by the positivity hypotheses,
results described in this paper, GCT is  $P$-constructive over $\C$;
the story over a finite is expected to be similar \cite{GCT11}.

That there should exist such a  $P$-constructive  proof technique 
for the $P\not = NP$ conjecture may, however, seem 
paradoxical at the surface. Because  a $P$-constructive (constructible)
proof technique seems to go against 
the very philosophical essence of the $P\not = NP$ 
conjecture that discovery is harder than verification. 
This is akin to the paradox in the proof of G\"odel's incompleteness 
theorem: that the statement which says there exist unprovable true statements
is itself easy to prove.
Similarly, Hypothesis~\ref{hphflipformalnew} (c) says 
that the statement which says discovery
is harder than verification should  itself be easy to discover.

\subsection{General setting} \label{sgeneralsetting}
So far we have described $P$-verifiable and $P$-constructible proof 
techniques only in the context of the $P$ vs. $NP$ problem.
But these notions can be defined in  a much more general context,
as we now briefly indicate:

\begin{defn} \label{dverigeneral}
A  technique for proving a mathematical  property $Q(X)$,
where $X$ ranges over a class ${\cal C}$ 
of mathematical objects under consideration,
is $P$-verifiable 
if:

\begin{enumerate} 

\item  The technique proves, explicitly or implicitly,
existence of a ``proof-certificate'' $c(X)$, for every $X \in {\cal C}$,
which serves as a ``witness'' that the property $Q(X)$ holds.

\item There exists a short proof certificate for every $X\in {\cal C}$.
By short, we mean its size is $\poly(\bitlength{X})$, where $\bitlength{X}$
denotes the specification-complexity of $X$.

\item Verification of a proof-certificate $c(X)$ is easy; i.e., can be
done in $\poly(\bitlength{X},\bitlength{c(X)})$ time, where 
$\bitlength{c(X)}$ denotes the bitlength of $c(X)$.
\end{enumerate} 
\end{defn} 

Again, we cannot formally define 
what a proof-certificate means. In what follows, we only consider
proof techniques wherein the notion of a proof-certificate is well
defined.
The specification complexity $\bitlength{X}$ here depends
on the problem under consideration, as  we shall see  
in the examples below.

\begin{defn} 
A $P$-verifiable proof technique is called $P$-constructible if
there exists an algorithm which,
given $X\in {\cal C}$, can construct a proof certificate $c(X)$ 
in $\poly(\bitlength{X})$ time.
\end{defn} 
But the proof technique itself need not  give such an algorithm explicitly.

\begin{defn}
A $P$-constructible proof technique is called $P$-constructive if, in addition,
it yields an algorithm that can  construct a proof-certificate $c(X)$ in 
$\poly(\bitlength{X})$ time.
\end{defn}

We can now state an informal working hypothesis:

\begin{hypo} {\bf (The $P$-hypothesis) (informal)}  \label{ppconstructible}

\noindent (a) Feasible $P$-verifiable proof techniques--that is the
$P$-verifiable techniques  that can actually be used to prove the
properties $Q(X)$  in practice--are usually 
$P$-constructible, though proving $P$-constructibility may 
turn out to be  nontrivial, and may only be done a posteriori. 

\noindent (b) Conversely, if a $P$-verifiable technique is $P$-constructible
then under reasonable conditions it may  also be feasible, i.e., 
can be used to actually prove $Q(X)$.

\noindent (c) 
A major part of the effort in a $P$-constructible  proof technique 
usually goes towards development of a polynomial time algorithm 
for constructing a proof-certificate, though this may be done only implicitly,
and may  become clear only a posteriori.
That is, a $P$-constructible proof  can 
usually  be extended to a $P$-constructive proof  with 
a ``reasonable additional effort'', albeit a posteriori.

\noindent (d)  Mathematical complexity of a $P$-constructive proof
technique is intimately linked to  the computational complexity 
of the algorithm for explicit construction of a proof certificate 
underlying the technique.
\end{hypo}

As we have already remarked,
the relationship between $P$-constructible 
proof techniques and $P$-verifiable proof techniques is akin to the 
relationship between $P$ and $NP$. The class $P$ is usually regarded
as the subclass of {\em feasible} problems in $NP$. 
Hence, the $P$-hypothesis just says that $P$ usually 
means feasible in practice.

The reasonable conditions in (b) means: there is a polynomial time
algorithm for constructing a proof-certificate, which
has, furthermore,
a reasonably simple structure,  and which is efficient in practice.
That is,  the definition of $P$
as standing for feasible is not misused.

\begin{defn} 
A mathematical theorem, which says  a property $Q(X)$ holds 
for every  $X$ in  a class ${\cal C}$ 
of mathematical objects under consideration, 
is called $P$-verifiable if it has  a $P$-verifiable proof.

A $P$-constructible or a $P$-constructive theorem is defined similarly.
\end{defn}

\subsubsection{Examples} 
We now give a few examples to illustrate these notions.

\subsubsection*{$P$ vs $NP$ problem}
In this  context,
${\cal C}$  is the class of  tuples $(n,m(n))$, $m(n)=n^{c}$ for any
constant $c>0$,  over all $n$ (large enough).
The property $Q(X)$, $X=(n,m(n))$, just says that 
the explicit function $h(X)$ under consideration,
such as  $E(X)$ in \cite{GCT1}, cannot be computed by a circuit of
size $m(n)$ for every $n$ large enough. Here $\bitlength{X}=n+m$; i.e, 
we assume that $n$ and $m$ are given in unary.
Then the notions of $P$-verifiability, and $P$-constructibility here
coincide with the ones in Definitions~\ref{dpveri1} and \ref{dpconstr}.
When $m=n^{c}$, an obstruction would always 
exist, assuming that $h(X)$ is co-(NP)-complete,  $P\not = NP$ and
the technique is correct.
That is why Definition~\ref{dverigeneral}
would  coincide with Definition~\ref{dpveri1},
even if in the former there  is no mention of deciding if an obstruction
exists or not.
As per Hypothesis~\ref{hphflipformalnew},  GCT is $P$-constructive
over $\C$, and hence the $P \not = NP$ conjecture 
over $\C$ is also $P$-constructive;
the same can be hypothesized over a finite field \cite{GCT11}.

\subsubsection*{Hall's theorem}
In this  context, ${\cal C}$ 
is the class of  $d$-regular bipartite 
graphs. The property $Q(X)$ is that every $d$-regular bipartite graph
$X \in {\cal C}$ has a perfect matching. 
The bit length ${\bitlength X}$ is the bitlength of the specification of
$X$. The proof certificate 
$c(X)$ is  a perfect matching in $X$.
The problems of verifying and constructing 
a perfect matching belong to $P$, the former trivially.
Hence, Hall's theorem is $P$-constructive.
Hall's original proof is  $P$-constructible,
though not $P$-constructive, since
it does not explicitly  give a polynomial time  algorithm for constructing
a perfect matching.  
But it does contain major ingradients for  such
a polynomial time algorithm, which came only much later. This is 
consistent with the $P$-hypothesis.

\subsubsection*{Four colour theorem} 
In this context, 
${\cal C}$ is the collection of planar graphs. The property $Q(X)$ is
that any planar graph $X$ is four colourable. 
The bitlength ${\bitlength X}$ is the bit length of the specification of $X$.
The proof certificate $c(X)$ is a four colouring of $X$. 
The problems of verifying and constructing  a proof certificate belong to
$P$, the former trivially. Hence, any proof of the four colour theorem
is $P$-constructible, and the four colour theorem is $P$-constructive.
The actual proof in \cite{appel} is also (more or less)
$P$-constructive since it implicitly yields to a polynomial (quartic)
time algorithm for four colouring,
Indeed, major part of the effort in the proof implicitly goes 
towards development  of such an algorithm.
This is consistent with the $P$-hypothesis.

A simpler $P$-constructive proof was subsequently given in \cite{robertson2},
which gives a better quadratic algorithm for the same problem. This
too is consistent with the $P$-hypothesis (d).

\subsubsection*{Forbidden minor theorem}
Fix a genus $g$.
The forbidden minor theorem
\cite{robertson} says that a graph which does not contain a forbidden
minor from a finite  list of minors depending on $g$ can be
embedded on a genus $g$ surface. Here ${\cal C}$ is the class of
graphs that do not contain a forbidden minor, $Q(X)$ the property 
above, and $\bitlength{X}$ the bitlength of the specification of $X$.
The proof certificate $c(X)$ is just a description that tells  how to
embed $X$ on a genus $g$ surface.

The forbidden minor theorem is $P$-constructive.
Any proof technique for proving the forbidden minor theorem is 
$P$-constructible: it was known  \cite{miller}
even before \cite{robertson} that 
$c(X)$ can be constructed in polynomial $O(\bitlength{X}^{O(g)})$ time.
The proof of the forbidden minor theorem in \cite{robertson}
gave an  $O(f(g) {\bitlength X}^2)$ algorithm, where $f(g)$ depends only
on $g$. Indeed,
a major part of the effort in \cite{robertson} implicitly goes towards 
finding a polynomial time  algorithm whose running time is of the
form $O(f(g) {\bitlength X}^{O(1)})$; i.e., wherein the exponent of 
$\bitlength{X}$ does not depend on $g$. This is again consistent 
with the $P$-hypothesis (d).

\subsubsection*{The Poincare conjecture} 
Here we can let ${\cal C}$ be the set of simplicial decompositions 
of compact three dimensional combinatorial manifolds that are simply
connected. The property $Q(X)$ says that $X$ is a (combinatorial) sphere.
The bitlength $\bitlength{X}$ is the bitlength of specifying $X$.
The article \cite{sphere} says that the sphere recognition problem
is in $NP$. That is,
there is a proof-certificate $c(X)$, verifiable in polynomial
time, which certifies that $X$ is a sphere. 
It is interesting to know here if the problem of constructing 
a proof certificate $c(X)$, for a given $X \in {\cal C}$, belongs to $P$.
It is plausible that the proof technique in \cite{perelman} can 
be extended/transformed  (in the combinatorial setting) 
to get  a polynomial time algorithm which constructs 
a proof-certificate in this spirit, though not exactly the one in 
\cite{sphere}. If that happens, it  would mean that the Poincare 
conjecture is $P$-constructible ($P$-constructive), and 
that the major effort in \cite{perelman} implicitly went towards 
getting a polynomial time algorithm for this problem.
This would provide support for the $P$-hypothesis (c).

Thus a major part of the effort in  the $P$-verifiable proofs
above indeed seems to go towards 
developing a polynomial time or a better polynomial time algorithm 
for constructing a proof-certificate, as per the $P$-hypothesis (c),
though this goal may  not be stated explicitly in the proofs.
In the flip, $P$-constructivity as a 
goal is explicitly spelled out right in the beginning,
given the complexity-theoretic significance  of the $P$ vs. $NP$ problem.
But just as in the examples above, 
it may  not be necessary to prove PHflip 
(Hypothesis~\ref{hphflipformalnew}) fully to prove $P\not = NP$ 
over $\C$. That is, it may suffice to develop only a part 
of all ingradients needed to put the required problems in $P$,
and the remaining part  can be left to posterity.
In this context, the basic minimum that seems to  be needed is PH1
(more or less).

\subsection{Explicit construction complexity} 
We will now try to formalize the intuition behind the $P$-hypothesis (d).
Towards that end
we wish to associate a measure of {\em proof-complexity} with a $P$-verifiable
proof technique. This is quite different, for example, from
Kolmogrov proof-complexity.

\begin{defn} \label{dexpcomplexity1}
Explicit construction complexity of a $P$-constructive technique is
the computational complexity of the algorithm underlying that technique
for explicit construction of a proof-certificate.
\end{defn} 

By computational complexity, we mean the usual measures such as depth 
and size of the corresponding computational circuit.
If a $P$-verifiable 
technique is not explicitly $P$-constructive but naturally leads 
to an algorithm for  construction of obstructions, with
additional effort, we agree to take the computational 
complexity of this algorithm to be explicit construction complexity of
the technique, albeit only a posteriori. 

\begin{defn} \label{dvericlass}

\noindent (a) Verification (complexity)  class  of a $P$-verifiable proof technique
is the abstract computational complexity class of the problem of verifying a 
proof-certificate (as per that technique).

\noindent (b) 
Explicit construction (complexity)
class  of a $P$-verifiable proof technique is
the computational complexity class of the problem of 
explicit construction of a proof-certificate as per that technique.
\end{defn}

A computational complexity class here  means an abstract computation
complexity class such as $P$, $NC$, $NC^k$, $AC$, $Dtime(N)$  etc.
The verification and explicit construction  classes of a
$P$-verifiable technique are well defined 
regardless of whether the technique shows how to construct a 
proof-certificate explicitly or not. But what these classes are
may become clear only a posteriori, possibly
after extending the proof technique to a get an efficient algorithm
for construction of a proof certificate therein.

The
complexity measures and classes above  are meaningful only for
$P$-verifiable proof techniques. They would  not make any
sense for nonconstructive or estimate-based techniques in analysis,
number theory and so forth, unless it is possible to define
a specification complexity $\bitlength{X}$ and  a proof-certificate
that  is polynomial time verifiable with this definition
of $\bitlength{X}$ naturally.

This following gives a notion 
of  {\em theorem complexity} for $P$-verifiable theorems.

\begin{defn} 
Explicit construction complexity (class) of a $P$-verifiable theorem
is the minimum explicit construction complexity (class) over all
$P$-verifiable proofs of the theorem. 
Verification complexity (class) is defined similarly.
\end{defn}

The explicit construction complexity seems to be a good 
measure of complexity for $P$-verifiable proof techniques and theorems.
We shall discuss the examples above a bit more in this context.

\subsubsection*{\bf Halls' theorem}
Verification class here is $AC$ (constant depth circuits), since 
a perfect matching  can be verified in constant depth. 
A perfect matching in a bipartite graph can be computed, if one exists, in 
$O(m \log n)$ time. This 
problem also belongs to  $RNC$ \cite{kwig,vazirani}.
Hence, the sequential explicit construction class 
of Hall's theorem  is $Dtime(m\log n)$. The parallel 
explicit construction class is  $RNC$; possibly  even $NC$.

\subsubsection*{Four colour theorem}
Verification class here 
is  $AC$. Explicit construction complexity of the proof in \cite{appel} 
is $O(n^4)$, whereas that of the proof in \cite{robertson2} 
is $O(n^2)$ \cite{robertson2}. Thus 
a proof technique with lower explicit construction complexity has indeed
lower proof-complexity.
The sequential explicit construction class of the four colour theorem
is thus $Dtime(n^2)$, or lower. The parallel explicit construction class
is possibly $NC$, in view of the parallel algorithms for four colouring in
special cases \cite{he}.

\subsubsection*{Forbidden minor theorem}
Verification class is AC. Explicit construction complexity of the
proof in 
\cite{robertson} is $O(f(g) n^2)$, where $g$ is an explicit function of the
genus $g$. The sequential 
explicit construction class of the forbidden minor theorem is
thus $Dtime(O(n^2))$; it may be $Dtime(n)$.
The parallel explicit construction class may be $NC$, since planarity
testing is in $NC$ \cite{simon}.

\subsubsection*{Poincare's conjecture} 
Verification class of the Poincare conjecture 
is $P$ \cite{sphere}, assuming that the proof technique in
\cite{perelman} is $P$-verifiable. It may be smaller. $NC$?.
Explicit construction class may be  $P$, plausibly smaller.  $NC$?

\subsubsection*{Trivial example}
We now give a trivial example to illustrate why $P$-verifiability
is essential for the complexity measures here to make sense.
Take a trivial mathematical theorem: that an integer $n$ has at most 
$\log n$ factors. An obvious proof-certificate, for a given $n$,
is the number of its factors, which shows that it is less than $\log n$.
But  verification of this proof requires factoring and hence is hard. 
Thus if $n$ is specified in binary, this theorem should  not be $P$-verifiable.
That is why explicit construction complexity of this proof-certificate
says nothing of the actual (trivial) proof-complexity of the theorem.
Similarly,  explicit construction complexity is not  meaningful 
for  estimate-centred proof techniques in mathematics. The article 
\cite{rudich} roughly says that such techniques are not expected to work in
the context of the $P$ vs. $NP$ problem since they tend to be applicable to
a large fraction of functions.

\ignore{
If  we specify $n$ here in unary, then this this theorem is $P$-verifiable
and  its  explicit construction complexity seems to make some sense.
But  whether this is really  reasonable in this
and  other contexts in mathematics is not clear.}

In the context of the $P$ vs. $NP$ problem, 
Definitions~\ref{dexpcomplexity1} and \ref{dvericlass}
become:

\begin{defn} \label{dexpcomplexity2}
Explicit construction complexity  of a $P$-constructive technique for
the $P\not = NP$ conjecture is
the computational complexity  of the algorithm underlying
that technique
for explicit construction of a proof-certificate (as per that technique).
\end{defn}

\begin{defn} 

\noindent (a) Verification complexity class 
of a $P$-verifiable proof technique
for the $P\not = NP$ conjecture
is the computational complexity class  of the problem of verifying an
obstruction as per that technique.

\noindent (b) 
Its explicit construction complexity class  is
the computational complexity class of the problem of 
explicit construction of an obstruction.
\end{defn}

Again these classes are well-defined 
regardless of whether the technique shows how to construct an obstruction
explicitly or not, once the notion of an obstruction in the proof technique is
well-defined.

One may also define {\em existential complexity class} of a $P$-verifiable
proof technique (for the $P$ vs. $NP$ problem): this is the computational
complexity class of the problem of deciding if there exists an obstruction for
a given $n$ and circuit size $m$.

The existence-vs-construction principle \cite{kwig} says that 
computational complexity of a construction problem is comparable to that
of the associated existence problem under natural conditions.
This means, under natural conditions,
existential and explicit-construction complexity classes  should coincide.
Hence, we shall not worry about
existential complexity anymore.

It is illuminating to compare the verification complexity of 
the $P$ vs. $NP$ problem with  the other problems we considered.
The verification complexity class of Halls' theorem, four colour theorem, or
forbidden minor theorem is AC.
For Poincare's conjecture, $P$-verifiability is 
quite  nontrivial \cite{sphere}. But 
fortunately the proof is not very  complex.

In contrast, $P$-verifiability is  already a formidable 
issue in the context of the  $P$ vs. $NP$ problem.

\subsection{Is there a simpler proof technique?}
Now we ask if there is a $P$-verifiable 
proof technique towards the $P\not = NP$
conjecture that is substantially ``easier'' than GCT. By easier
we  mean, with lower  verification and explicit construction complexity 
(classes).
Since GCT is $P$-verifiable and also $P$-constructive over $\C$ as per 
Hypothesis~\ref{hphflipformalnew},
$P\not = NP$ conjecture is conjecturally $P$-verifiable and
also $P$-constructive  over $C$.
The same can be conjectured over $F_p$ or $\bar F_p$ as well \cite{GCT10}.
Assuming this, it is meaningful to talk of its verification
and explicit construction classes. So  we can ask:

\begin{question} 
What are  the (smallest) verification and explicit construction complexity 
classes of the $P\not = NP$ conjecture? 
\end{question} 

The best  and the most natural answer that  one can expect here is $P$.
It would really be unsettling if the answer were, say, $NC$.
Specifically,  the problems of verification and
explicit construction of obstructions 
in any $P$-verifiable approach to the $P\not = NP$ conjecture 
should be at least as hard as $P$-complete problems.
This is   supported 
by the presence of  linear programming,
which is $P$-complete,  in the
algorithms for the basic decision problems in Theorem~\ref{tmaincomplexity}. 

If so,  GCT may be among the ``easiest''  $P$-verifiable approaches to
the $P\not = NP$ conjecture over $\C$.
The story over $F_p$ may  be similar;  cf. \cite{GCT11}.


\begin{thebibliography}{[Welzl]}

\bibitem[A1]{agrawalprime} M. Agrawal, N. Kayal, N. Saxena, Primes is in P,
Annals of Mathematics, 160 (2): 781-793, 2004.


\bibitem[A2]{agrawal} M. Agrawal, Proving lower bounds via pseudo-random 
generators, Proceedings of FSTTCS 2005, 92-105, 2005.

\bibitem[Ak]{akhiezer}D. Akhiezer,  Homogeneous complex manifolds, Encyclopaedia of
mathematical sciences, volume 10, Springer-Verlag. 1986.

\bibitem[AH]{appel} K. Appel and W. Haken, Every planar map is four 
colorable, A.M.S. Contemporary Math. 98 (1989). MR 91m:05079.

\bibitem[B]{babai} L. Babai, private communication.

\bibitem[BGS]{solovay} T. Baker, J. Gill, R. Soloway, Relativization of the $P=?NP$ question, 
SIAM J. Comput. 4, 431-442, 1975.



\bibitem[BBD]{beilinson} A. Beilinson, J. Bernstein, P. Deligne, Faisceaux pervers, Ast\'erisque 
100, (1982), Soc. Math. France.


\bibitem[BZ]{berenstein} A. Berenstein, A. Zelevinsky, Tensor product multiplicities and 
convex polytopes in partition space, J. Geom. Phys. 5(3): 453-472, 1988.

\ignore{
\bibitem[BL]{lakshmibai} S. Billey, V. Lakshmibai, Singular Loci of Schubert 
varieties, Birkh\"auser, 2000.}

\bibitem[BS]{sipser} R. Boppana, M. Sipser: The complexity of finite functions,
Handbook of Theoretical Computer Science, vol. A,
Edited by
J. van Leeuwen, 
North Holland, Amsterdam, 1990, 757--804. 


\bibitem[Bou]{boutot} J. Boutot, Singularit'es rationelles et quotients par les 
groupes r'eductifs,
Invent. Math. 88, (1987), 65-68.

\bibitem[Co]{cook}  S. Cook: The complexity of theorem-proving procedures. 
Proceedings of the third annual ACM Symposium on Theory of Computing.
151-158. (1971). 

\bibitem[Dh]{dehy} R. Dehy, Combinatorial results on Demazure modules, J. of Algebra 205, 505-524 
(1998). 


\bibitem[Dl1]{deligne1} P. Deligne, J. Milne, Tannakien categories, Lecture 
notes in Math 900.

\bibitem[Dl2]{weil2} P. Deligne, La conjecture de Weil II, Publ. Math. Inst. Haut. \'Etud. Sci. 52,
(1980) 137-252. 



\bibitem[DM1]{deloeravertices} J. De Loera, T. McAllister, 
Vertices of Gelfand-Tsetlin polytopes, math.CO/0309329, Sept. 2003.

\bibitem[DM2]{loera} J. De Loera, T. McAllister, On the computation of 
Clebsch-Gordon coefficients and the dilation effect,
Experiment Math. 15, (2006), no. 1, 7-20. 



\bibitem[Der]{derkesen} H. Derkesen, J. Weyman, On the Littlewood-Richardson 
polynomials, J. Algebra 255 (2002), no. 2, 247-257.

\bibitem[Dri]{drinfeld} V. Drinfeld, Quantum groups, Proc. Int. Congr. Math. Berkeley, 1986,
vol. 1, Amer. Math. Soc. 1988, 798-820.


\bibitem[Ed]{edmonds} J. Edmonds, Maximum matching and a polyhedron with
$0-1$ vertices, Journal of Research of the National Bureau of Standards 
B 69, (1965), 125-130.

\bibitem[FMR]{miller} I. Filotti, G. Miller, J. Reif, On determining the genus of 
a graph in $O(v^{O(g)})$ steps, Proc. of the eleventh ACM Symposium on
the Theory of Computing, 1979.

\bibitem[Fl]{flenner} H. Flenner, Rationale quasihomogene Singularit\"aten,
Arch. Math. 36 (1981), 35-44.

\bibitem[F1]{fultonyoung} W. Fulton, Young tableaux, Cambridge University
Press, 1997.


\bibitem[F2]{fultonkly} W. Fulton, Eigenvalues of sums of Hermitian matrices (after 
A. Klyachko), S\'eminaire Bourbaki, vol. 1997/98. Ast\'erisque No. 2523
(1998), Exp. No. 845, 5, 255-269. 


\bibitem[FH]{fultonrepr} W. Fulton, J. Harris, Representation theory,
 A first course, Springer, 1991.

\bibitem[GCTabs]{GCTAbs} K. Mulmuley, Geometric complexity theory,
abstract, 
technical report TR-2007-12,
computer science dept., The university of Chicago, Sept. 2007.
Available at: http://ramakrishnadas.cs.uchicago.edu.


\bibitem[GCTconf]{GCTabs} K. Mulmuley, M. Sohoni, Geometric complexity theory,
P vs. NP and explicit obstructions, in ``Advances in Algebra and Geometry'', 
Edited by C. Musili, the proceedings of the International Conference on 
Algebra and Geometry, Hyderabad, 2001.


\bibitem[GCTflip2]{GCTflip2} K. Mulmuley,
 On P vs. NP, geometric complexity theory,
and the flip II, under preparation.

\bibitem[GCTintro]{GCTintro} K. Mulmuley, M. Sohoni,
Geometric complexity theory: introduction, 
technical report TR-2007-16,
computer science dept., The university of Chicago, Sept. 2007.
Available at: http://ramakrishnadas.cs.uchicago.edu.



\bibitem[GCT1]{GCT1} K. Mulmuley, M. Sohoni, Geometric complexity theory I:
an approach to the $P$ vs. $NP$ and related problems, 
SIAM J. Comput., vol 31, no 2, pp 496-526, 2001.

\bibitem[GCT2]{GCT2} K. Mulmuley, M. Sohoni, Geometric complexity theory II: 
towards explicit obstructions for embeddings among class varieties, 
to appear in SIAM J. Comput., 
cs. ArXiv preprint cs. CC/0612134, December 25, 2006.
Available at: http://ramakrishnadas.cs.uchicago.edu


\bibitem[GCT3]{GCT3} K. Mulmuley, M. Sohoni, Geometric complexity theory III,
on deciding positivity of Littlewood-Richardson coefficients, cs. ArXiv 
preprint cs. CC/0501076 v1 26 Jan 2005.


\bibitem[GCT4]{GCT4} K. Mulmuley, M. Sohoni, Geometric complexity theory IV: 
quantum group for the Kronecker problem, cs. ArXiv preprint cs. CC/0703110,
March, 2007. Available at:
http://ramakrishnadas.cs.uchicago.edu

\bibitem[GCT5]{GCT5} K. Mulmuley, H. Narayanan, Geometric complexity theory V:
on deciding nonvanishing of a generalized Littlewood-Richardson coefficient,
Technical report  TR-2007-05, Comp. Sci. Dept. The university of chicago,
May, 2007. 


\bibitem[GCT6]{GCT6} K. Mulmuley, Geometric complexity theory VI:
the flip via saturated  and 
positive integer programming in representation theory and algebraic 
geometry,  Technical report TR 2007-04, Comp. Sci. Dept., 
The University of Chicago, May, 2007.
Available at: http://ramakrishnadas.cs.uchicago.edu.
Revised version to be available here.

\bibitem[GCT7]{GCT7} K. Mulmuley, Geometric complexity theory VII:
a quantum group for the  plethysm problem,
technical report TR-2007-14,
computer science dept., The university of Chicago, Sept. 2007.
Available at: http://ramakrishnadas.cs.uchicago.edu.




\bibitem[GCT8]{GCT8} K. Mulmuley, Geometric complexity theory VIII: 
On canonical bases for the nonstandard quantum groups,
technical report TR-2007-15,
computer science dept., The university of Chicago, Sept. 2007.
Available at: http://ramakrishnadas.cs.uchicago.edu.



\bibitem[GCT9]{GCT9}  B. Adsul, M. Sohoni, K. Subrahmanyam,
Geometric complexity theory IX: algbraic and combinatorial
aspects of the Kronecker problem, under preparation.


\bibitem[GCT10]{GCT10} K. Mulmuley, Geometric complexity theory X:
On  class varieties, and the natural proof barrier, under preparation.

\bibitem[GCT11]{GCT11} K. Mulmuley, Geometric complexity theory XI:
on the flip over  finite or algebraically closed fields of  positive characteristic,
under preparation.

\bibitem[GL]{gro} I. Grojnowski, G. Lusztig, A comparison of bases of
quantized enveloping algebras, Contemp. Math. 153 (1993), 11-19.


\bibitem[GLS]{lovasz} M. Gr\"otschel, L. Lov\'asz, A. Schrijver, 
Geometric algorithms and combinatorial optimzation, Springer-Verlag,
1993.


\bibitem[Gu]{gurvits} L. Gurvits, private communication.

\bibitem[Ha]{hartshorne} R. Hartshorne, Algebraic geometry, Springer, 1997.

\bibitem[He]{he} X. He, Efficient parallel and sequential algorithms for
$4$-coloring perfect planar graphs, Algorithmica (1990) 5: 545-559.


\bibitem[Hi]{hironaka} H. Hironaka, Resolution of singularities of an 
algebraic variety over a field of characteristic zero, Annal of math. 79
(1964). I: 109-203, II: 205-326. 

\bibitem[IW]{wigderson} R. Impagliazzo, A. Wigderson, P=BPP if E requires
exponential size circuits: Derandomizing the XOR lemma, in proceedings of 
Annual ACM Symposium on the Theory of Computing, pages 220-229, 1997. 

\bibitem[JS]{simon} J. J\'aJ\'a, J. Simon., Parallel algorithms in graph
theory: planarity testing, SIAM J. Computing, 11 (2): 314-328, 1982.


\bibitem[Ji]{jimbo} M. Jimbo, A $q$-difference analogue of $U({\cal G})$
 and the Yang-Baxter equation,
Lett. Math. Phys. 10 (1985), 63-69.

\bibitem[KI]{kabanets} V. Kabanets, R. Impagliazzo, Derandomizing polynomial
identity tests means proving circuit lower bounds, in proceedings of Annual
ACM Symposium on the Theory of Computing, 355-364, 2003.


\bibitem[KB]{kannan} R. Kannan, A. Bachem, Polynomial algorithms for 
computing the Smith and Hermite normal forms of an integer matrix, 
SIAM J. comput., 8 (1979) 499-507.



\bibitem[Ka]{karp} R. Karp: Reducibility among combinatorial problems. 
R. E. Miller and J. W. Thatcher (eds.) Complexity of computer computations,
Plenum Press, New York, 1972, 85-103. 

\bibitem[KUW]{kwig} R. Karp, E. Upfal, A. Wigderson, Are search and 
decision problems computationally equivalent? Seventeenth Annual Symp. 
on Theory of Computing (1985). 



\bibitem[Kas1]{kashiwara1} M. Kashiwara, Crystalizing the $q$-analogue of universal enveloping
algebras, Comm. Math. Phys. 133 (1990), 249-260.

\bibitem[Kas2]{kashiwara2} M. Kashiwara, On crystal bases of the $q$-analogue of
 universal enveloping algebras, Duke Math. J. 63 (1991), 465-516.

\bibitem[Kas3]{kashiwaraglobal} M. Kashiwara, Global crystal bases of quantum groups, 
Duke Mathematical Journal, vol. 69, no.2, 455-485.

\bibitem[KL1]{kazhdan} D. Kazhdan, G. Lusztig, Representations of Coxeter groups
and Hecke algebras, Invent. Math. 53 (1979), 165-184.

\bibitem[KL2]{kazhdan1} D. Kazhdan, G. Lusztig, Schubert varieties and Poincare
duality, Proc. Symp. Pure Math., AMS, 36 (1980), 185-203. 



\bibitem[Ke]{kempf} G. Kempf, Vanishing theorems for flag manifolds,
American Journal of mathematics, vol. 98, No.2, pp-325-331.





\bibitem[Kh]{khachian}  L. Khachian, 
A polynomial algorithm in linear programming
(in Russian), Doklady Akad. Nauk SSSR 1979, t. 244, No. 5, 1093--1096.



\bibitem[KTT]{king} R. King, C. Tollu, F. Toumazet
Stretched Littlewood-Richardson coefficients and Kostka coefficients.
In, Winternitz, P., Harnard, J., Lam, C.S. and Patera, J. (eds.) 
Symmetry in Physics: In Memory of Robert T. Sharp. Providence, USA,
AMS  OUP, 99-112., CRM Proceedings and Lecture Notes 34, 2004.


\bibitem[Ki]{kirillov} A. Kirillov, An invitation to the generalized saturation
conjecture, math. CO/0404353., 20 Apr. 2004.


\bibitem[Kli]{klimyk} A. Klimyk, K. Schm\"udgen, Quantum groups and their representations, 
Springer, 1997.

\bibitem[Kl]{kly} A. Klyachko, Stable vector bundles and Hermitian operators, IGM, 
University of Marne-la-Vallee preprint (1994).


\bibitem[KT1]{knutson} A. Knutson, T. Tao, The Honeycomb model of $GL_n(\C)$
tensor products I: proof of the saturation conjecture, J. Amer. Math. Soc, 12,
1999, pp. 1055-1090.

\bibitem[KT2]{knutson2} A. Knutson, T. Tao, Honeycombs and sums of Hermitian matrices, 
Notices Amer. Math. Soc. 48, (2001) No. 2, 175186.

\bibitem[KM]{mayr2} K. K\"uhnle, E. Mayr, Exponential space computation of 
Gr\"obner bases, preprint, Technische Universit\"at M\"unchen, January 1996.



\bibitem[LLM]{smt} V. Lakshmibai, P. Littlemann, P. Magyar, Standard monomial 
theory and applications, in Representation theories and algebraic geometry,
(Ed.  A. Broer),  Kluwer Academic Publishers, (1997), 319-364.

\bibitem[LP]{larsen} M. Larsen, R. Pink,
Determining representations from invariant dimensions, Invent. math. 102,
377-398 (1990).


\bibitem[LLL]{lenstra} A. Lenstra, H. Lenstra, Jr., L. Lov'asz, Factoring
polynomials with rational coefficients, Mathematische Annalen 261 (1982),
515-534.



\bibitem[Le]{levin} A. Levin: Universal sequential search problems. 
Problems of information transmission  (translated from 
Problemy Peredachi Informatsii (Russian)) 9 (1973). 


\bibitem[Li]{littelmann} 
 P. Littelmann, Paths and root operators in representation theory,
Ann. of Math. 142 (1995), 499-525.

\bibitem[Lb]{lubot} A. Lubotzky, Discrete groups, expanding graphs, 
and invariant measures, Progress in mathematics, Boston,
Birkh\"auser, 1994.


\bibitem[LPS]{sarnak} A. Lubotzky, R. Phillips, P. Sarnak: Ramanujan graphs,
Combinatorica 8 (1988), 261-277. 

\bibitem[LV]{lunavust} D. Luna, Th. Vust, Plongements d'espaces homogenes,
Comment. Math. Helv.  58, 186 (1983). 


\bibitem[Lu1]{lusztigcanonical} G. Lusztig, Canonical bases arising from 
quantized enveloping algebras, J. Amer. Math. Soc. 3, (1990), 447-498.



\bibitem[Lu2]{lusztigbook} G. Lusztig,
Introduction to quantum groups, Birkh\"auser, 1993.



\bibitem[Mc]{macdonald} I. Macdonald, Symmetric functions and Hall polynomials,
Oxford science publications, Clarendon press, 1995.

\bibitem[Ma]{margulis} G. Margulis: Explicit constructions of 
concentrators, Problemy Inf. Trans. 9 (1973), 325-332.

\bibitem[MM]{mayr} E. Mayr, and A. Meyer, The complexity of the word
problems for commutative semigroups and polynomial ideals, 
Advances in mathematics, 46 (3): 305-329, 1982.

\bibitem[Mu1]{PRAM} K. Mulmuley, Lower bounds in a parallel model 
without bit operations, SIAM J. Comput., 28 (1999), 1460-1509.


\bibitem[MVV]{vazirani} K. Mulmuley, U. Vazirani, V. Vazirani, 
Matching is as easy as matrix inversion, Combinatorica, vol 7, no. 1, 1987,
pp 105-113.


\bibitem[Mm1]{mumford} D. Mumford, Algebraic Geometry I: complex projective
varieties, Springer, 1995.

\bibitem[Mm2]{mumford2}  D. Mumford, J. Fogarty, F. Kirwan: 
Geometric invariant theory. Springer-Verlag, 1994. 


\bibitem[N]{hari} H. Narayanan,
On the complexity of computing Kostka numbers and 
Littlewood-Richardson coefficients, J. of Algebraic combinatorics, vol. 24, issue 3, Nov. 2006.


\bibitem[NW]{nisan} N. Nisan, A. Wigderson, Hardness vs. randomness,
J. Comput. Sys. Sci., 49 (2): 149-167, 1994.

\bibitem[O]{orlin} J. Orlin, A faster strongly polynomial minimum cost flow
algorithm, proceedings of the twentieth Annual Symposium on Theory of 
Computing, 1988. 

\bibitem[Pe]{perelman} G. Perelman, The entropy formula for the Ricci flow
and its geometric applications, arXiv:math.DG/0211159.


\bibitem[Rs]{rassart} E. Rassart, A polynomiality property for
 Littlewood-Richardson coefficients, arXiv:math.CO/0308101, 16 Aug. 2003.

\bibitem[RR]{rudich} A. Razborov, S. Rudich, Natural proofs, J.
 Comput. System Sci., 55 (1997), pp. 24-35. 

\bibitem[Re]{reingold} O. Reingold, Undirected s-t-connectivity in logspace,
in proceedings of Annual ACM symposium on the Theory of Computing, 376-385,
2005.

\bibitem[RVW]{reingold2} O. Reingold, S. Vadhan, A. Wigderson, Entropy
waves, the zig-zag graph product and new constant-degree, Ann. of Math (2).
vol 155 (2002), no. 1, 157-187.


\bibitem[RTF]{rtf} N. Reshetikhin, L. Takhtajan, L. Faddeev, Quantization of Lie groups 
and Lie algebras, Leningrad Math. J., 1 (1990), 193-225. 

\bibitem[RSST]{robertson2} N. Robertson, D. Sanders, P. Seymour, R. Thomas,
A new proof of the four colour theorem, Electronic research announcement of 
the american mathematical society, vol. 2, number 1, August, 1996.


\bibitem[RS]{robertson} N. Robertson, P. Seymour, Graph minors. I. Excluding
a forest, Journal of Combinatorial theory, Series B 35 (1): 39-61. 



\bibitem[Sr]{sarnakbook} P. Sarnak, Some applications of modular forms,
Cambridge U. Press (1990).



\bibitem[Sc]{sphere} S. Schleimer, Sphere recognition lies in NP, 
arXiv:math/0407047v1, Jul, 2004.


\bibitem[Sc]{schrijver} A. Schrijver, Combinatorial optimization, Vol. A-C,
Springer, 2004.

\bibitem[Sp]{springer} T. Springer, 
Linear algebraic groups, in Algebraic Geometry IV,
Encyclopaedia of Mathematical Sciences, Springer-Verlag, 1989.



\bibitem[St1]{stanleyenu} R. Stanley, Enumerative combinatorics, vol. 1, 
Wadsworth and Brooks/Cole, Advanced Books and Software, 1986.




\bibitem[St4]{stanleypos} R. Stanley, Positivity problems and conjectures in algebraic 
combinatorics, manuscript, to appear in Mathematics: Frontiers and Perpsectives, 1999.





\bibitem[Ta]{tardos} E. Tardos, A strongly polynomial algorithm to solve
combinatorial linear programs, Operations Research 34 (1986), 250-256.

\bibitem[Sh]{shamir} A. Shamir, IP=PSPACE, Journal of the ACM, vol. 39, 
issue 4 ,October 1992.

\bibitem[St]{sturmfels} B. Sturmfels, Algorithms in invariant theory, 
Springer-Verlag, 1993.

\bibitem[Vd]{vaidya} P. Vaidya, A new algorithm for minimizing convex 
functions over convex sets, Mathematical Programming 73 (1996) 291-341.

\bibitem[V]{valiant} L. Valiant, The complexity of computing the permanent,
 Theoretical Computer
Science 8, pp 189-201, 1979.

\bibitem[Ve]{vempala} S. Vempala, Private communication.

\bibitem[W]{weyl} H. Weyl, Classical groups, Princeton University Press, 
1939.

\bibitem[Wo]{wor1} S. Woronowicz: Compact matrix pseudogroups, Commun. Math. Phys. 111 (1987), 613-665.

\bibitem[Z]{zelevinsky} A. Zelevinsky, Littlewood-Richardson semigroups, 
arXiv:math.CO/9704228 v1 30 Apr 1997.

\end{thebibliography}
\end{document}